\documentclass[a4paper,11pt]{article}
\pdfoutput=1 

\usepackage{jheppub} 

\usepackage[T1]{fontenc} 
\usepackage[english]{babel}
\usepackage{graphicx}			    	
\usepackage{slashed}
\usepackage{caption}
\usepackage{amsmath}
\usepackage{amsfonts}
\usepackage{mathtools}
\usepackage{subcaption}
\usepackage{tikz} 
\usepackage{tikz-feynman}
\usepackage{tikz-3dplot}
\tikzfeynmanset{compat=1.1.0}
\usepackage{placeins}
\usepackage{scalerel}
\usepackage{verbatim}
\usetikzlibrary{shapes.misc}
\tikzset{cross/.style={cross out, draw=black, minimum size=2*(#1-\pgflinewidth), inner sep=0pt, outer sep=0pt},
	cross/.default={5pt}}

\DeclareMathOperator{\Det}{Det}
\DeclareMathOperator{\Tr}{Tr}

\newcommand*\Laplace{\mathop{}\!\mathbin\bigtriangleup}

\makeatletter
\def\@fpheader{\relax}
\makeatother

\title{$n$-point correlators of twist-$2$ operators in $SU(N)$ Yang-Mills theory to the lowest perturbative order}

\author[a]{Marco Bochicchio}
\author[b,a]{Mauro Papinutto}
\author[b,a]{Francesco Scardino}

\affiliation[a]{Physics Department, INFN Roma1, \\
Piazzale A. Moro 2, Roma, I-00185, Italy}
\affiliation[b]{Physics Department, Sapienza University,\\Piazzale A. Moro 2, Roma, I-00185, Italy}

\emailAdd{marco.bochicchio@roma1.infn.it}
\emailAdd{mauro.papinutto@roma1.infn.it}
\emailAdd{francesco.scardino@roma1.infn.it}

\abstract{We compute, to the lowest perturbative order in $SU(N)$ Yang-Mills theory, $n$-point correlators in the coordinate and momentum representation
of the gauge-invariant twist-$2$ operators with maximal spin along the $p_+$ direction, both in Minkowskian and -- by analytic continuation -- Euclidean space-time.
We also construct the corresponding generating functionals. Remarkably, they have the structure of the logarithm of a functional determinant of the identity plus a term involving the effective propagators that act on the appropriate source fields.}

\begin{document} 

\definecolor{c969696}{RGB}{150,150,150}
\maketitle	
\flushbottom

\section{Introduction and physics motivations} \label{0}

In the present paper we compute, to the lowest perturbative order in $SU(N)$ Yang-Mills (YM) theory, $n$-point connected correlators, $G^{(n)}_{conf}(x_1,\ldots,x_n)$, in the coordinate representation of the gauge-invariant twist-$2$ operators with maximal spin along the $p_+$ direction, both in Minkowskian and -- by analytic continuation -- Euclidean space-time. \par
In fact, our computation matches and extends the previous lowest-order perturbative computation of $2$- and $3$-point gluonic correlators of twist-$2$ operators in $\mathcal{N}=4$ SUSY YM theory \cite{Kazakov:2012ar}, by including the unbalanced  \footnote{In our terminology 'unbalanced' and 'balanced' refers to either the different or  the equal number of dotted and undotted indices that the aforementioned operators possess in the spinorial representation respectively. Unbalanced operators are referred to as 'asymmetric' in \cite{Beisert:2004fv} and 'anisotropic' in \cite{Robertson:1990bf}.} operators with collinear twist $2$ in pure YM theory and, most importantly, by calculating all the $n$-point correlators in the balanced and unbalanced sectors separately, and the $3$-point correlators in the mixed sector as well.\par
Our physics motivation is threefold: \par
Firstly, our lowest-order computation has an intrinsic interest in YM theory, and -- according to \cite{Kazakov:2012ar} -- in theories that extends it, such as its supersymmetric versions and QCD.\par
Secondly, our computation is preliminary to work out the ultraviolet (UV) asymptotics \cite{Bochicchio:2013tfa, Bochicchio:2013eda} -- based on the renormalization-group (RG) improvement of perturbation theory -- of the above Euclidean $n$-point correlators.\par
Thirdly, our computation is an essential ingredient to test the prediction in section $3$ of \cite{Bochicchio:2016toi} that, by fundamental principles of the large-$N$ 't Hooft expansion, the generating functional of the nonperturbative leading nonplanar contributions to the aforementioned Euclidean correlators must have the structure of the logarithm of a functional determinant \cite{Bochicchio:2016toi} that sums the glueball one-loop diagrams.\par
Indeed, according to the philosophy of the asymptotically free bootstrap outlined in \cite{Bochicchio:2016toi}, the RG-improved correlators mentioned above must be asymptotic in the UV \cite{Bochicchio:2016toi}
to the corresponding nonperturbative correlators involving glueballs. Therefore, to the leading nonplanar order, the generating functional of the former must share with the one of the latter the very same structure of the logarithm of a functional determinant. \par
As an intermediate step for the program above, we construct the generating functionals of the aforementioned lowest-order $n$-point correlators.\par
Remarkably, they have the structure of the logarithm of a functional determinant of the identity plus a term involving the effective propagators that act on the appropriate source fields.\par
Hence, according to the argument above, our formulas for the generating functionals are as simple as they can be.\par
Incidentally, the generating functionals also allow us to compute straightforwardly the $n$-point correlators in the momentum representation, whose structure is slightly simpler than in the coordinate representation.

\section{Main results} \label{00}

\subsection{Balanced and unbalanced twist-$2$ conformal operators}

We describe our calculation and the operators that enter it. We compute, to the lowest perturbative order in $SU(N)$ YM theory, $n$-point connected correlators in Minkowskian space-time of the gauge-invariant twist-$2$ operators with maximal spin along the $p_+$ direction:
\begin{equation}
\langle \mathcal{O}_1(x_1)\ldots\mathcal{O}_n(x_n)\rangle_{lowest \, order} = G^{(n)}_{conf}(x_1,\ldots,x_n)
\end{equation}
It has been known for some time that, to the lowest order and the next one \footnote{In fact, to the order of $g^2$, in the conformal subtraction scheme \cite{Braun:2003rp}.}, YM theory is conformal invariant \cite{Braun:2003rp},
since the beta function only affects the solution of the Callan-Symanzik equation starting from the order of $g^4$. 
More recently, the exact conformal symmetry of QCD at the Wilson-Fisher critical point in $d = 4 - 2 \epsilon$ dimensions has been exploited \cite{Braun2,Braun3} \footnote{and references therein.} as a computational tool in higher orders of perturbation theory. \par
Therefore, following \cite{Braun:2003rp} we employ operators that have nice transformation properties with respect to the collinear conformal subgroup involving the coordinate $x^+$. Alternatively, operators can be also constructed with nice transformation properties \cite{Ohrndorf:1981qv,conformalops} with respect to the conformal group, whose suitably chosen components restrict to the aforementioned representations of the collinear conformal subgroup. \par
Primary conformal operators $\mathcal{O}_{j}(x)$, with collinear conformal spin $j= s+\frac{\tau}{2}$, where $\tau$ is the collinear twist and $s$ the collinear spin, i.e., the spin projected along the $p_+$ direction, transform under the action of the generators \cite{Braun:2003rp} of the collinear conformal algebra $SL(2,\mathbb{R})$:
\begin{align}
&\left[L_0,L_\mp\right]=\mp L_\mp  \\\nonumber
&\left[L_-,L_+\right]=-2L_0
\end{align}
according \cite{Braun:2003rp} to:
\begin{align}
\label{confcom}
&\left[L_+,\mathcal{O}_j(x)\right] = -\partial_{+}\mathcal{O}_j(x)\\\nonumber
&\left[L_-,\mathcal{O}_j(x)\right] = (x^{+\,2}\partial_++2jx^+)\mathcal{O}_j(x)\\\nonumber
&\left[L_0,\mathcal{O}_j(x)\right] = (x^+ \partial_++j)\mathcal{O}_j(x)
\end{align}
where in eq. \eqref{confcom} $x=(x^+,x^-,x^1,x^2)$ is restricted \cite{Braun:2003rp} to the line $x^-=x^1=x^2=0$. Their conformal descendants, $\partial_+^i \mathcal{O}_j(x)$, are obtained by taking derivatives with respect to $x^+$, and have the same $\tau$.\par  For a given canonical dimension $d=\tau+s$, the quasi-partonic \cite{BUKHVOSTOV1985601} operators have minimum $\tau$ and maximum $s$, with nice mixing under renormalization and conformal properties as above \cite{Ohrndorf:1981qv, conformalops, BUKHVOSTOV1985601, Braun:2003rp, Belitsky:2003sh, Belitsky:2004sc, Braun:2008ia}.
Their collinear twist $\tau$ does not necessarily coincide \cite{Braun:2003rp} with the twist $\mathcal T$ -- defined by $d=\mathcal T+S$, where $S$ is the spin -- that refers to the conformal group \cite{conformalops,Ohrndorf:1981qv} instead of the collinear subgroup.\par
In general, local gauge-invariant operators with $\mathcal T=2 $ provide the leading contribution to the OPE of two vector currents in massless QCD-like theories \footnote{By massless QCD-like theories we mean asymptotically free gauge theories that are massless to all orders of perturbation theory, such as QCD with massless quarks.} \cite{Braun1} \footnote{and references therein.} in Minkowskian space-time near the light-cone.\par
An infinite family of quasi-partonic operators is constructed as follows.
A composite gauge-covariant primary conformal operator, built by two elementary \footnote{In the present paper, we refer to the operators $\Phi_{j}(x)$ as elementary, since they play the role of elementary constituents, though they may actually be composite operators.}
gauge-covariant primary conformal operators $\Phi_{j_1},\Phi_{j_2}$, with collinear conformal spins $j_1, j_2$, has the form \cite{Braun:2003rp}:
\begin{equation}
\label{maximum2}
\mathcal{O}_l^{j_1 j_2}(x) = \Phi_{j_1}(x)(i\overrightarrow{D}_++i\overleftarrow{D}_+)^l P_l^{(2j_1-1,2j_2-1)}\left(\frac{\overrightarrow{D}_+-\overleftarrow{D}_+}{\overrightarrow{D}_++\overleftarrow{D}_+}\right)\Phi_{j_2}(x)
\end{equation}
where $P_l^{(2j_1-1,2j_2-1)}$ are Jacobi polynomials (appendix \ref{appB}), $D_+$ is the covariant derivative along the $p_+$ direction (appendix \ref{appN}), and the arrows denote the action of the derivative on the right or the left. The corresponding gauge-invariant object is obtained by taking the color trace.\par
The collinear conformal spin, $j$, of the operator, $\mathbb{O}_l^{j_1 j_2}(x)$, is $j =j_1+j_2+l$, where $l$ is the power of the derivative in eq. \eqref{maximum2}.
By working out the definition in eq \eqref{maximum2}, we get:
\begin{align}
\label{confexp}
\nonumber
\mathcal{O}_l^{j_1j_2}(x) 
&=\sum_{k = 0}^{l} {l+2j_1-1\choose k}{l+2j_2-1\choose k+2j_2-1}(-1)^{l-k} \Phi_{j_1}(x) \overleftarrow{D}_+^{l-k} \overrightarrow{D}_+^k\Phi_{j_2}(x)\\
&=\sum_{k = 0}^{l} \mathcal{O}_{l k}^{j_1 j_2}(x)
\end{align}
thus realizing the conformal operator $\mathcal{O}_l^{j_1j_2}(x)$ as a sum of $l+1$ operators, $\mathcal{O}_{l k}^{j_1 j_2}(x)$, that are not necessarily conformal.\par
Hence, the composite operators depend on a choice of the elementary conformal operators $\Phi_{j_1},\Phi_{j_2}$.
We define the standard conformal basis for primary operators with collinear twist $2$, where the elementary operators are $f_{11},f_{\dot{1}\dot{1}}$ (section \ref{standardb}) with conformal spin $j=\frac{3}{2}$. In the standard basis the gluonic operators are classified as in \cite{Ohrndorf:1981qv, conformalops,Beisert:2004fv}:
\begin{align} \label{OO}
\nonumber
&\mathbb{O}_{s} = \Tr f_{11}(x)(i\overrightarrow{D}_++i\overleftarrow{D}_+)^{s-2}C^{\frac{5}{2}}_{s-2}\left(\frac{\overrightarrow{D}_+-\overleftarrow{D}_+}{\overrightarrow{D}_++\overleftarrow{D}_+}\right) f_{\dot{1}\dot{1}}(x) \qquad s = 2,4,6,\ldots  \\\nonumber
&\tilde{\mathbb{O}}_{s} = \Tr f_{11}(x)(i\overrightarrow{D}_++i\overleftarrow{D}_+)^{s-2}C^{\frac{5}{2}}_{s-2}\left(\frac{\overrightarrow{D}_+-\overleftarrow{D}_+}{\overrightarrow{D}_++\overleftarrow{D}_+}\right) f_{\dot{1}\dot{1}}(x) \qquad s = 3,5,7,\ldots  \\\nonumber
&\mathbb{S}_{s} =\frac{1}{\sqrt{2}}\Tr f_{11}(x)(i\overrightarrow{D}_++i\overleftarrow{D}_+)^{s-2}C^{\frac{5}{2}}_{s-2}\left(\frac{\overrightarrow{D}_+-\overleftarrow{D}_+}{\overrightarrow{D}_++\overleftarrow{D}_+}\right)f_{11}(x) \qquad s = 2,4,6,\ldots\\
&\bar{\mathbb{S}}_{s} =\frac{1}{\sqrt{2}}\Tr f_{\dot{1}\dot{1}}(x)(i\overrightarrow{D}_++i\overleftarrow{D}_+)^{s-2}C^{\frac{5}{2}}_{s-2}\left(\frac{\overrightarrow{D}_+-\overleftarrow{D}_+}{\overrightarrow{D}_++\overleftarrow{D}_+}\right) f_{\dot{1}\dot{1}}(x)\qquad s = 2,4,6,\ldots
\end{align}
by restricting the appropriate conformal multiplet \cite{Ohrndorf:1981qv, conformalops} to the components along the $p_+$ direction, with $C^{\alpha}_{l}$ Gegenbauer polynomials (appendix \ref{appB}), which are a special case of Jacobi polynomials.\par
$\mathbb{O}_{s}$ and $\tilde{\mathbb{O}}_s$ are Hermitian balanced operators with $\tau=\mathcal T=2$. They have an equal number of undotted and dotted spinor indices (appendices \ref{appA200} and \ref{appA2}):
\begin{align}
\nonumber
&\mathbb{O}_s = \mathbb{O}_{1\dot{1}\ldots 1\dot{1}}\\
&\tilde{\mathbb{O}}_s = \tilde{\mathbb{O}}_{1\dot{1} \ldots 1 \dot{1}}
\end{align}
$\mathbb{S}_{s}$ and its Hermitian conjugate, $\bar{\mathbb{S}}_{s}$, denoted by the bar superscript, are unbalanced operators with $\tau=2$. 
They have a different number of undotted and dotted spinor indices:
\begin{align}
\nonumber
&\mathbb{S}_s = \mathbb{S}_{1111\ldots 1\dot{1}}\\
&\bar{\mathbb{S}}_s= \mathbb{S}_{\dot{1}\dot{1}\dot{1}\dot{1} \ldots \dot{1} 1}
\end{align}
Besides, we also define the extended conformal basis for primary operators with collinear twist $2$, where the elementary operators are $D_{+}^{-1}f_{11}, D_{+}^{-1}f_{\dot{1}\dot{1}}$, with conformal spin $j=\frac{1}{2}$, which are nonlocal in general, but local (appendix \ref{appC}) in the light-cone gauge $A_+=0$. Clearly, gauge invariance ensures that all their correlators are local, as we verify explicitly. \par
The extended basis is natural in SUSY calculations \cite{Belitsky:2004sc}, and includes (nonlocal) operators with $\tau=2$ and $s=0,1$. We have chosen it in YM theory because of the simplicity of the results for the correlators.
In the extended basis (section \ref{nonstandardb}) the gluonic operators are:
\begin{align} \label{AA}
\nonumber
&\mathbb{A}_{s} = \Tr D_{+}^{-1}f_{11}(x)\,(i\overrightarrow{D}_++i\overleftarrow{D}_+)^{s}C^{\frac{1}{2}}_{s}\left(\frac{\overrightarrow{D}_+-\overleftarrow{D}_+}{\overrightarrow{D}_++\overleftarrow{D}_+}\right)D_{+}^{-1} f_{\dot{1}\dot{1}}(x)  \qquad s = 0,2,4,\ldots\\\nonumber
&\tilde{\mathbb{A}}_{s} = \Tr D_{+}^{-1}f_{11}(x)\,(i\overrightarrow{D}_++i\overleftarrow{D}_+)^{s}C^{\frac{1}{2}}_{s}\left(\frac{\overrightarrow{D}_+-\overleftarrow{D}_+}{\overrightarrow{D}_++\overleftarrow{D}_+}\right) D_{+}^{-1} f_{\dot{1}\dot{1}}(x)  \qquad s = 1,3,5,\ldots\\\nonumber
&\mathbb{B}_{s} =\frac{1}{\sqrt{2}}\Tr D_{+}^{-1}f_{11}(x)\,(i\overrightarrow{D}_++i\overleftarrow{D}_+)^{s}C^{\frac{1}{2}}_{s}\left(\frac{\overrightarrow{D}_+-\overleftarrow{D}_+}{\overrightarrow{D}_++\overleftarrow{D}_+}\right)D_{+}^{-1}f_{11}(x)\qquad s = 0,2,4,\ldots\\
&\bar{\mathbb{B}}_{s} =\frac{1}{\sqrt{2}}\Tr D_{+}^{-1} f_{\dot{1}\dot{1}}(x)\,(i\overrightarrow{D}_++i\overleftarrow{D}_+)^{s}C^{\frac{1}{2}}_{s}\left(\frac{\overrightarrow{D}_+-\overleftarrow{D}_+}{\overrightarrow{D}_++\overleftarrow{D}_+}\right)D_{+}^{-1} f_{\dot{1}\dot{1}}(x) \qquad s = 0,2,4,\ldots
\end{align}

\subsection{Minkowskian $n$-point correlators in the coordinate representation}

\subsubsection{Standard basis}

We have normalized our operators in such a way that the $2$-point correlators in the standard basis are equal for even $s$:
\begin{align}
&\langle {\mathbb{O}}_{s_1}(x) {\mathbb{O}}_{s_2}(y)\rangle = 
\langle \mathbb{S}_{s_1}(x)\bar{\mathbb{S}}_{s_2}(y)\rangle  =  \mathcal{C}_{s_1}(x,y) \delta_{s_1 s_2} 
\end{align}
and for odd $s$:
\begin{align}
\langle \tilde{\mathbb{O}} _{s_1}(x)\tilde{\mathbb{O}} _{s_2}(y)\rangle =  \mathcal{C}_{s_1}(x,y) \delta_{s_1 s_2}
\end{align}
with:
\begin{align}
\nonumber
\mathcal{C}_{s}(x,y) =&\frac{1}{(4\pi^2)^2} \frac{N^2-1}{4} \frac{2^{2s+2}i^{2s-4}}{(4!)^2}(s+1)^2(s+2)^2
\frac{(x-y)_+^{2s}}{(\rvert x-y\rvert^2)^{2s+2}}\\\nonumber
&\sum_{k_1 = 0}^{s-2}\sum_{k_2 = 0}^{s-2}{s\choose k_1}{s\choose k_1+2}{s\choose k_2}{s\choose k_2+2}(-1)^{s-k_2+k_1}\\\nonumber
&(s-k_1+k_2)!(s+k_1-k_2)! \\\nonumber
=&\frac{1}{(4\pi^2)^2} \frac{N^2-1}{4}\frac{2^{2s+2}i^{2s-4}}{(4!)^2}(s+1)^2(s+2)^2
 (2s)! \frac{(x-y)_+^{2s}}{(\rvert x-y\rvert^2)^{2s+2}}\\
&\sum_{k_1 = 0}^{s-2}\sum_{k_2 = 0}^{s-2}{s\choose k_1}{s\choose k_1+2}{s\choose k_2}{s\choose k_2+2}(-1)^{k_2+k_1} \frac{1}{{2s\choose k_1+k_2+2}}
\end{align}
where we omit the $i\epsilon$ prescription in the propagators in the coordinate representation, in such a way that (appendix \ref{appN}): 
\begin{equation}
\frac{1}{\rvert x\rvert^2} 
\end{equation}
should be read (appendix \ref{appA1}):
\begin{equation}
\frac{1}{\rvert x\rvert^2-i\epsilon}
\end{equation}
The very same correlators are evaluated by a trick \cite{Kazakov:2012ar} (appendix \ref{appA3}):
\begin{align}
\label{c2intro}
\mathcal{C}_{s}(x,y) = &\frac{1}{(4\pi^2)^2} \frac{N^2-1}{4} \frac{2^{2s+2}}{(4!)^2}(-1)^s(s-1)s(s+1)(s+2)(2s)!
\frac{(x-y)_+^{2s}}{(\rvert x-y\rvert^2)^{2s+2}}
\end{align}
Therefore, we have discovered the following -- seemingly nontrivial -- identity (section \ref{coordinates}):
\begin{align}
\frac{s_1(s_1-1)}{(s_1+1)(s_1+2)}\delta_{s_1 s_2}=
\sum_{k_1 = 0}^{s_1-2}\sum_{k_2 = 0}^{s_2-2}{s_1\choose k_1}{s_1\choose k_1+2}{s_2\choose k_2}{s_2\choose k_2+2}(-1)^{k_2+k_1}\frac{1}{{s_1+s_2\choose k_1+k_2+2}}
\end{align}  
We have not found a direct proof of the above identity, but we have verified it numerically. \par
Moreover, the only nonvanishing $3$-point correlators are:
\begin{align}
\label{defc3}
\langle {\mathbb{O}}_{s_1}(x){\mathbb{O}}_{s_2}(y){\mathbb{O}}_{s_3}(z)\rangle=\langle {\mathbb{O}}_{s_1}(x){\mathbb{S}}_{s_2}(y)\bar{\mathbb{S}}_{s_3}(z)\rangle  = \mathcal{C}_{s_1s_2s_3}(x,y,z)
\end{align}
and:
\begin{align}
\label{defc3}
\langle {\mathbb{O}}_{s_1}(x)\tilde{\mathbb{O}}_{s_2}(y)\tilde{\mathbb{O}}_{s_3}(z)\rangle = \mathcal{C}_{s_1s_2s_3}(x,y,z)
\end{align}
with:
\begin{align} \label{312}
\nonumber
 \mathcal{C}_{s_1s_2s_3}(x,y,z) 
=&-\frac{1}{(4\pi^2)^3} (1+(-1)^{s_1+s_2+s_3})   \left(\frac{2}{4!}\right)^3\frac{N^2-1}{8}i^{s_1+s_2+s_3}2^{s_1+s_2+s_3}\\\nonumber
& (s_1+1)(s_1+2)(s_2+1)(s_2+2)(s_3+1)(s_3+2)\\\nonumber
&\sum_{k_1 = 0}^{s_1-2}\sum_{k_2 = 0}^{s_2-2}\sum_{k_3 = 0}^{s_3-2}{s_1\choose k_1}{s_1\choose k_1+2}{s_2\choose k_2}{s_2\choose k_2+2}{s_3\choose k_3}{s_3\choose k_3+2}\\\nonumber
&(s_1-k_1+k_2)!(s_2-k_2+k_3)!(s_3-k_3+k_1)!  \\
&\frac{(x-y)^{s_1-k_1+k_2}_+}{(\rvert x-y\rvert^2)^{s_1+1-k_1+k_2}}\frac{(y-z)^{s_2-k_2+k_3}_+}{(\rvert y-z\rvert^2)^{s_2+1-k_2+k_3}}\frac{(z-x)^{s_3-k_3+k_1}_+}{(\rvert z-x\rvert^2)^{s_3+1-k_3+k_1}}
\end{align}
We also compute the $n$-point correlators. In the balanced sector, we get:
\begin{align} \label{O0}
\nonumber
&\langle \mathbb{O}_{s_1}(x_1)\ldots \mathbb{O}_{s_n}(x_n)\rangle_{conn}=\frac{1}{(4\pi^2)^n}\frac{N^2-1}{2^n}2^{\sum_{l=1}^n s_l}i^{\sum_{l=1}^n s_l} \\\nonumber
&\frac{\Gamma(3)\Gamma(s_1+3)}{\Gamma(5)\Gamma(s_1+1)}\ldots \frac{\Gamma(3)\Gamma(s_n+3)}{\Gamma(5)\Gamma(s_n+1)}\sum_{k_1=0}^{s_1-2}\ldots \sum_{k_n = 0}^{s_n-2}{s_1\choose k_1}{s_1\choose k_1+2}\ldots {s_n\choose k_n}{s_n\choose k_n+2}\\\nonumber
&\frac{(-1)^n}{n}\sum_{\sigma\in P_n}(s_{\sigma(1)}-k_{\sigma(1)}+k_{\sigma(2)})!\ldots(s_{\sigma(n)}-k_{\sigma(n)}+k_{\sigma(1)})!\\
&\frac{(x_{\sigma(1)}-x_{\sigma(2)})_+^{s_{\sigma(1)}-k_{\sigma(1)}+k_{\sigma(2)}}}{\left(\rvert x_{\sigma(1)}-x_{\sigma(2)}\rvert^2\right)^{s_{\sigma(1)}-k_{\sigma(1)}+k_{\sigma(2)}+1}}\ldots\frac{(x_{\sigma(n)}-x_{\sigma(1)})_+^{s_{\sigma(n)}-k_{\sigma(n)}+k_{\sigma(1)}}}{\left(\rvert x_{\sigma(n)}-x_{\sigma(1)}\rvert^2\right)^{s_{\sigma(n)}-k_{\sigma(n)}+k_{\sigma(1)}+1}}
\end{align}
The very same formula holds for an even number of operators $\tilde{\mathbb{O}}_s$, otherwise the correlators vanish.
The nonvanishing correlators in the balanced sector are:
\begin{align} \label{O}
\nonumber
&\langle \mathbb{O}_{s_1}(x_1)\ldots \mathbb{O}_{s_n}(x_n)\tilde{\mathbb{O}}_{s_{n+1}}(x_{n+1})\ldots \tilde{\mathbb{O}}_{s_{n+2m}}(x_{n+2m})\rangle_{conn}\\\nonumber
& =\frac{1}{(4\pi^2)^{n+2m}}\frac{N^2-1}{2^{n+2m}}2^{\sum_{l=1}^{n+2m} s_l}i^{\sum_{l=1}^{n+2m} s_l}\frac{\Gamma(3)\Gamma(s_1+3)}{\Gamma(5)\Gamma(s_1+1)} \ldots\frac{\Gamma(3)\Gamma(s_{n+2m}+3)}{\Gamma(5)\Gamma(s_{n+2m}+1)}\\\nonumber
&\sum_{k_1=0}^{s_1-2}\ldots \sum_{k_{n+2m} = 0}^{s_{n+2m}-2}{s_1\choose k_1}{s_1\choose k_1+2}\ldots{s_{n+2m}\choose k_{n+2m}}{s_{n+2m}\choose k_{n+2m}+2}\\\nonumber
&\frac{(-1)^{n+2m}}{n+{2m}}\sum_{\sigma\in P_{n+2m}}(s_{\sigma(1)}-k_{\sigma(1)}+k_{\sigma(2)})!\ldots(s_{\sigma(n+{2m})}-k_{\sigma(n+{2m})}+k_{\sigma(1)})!\\
&\frac{(x_{\sigma(1)}-x_{\sigma(2)})_+^{s_{\sigma(1)}-k_{\sigma(1)}+k_{\sigma(2)}}}{\left(\rvert x_{\sigma(1)}-x_{\sigma(2)}\rvert^2\right)^{s_{\sigma(1)}-k_{\sigma(1)}+k_{\sigma(2)}+1}}\ldots\frac{(x_{\sigma(n+2m)}-x_{\sigma(1)})_+^{s_{\sigma(n+{2m})}-k_{\sigma(n+{2m})}+k_{\sigma(1)}}}{\left(\rvert x_{\sigma(n+{2m})}-x_{\sigma(1)}\rvert^2\right)^{s_{\sigma(n+{2m})}-k_{\sigma(n+{2m})}+k_{\sigma(1)}+1}}\,
\end{align}
In the unbalanced sector, we get:
\begin{align} \label{S}
\nonumber
&\langle \mathbb{S}_{s_1}(x_1)\ldots \mathbb{S}_{s_n}(x_n)\bar{\mathbb{S}}_{{s'\!\!}_1}(y_1)\ldots \bar{\mathbb{S}}_{{s'\!\!}_n}(y_n)\rangle=\frac{1}{(4\pi^2)^{2n}}\frac{N^2-1}{2^{2n}}2^{\sum_{l=1}^n s_l+{s'\!\!}_l}i^{\sum_{l=1}^n s_l+{s'\!\!}_l}\\\nonumber
&\frac{\Gamma(3)\Gamma(s_1+3)}{\Gamma(5)\Gamma(s_1+1)}\ldots \frac{\Gamma(3)\Gamma(s_n+3)}{\Gamma(5)\Gamma(s_n+1)}\frac{\Gamma(3)\Gamma({s'\!\!}_1+3)}{\Gamma(5)\Gamma({s'\!\!}_1+1)}\ldots \frac{\Gamma(3)\Gamma({s'\!\!}_n+3)}{\Gamma(5)\Gamma({s'\!\!}_n+1)}\\\nonumber
&\sum_{k_1=0}^{s_1-2}\ldots \sum_{k_n = 0}^{s_n-2}{s_1\choose k_1}{s_1\choose k_1+2}\ldots {s_n\choose k_n}{s_n\choose k_n+2}\\\nonumber
&\sum_{{k'\!\!}_1=0}^{{s'\!\!}_1-2}\ldots \sum_{{k'\!\!}_n = 0}^{{s'\!\!}_n-2}{{s'\!\!}_1\choose {k'\!\!}_1}{{s'\!\!}_1\choose {k'\!\!}_1+2}\ldots {{s'\!\!}_n\choose {k'\!\!}_n}{{s'\!\!}_n\choose {k'\!\!}_n+2}\\\nonumber
&\frac{2^{n-1}}{n}\sum_{\sigma\in P_n}\sum_{\rho\in P_n}
(s_{\sigma(1)}-k_{\sigma(1)}+{k'\!\!}_{\rho(1)})!({s'\!\!}_{\rho(1)}-{k'\!\!}_{\rho(1)}+k_{\sigma(2)})!\\\nonumber
&\ldots(s_{\sigma(n)}-k_{\sigma(n)}+{k'\!\!}_{\rho(n)})!({s'\!\!}_{\rho(n)}-{k'\!\!}_{\rho(n)}+k_{\sigma(1)})!\\\nonumber
&\frac{(x_{\sigma(1)}-y_{\rho(1)})_+^{s_{\sigma(1)}-k_{\sigma(1)}+{k'\!\!}_{\rho(1)}}}{\left(\rvert x_{\sigma(1)}-y_{\rho(1)}\rvert^2\right)^{s_{\sigma(1)}-k_{\sigma(1)}+{k'\!\!}_{\rho(1)}+1}}\frac{(y_{\rho(1)}-x_{\sigma(2)})_+^{{s'\!\!}_{\rho(1)}-{k'\!\!}_{\rho(1)}+k_{\sigma(2)}}}{\left(\rvert y_{\rho(1)}-x_{\sigma(2)}\rvert^2\right)^{{s'\!\!}_{\rho(1)}-{k'\!\!}_{\rho(1)}+k_{\sigma(2)}+1}}\\
&\ldots\frac{(x_{\sigma(n)}-y_{\rho(n)})_+^{s_{\sigma(n)}-k_{\sigma(n)}+{k'\!\!}_{\rho(n)}}}{\left(\rvert x_{\sigma(n)}-y_{\rho(n)}\rvert^2\right)^{s_{\sigma(n)}-k_{\sigma(n)}+{k'\!\!}_{\rho(n)}+1}}
\frac{(y_{\rho(n)}-x_{\sigma(1)})_+^{{s'\!\!}_{\rho(n)}-{k'\!\!}_{\rho(n)}+k_{\sigma(1)}}}{\left(\rvert y_{\rho(n)}-x_{\sigma(1)}\rvert^2\right)^{{s'\!\!}_{\rho(n)}-{k'\!\!}_{\rho(n)}+k_{\sigma(1)}+1}}
\end{align}

\subsubsection{Extended basis}

We normalize our operators in such a way that the $2$-point correlators in the extended basis are equal for even $s$:
\begin{align}
&\langle {\mathbb{A}}_{s_1}(x) {\mathbb{A}}_{s_2}(y)\rangle = \langle \bar{\mathbb{B}}_{s_1}(x)\bar{\mathbb{B}} _{s_2}(y)\rangle
 =  \mathcal{A}_{s_1}(x,y) \delta_{s_1 s_2} 
\end{align}
and for odd $s$:
\begin{align}
\langle \tilde{\mathbb{A}}_{s_1}(x)\tilde{\mathbb{A}}_{s_2}(y)\rangle  =  \mathcal{A}_{s_1}(x,y) \delta_{s_1 s_2}
\end{align}
with:
\begin{align}
\nonumber
\mathcal{A}_{s}(x,y) =& \frac{1}{(4\pi^2)^2} \frac{N^2-1}{4} 2^{2s}i^{2s}
\frac{(x-y)_+^{2s}}{(\rvert x-y\rvert^2)^{2s+2}}\\\nonumber
&\sum_{k_1 = 0}^{s}\sum_{k_2 = 0}^{s}{s\choose k_1}{s\choose k_1}{s\choose k_2}{s\choose k_2}(-1)^{s-k_2+k_1}
(s-k_1+k_2)!(s+k_1-k_2)!\\\nonumber
=&\frac{1}{(4\pi^2)^2} \frac{N^2-1}{4} 2^{2s}i^{2s}
 (2s)! \frac{(x-y)_+^{2s}}{(\rvert x-y\rvert^2)^{2s+2}}\\
&\sum_{k_1 = 0}^{s}\sum_{k_2 = 0}^{s}{s\choose k_1}{s\choose k_1}{s\choose k_2}{s\choose k_2}(-1)^{k_2+k_1}\frac{1}{{2s\choose k_1+k_2}}
\end{align}
The very same correlators are evaluated by a trick \cite{Kazakov:2012ar} (appendix \ref{appA3}):
\begin{align}
 \mathcal{A}_{s}(x,y)
= &\frac{1}{(4\pi^2)^2} \frac{N^2-1}{4} 2^{2s} (-1)^s (2s)!
\frac{(x-y)_+^{2s}}{(\rvert x-y\rvert^2)^{2s+2}}
\end{align}
Therefore, we have discovered the following -- seemingly nontrivial -- identity (section \ref{coordinates}):
\begin{align}
\delta_{s_1s_2}= 
\sum_{k_1 = 0}^{s_1}\sum_{k_2 = 0}^{s_2}{s_1\choose k_1}{s_1\choose k_1}{s_2\choose k_2}{s_2\choose k_2}(-1)^{k_2+k_1}\frac{1}{{s_1+s_2\choose k_1+k_2}}
\end{align}
We have not found a direct proof of the above identity, but we have verified it numerically. \par
Moreover, the only nonvanishing $3$-point correlators are:
\begin{align}
\label{defc3}
\langle {\mathbb{A}}_{s_1}(x){\mathbb{A}}_{s_2}(y){\mathbb{A}}_{s_3}(z)\rangle=\langle {\mathbb{A}}_{s_1}(x){\mathbb{B}}_{s_2}(y)\bar{\mathbb{B}}_{s_3}(z)\rangle= \mathcal{A}_{s_1s_2s_3}(x,y,z)
\end{align}
and:
\begin{align}
\label{defc3}
\langle {\mathbb{A}}_{s_1}(x)\tilde{\mathbb{A}}_{s_2}(y)\tilde{\mathbb{A}}_{s_3}(z)\rangle  = \mathcal{A}_{s_1s_2s_3}(x,y,z)
\end{align}
with:
\begin{align}
\nonumber
 \mathcal{A}_{s_1s_2s_3}(x,y,z)
 =&-\frac{1}{(4\pi^2)^3} (1+(-1)^{s_1+s_2+s_3})  \frac{N^2-1}{8}i^{s_1+s_2+s_3}2^{s_1+s_2+s_3}\\\nonumber
&\sum_{k_1 = 0}^{s_1}\sum_{k_2 = 0}^{s_2}\sum_{k_3 = 0}^{s_3}{s_1\choose k_1}{s_1\choose k_1}{s_2\choose k_2}{s_2\choose k_2}{s_3\choose k_3}{s_3\choose k_3}\\\nonumber
&(s_1-k_1+k_2)!(s_2-k_2+k_3)!(s_3-k_3+k_1)!  \\\nonumber
&\frac{(x-y)^{s_1-k_1+k_2}_+}{(\rvert x-y\rvert^2)^{s_1+1-k_1+k_2}}\frac{(y-z)^{s_2-k_2+k_3}_+}{(\rvert y-z\rvert^2)^{s_2+1-k_2+k_3}}\frac{(z-x)^{s_3-k_3+k_1}_+}{(\rvert z-x\rvert^2)^{s_3+1-k_3+k_1}}\nonumber \\
\end{align}
We also compute the $n$-point correlators. In the balanced sector, we get:
\begin{align} \label{A0}
\nonumber
&\langle \mathbb{A}_{s_1}(x_1)\ldots \mathbb{A}_{s_n}(x_n)\rangle_{conn} =\frac{1}{(4\pi^2)^n}\frac{N^2-1}{2^n}2^{\sum_{l=1}^n s_l}i^{\sum_{l=1}^n s_l}\\\nonumber
&\sum_{k_1=0}^{s_1}\ldots \sum_{k_n = 0}^{s_n}{s_1\choose k_1}^2\ldots {s_n\choose k_n}^2\frac{(-1)^n}{n}\sum_{\sigma\in P_n}(s_{\sigma(1)}-k_{\sigma(1)}+k_{\sigma(2)})!\ldots(s_{\sigma(n)}-k_{\sigma(n)}+k_{\sigma(1)})!\\
&\frac{(x_{\sigma(1)}-x_{\sigma(2)})_+^{s_{\sigma(1)}-k_{\sigma(1)}+k_{\sigma(2)}}}{\left(\rvert x_{\sigma(1)}-x_{\sigma(2)}\rvert^2\right)^{s_{\sigma(1)}-k_{\sigma(1)}+k_{\sigma(2)}+1}}\ldots\frac{(x_{\sigma(n)}-x_{\sigma(1)})_+^{s_{\sigma(n)}-k_{\sigma(n)}+k_{\sigma(1)}}}{\left(\rvert x_{\sigma(n)}-x_{\sigma(1)}\rvert^2\right)^{s_{\sigma(n)}-k_{\sigma(n)}+k_{\sigma(1)}+1}}
\end{align}
The very same formula holds for an even number of operators $\tilde{\mathbb{A}}_s$, otherwise the correlators vanish. 
The nonvanishing correlators in the balanced sector are:
\begin{align} \label{A}
\nonumber
&\langle \mathbb{A}_{s_1}(x_1)\ldots \mathbb{A}_{s_n}(x_n)\tilde{\mathbb{A}}_{s_{n+1}}(x_{n+1})\ldots \tilde{\mathbb{A}}_{s_{n+2m}}(x_{n+2m})\rangle_{conn} \\\nonumber
&=\frac{1}{(4\pi^2)^{n+2m}}\frac{N^2-1}{2^{n+2m}}2^{\sum_{l=1}^{n+2m} s_l}i^{\sum_{l=1}^{n+2m} s_l}\sum_{k_1=0}^{s_1}\ldots \sum_{k_{n+2m} = 0}^{s_{n+2m}}{s_1\choose k_1}^2\ldots{s_{n+2m}\choose k_{n+2m}}^2\\\nonumber
&\frac{(-1)^{n+2m}}{n+{2m}}\sum_{\sigma\in P_{n+2m}}(s_{\sigma(1)}-k_{\sigma(1)}+k_{\sigma(2)})!\ldots(s_{\sigma(n+{2m})}-k_{\sigma(n+{2m})}+k_{\sigma(1)})!\\
&\frac{(x_{\sigma(1)}-x_{\sigma(2)})_+^{s_{\sigma(1)}-k_{\sigma(1)}+k_{\sigma(2)}}}{\left(\rvert x_{\sigma(1)}-x_{\sigma(2)}\rvert^2\right)^{s_{\sigma(1)}-k_{\sigma(1)}+k_{\sigma(2)}+1}}\ldots\frac{(x_{\sigma(n+2m)}-x_{\sigma(1)})_+^{s_{\sigma(n+{2m})}-k_{\sigma(n+{2m})}+k_{\sigma(1)}}}{\left(\rvert x_{\sigma(n+{2m})}-x_{\sigma(1)}\rvert^2\right)^{s_{\sigma(n+{2m})}-k_{\sigma(n+{2m})}+k_{\sigma(1)}+1}}
\end{align}
In the unbalanced sector, we get:
\begin{align} \label{B}
\nonumber
&\langle \mathbb{B}_{s_1}(x_1)\ldots \mathbb{B}_{s_n}(x_n)\bar{\mathbb{B}}_{{s'\!\!}_1}(y_1)\ldots \bar{\mathbb{B}}_{{s'\!\!}_n}(y_n)\rangle=\frac{1}{(4\pi^2)^{2n}}\frac{N^2-1}{2^{2n}}2^{\sum_{l=1}^n s_l+{s'\!\!}_l}i^{\sum_{l=1}^n s_l+{s'\!\!}_l}\\\nonumber
&\sum_{k_1=0}^{s_1}\ldots \sum_{k_n = 0}^{s_n}\sum_{{k'\!\!}_1=0}^{{s'\!\!}_1-2}\ldots \sum_{{k'\!\!}_n = 0}^{{s'\!\!}_n}{s_1\choose k_1}^2\ldots {s_n\choose k_n}^2{{s'\!\!}_1\choose {k'\!\!}_1}^2\ldots {{s'\!\!}_n\choose {k'\!\!}_n}^2\\\nonumber
&\frac{2^{n-1}}{n}\sum_{\sigma\in P_n}\sum_{\rho\in P_n}
(s_{\sigma(1)}-k_{\sigma(1)}+{k'\!\!}_{\rho(1)})!({s'\!\!}_{\rho(1)}-{k'\!\!}_{\rho(1)}+k_{\sigma(2)})!\\\nonumber
&\ldots(s_{\sigma(n)}-k_{\sigma(n)}+{k'\!\!}_{\rho(n)})!({s'\!\!}_{\rho(n)}-{k'\!\!}_{\rho(n)}+k_{\sigma(1)})!\\\nonumber
&\frac{(x_{\sigma(1)}-y_{\rho(1)})_+^{s_{\sigma(1)}-k_{\sigma(1)}+{k'\!\!}_{\rho(1)}}}{\left(\rvert x_{\sigma(1)}-y_{\rho(1)}\rvert^2\right)^{s_{\sigma(1)}-k_{\sigma(1)}+{k'\!\!}_{\rho(1)}+1}}\frac{(y_{\rho(1)}-x_{\sigma(2)})_+^{{s'\!\!}_{\rho(1)}-{k'\!\!}_{\rho(1)}+k_{\sigma(2)}}}{\left(\rvert y_{\rho(1)}-x_{\sigma(2)}\rvert^2\right)^{{s'\!\!}_{\rho(1)}-{k'\!\!}_{\rho(1)}+k_{\sigma(2)}+1}}\\
&\ldots\frac{(x_{\sigma(n)}-y_{\rho(n)})_+^{s_{\sigma(n)}-k_{\sigma(n)}+{k'\!\!}_{\rho(n)}}}{\left(\rvert x_{\sigma(n)}-y_{\rho(n)}\rvert^2\right)^{s_{\sigma(n)}-k_{\sigma(n)}+{k'\!\!}_{\rho(n)}+1}}
\frac{(y_{\rho(n)}-x_{\sigma(1)})_+^{{s'\!\!}_{\rho(n)}-{k'\!\!}_{\rho(n)}+k_{\sigma(1)}}}{\left(\rvert y_{\rho(n)}-x_{\sigma(1)}\rvert^2\right)^{{s'\!\!}_{\rho(n)}-{k'\!\!}_{\rho(n)}+k_{\sigma(1)}+1}}
\end{align}

 \subsection{Euclidean $n$-point correlators in the coordinate representation}

 \subsubsection{Standard basis}

 After the Wick rotation (appendix \ref{appN} and section \ref{8}), we obtain in the standard basis:
 \begin{align}
 \nonumber
 \mathcal{C}^E_{s}(x,y)
 = &\frac{1}{(4\pi^2)^2}\frac{N^2-1}{4} \frac{2^{2s+2}}{(4!)^2}(s+1)^2(s+2)^2
 \frac{(x-y)_{z}^{2s}}{((x-y)^2)^{2s+2}}\\\nonumber
 &\sum_{k_1 = 0}^{s-2}\sum_{k_2 = 0}^{s-2}{s\choose k_1}{s\choose k_1+2}{s\choose k_2+2}{s\choose k_2}(-1)^{s-k_2+k_1}\\
 &(s-k_1+k_2)!(s+k_1-k_2)!
  \end{align}
which is equivalent to:
 \begin{align}
 \mathcal{C}^E_{s}(x,y) = &\frac{1}{(4\pi^2)^2} \frac{N^2-1}{4} \frac{2^{2s+2}}{(4!)^2}(s-1)s(s+1)(s+2)(2s)!
 \frac{(x-y)_{z}^{2s}}{(( x-y)^2)^{2s+2}}
 \end{align}
 For the nonvanishing $3$-point correlators, we get:
 \begin{align}
\nonumber
 \mathcal{C}^E_{s_1s_2s_3}(x,y,z)=&\frac{1}{(4\pi^2)^3} (-1)^{s_1+s_2+s_3}(1+(-1)^{s_1+s_2+s_3})   \left(\frac{2}{4!}\right)^3\frac{N^2-1}{8}2^{s_1+s_2+s_3}\\\nonumber
 & (s_1+1)(s_1+2)(s_2+1)(s_2+2)(s_3+1)(s_3+2)\\\nonumber
 &\sum_{k_1 = 0}^{s_1-2}\sum_{k_2 = 0}^{s_2-2}\sum_{k_3 = 0}^{s_3-2}{s_1\choose k_1}{s_1\choose k_1+2}{s_2\choose k_2}{s_2\choose k_2+2}{s_3\choose k_3}{s_3\choose k_3+2}\\\nonumber
 &(s_1-k_1+k_2)!(s_2-k_2+k_3)!(s_3-k_3+k_1)!  \\
 &\frac{(x-y)^{s_1-k_1+k_2}_{z}}{(( x-y)^2)^{s_1+1-k_1+k_2}}\frac{(y-z)^{s_2-k_2+k_3}_{z}}{(( y-z)^2)^{s_2+1-k_2+k_3}}\frac{(z-x)^{s_3-k_3+k_1}_{z}}{(( z-x)^2)^{s_3+1-k_3+k_1}}
 \end{align}
Moreover, for the $n$-point correlators in the balanced sector, we obtain:
 \begin{align} \label{O0E}
 \nonumber
 &\langle \mathbb{O}^E_{s_1}(x_1)\ldots \mathbb{O}^E_{s_n}(x_n)\rangle_{conn} =\frac{1}{(4\pi^2)^n}\frac{N^2-1}{2^n}2^{\sum_{l=1}^n s_l}(-1)^{\sum_{l=1}^ns_l}\\\nonumber
 &\frac{\Gamma(3)\Gamma(s_1+3)}{\Gamma(5)\Gamma(s_1+1)}\ldots \frac{\Gamma(3)\Gamma(s_n+3)}{\Gamma(5)\Gamma(s_n+1)}\sum_{k_1=0}^{s_1-2}\ldots \sum_{k_n = 0}^{s_n-2}{s_1\choose k_1}{s_1\choose k_1+2}\ldots {s_n\choose k_n}{s_n\choose k_n+2}\\\nonumber
 &\frac{1}{n}\sum_{\sigma\in P_n}(s_{\sigma(1)}-k_{\sigma(1)}+k_{\sigma(2)})!\ldots(s_{\sigma(n)}-k_{\sigma(n)}+k_{\sigma(1)})!\\
 &\frac{(x_{\sigma(1)}-x_{\sigma(2)})_{z}^{s_{\sigma(1)}-k_{\sigma(1)}+k_{\sigma(2)}}}{\left(( x_{\sigma(1)}-x_{\sigma(2)})^2\right)^{s_{\sigma(1)}-k_{\sigma(1)}+k_{\sigma(2)}+1}}\ldots\frac{(x_{\sigma(n)}-x_{\sigma(1)})_{z}^{s_{\sigma(n)}-k_{\sigma(n)}+k_{\sigma(1)}}}{\left(( x_{\sigma(n)}-x_{\sigma(1)})^2\right)^{s_{\sigma(n)}-k_{\sigma(n)}+k_{\sigma(1)}+1}}
 \end{align}
The very same formula holds for an even number of operators $\tilde{\mathbb{O}}^E_s$, otherwise the correlators vanish. The nonvanishing correlators in the balanced sector are:
\begin{align} \label{OE}
\nonumber
&\langle \mathbb{O}^E_{s_1}(x_1)\ldots \mathbb{O}^E_{s_n}(x_n)\tilde{\mathbb{O}}^E_{s_{n+1}}(x_{n+1})\ldots \tilde{\mathbb{O}}^E_{s_{n+2m}}(x_{n+2m})\rangle_{conn}\\\nonumber
&=\frac{1}{(4\pi^2)^{n+2m}}\frac{N^2-1}{2^{n+2m}}2^{\sum_{l=1}^{n+2m} s_l}(-1)^{\sum_{l=1}^{n+2m}s_l}\frac{\Gamma(3)\Gamma(s_1+3)}{\Gamma(5)\Gamma(s_1+1)}\ldots\frac{\Gamma(3)\Gamma(s_{n+2m}+3)}{\Gamma(5)\Gamma(s_{n+2m}+1)}\\\nonumber
&\sum_{k_1=0}^{s_1-2}\ldots \sum_{k_{n+2m} = 0}^{s_{n+2m}-2}{s_1\choose k_1}{s_1\choose k_1+2}\ldots{s_{n+2m}\choose k_{n+2m}}{s_{n+2m}\choose k_{n+2m}+2}\\\nonumber
&\frac{1}{n+{2m}}\sum_{\sigma\in P_{n+2m}}(s_{\sigma(1)}-k_{\sigma(1)}+k_{\sigma(2)})!\ldots(s_{\sigma(n+{2m})}-k_{\sigma(n+{2m})}+k_{\sigma(1)})!\\
&\frac{(x_{\sigma(1)}-x_{\sigma(2)})_{z}^{s_{\sigma(1)}-k_{\sigma(1)}+k_{\sigma(2)}}}{\left(( x_{\sigma(1)}-x_{\sigma(2)})^2\right)^{s_{\sigma(1)}-k_{\sigma(1)}+k_{\sigma(2)}+1}}\ldots\frac{(x_{\sigma(n+2m)}-x_{\sigma(1)})_{z}^{s_{\sigma(n+{2m})}-k_{\sigma(n+{2m})}+k_{\sigma(1)}}}{\left(( x_{\sigma(n+{2m})}-x_{\sigma(1)})^2\right)^{s_{\sigma(n+{2m})}-k_{\sigma(n+{2m})}+k_{\sigma(1)}+1}}\,
\end{align}
In the unbalanced sector, we get:
\begin{align} \label{SE}
\nonumber
&\langle \mathbb{S}^E_{s_1}(x_1)\ldots \mathbb{S}^E_{s_n}(x_n)\bar{\mathbb{S}}^E_{{s'\!\!}_1}(y_1)\ldots \bar{\mathbb{S}}^E_{{s'\!\!}_n}(y_n)\rangle=\frac{1}{(4\pi^2)^{2n}}\frac{N^2-1}{2^{2n}}2^{\sum_{l=1}^n s_l+{s'\!\!}_l}(-1)^{\sum_{l=1}^ns_l+{s'\!\!}_l}\\\nonumber
&\frac{\Gamma(3)\Gamma(s_1+3)}{\Gamma(5)\Gamma(s_1+1)}\ldots \frac{\Gamma(3)\Gamma(s_n+3)}{\Gamma(5)\Gamma(s_n+1)}\frac{\Gamma(3)\Gamma({s'\!\!}_1+3)}{\Gamma(5)\Gamma({s'\!\!}_1+1)}\ldots \frac{\Gamma(3)\Gamma({s'\!\!}_n+3)}{\Gamma(5)\Gamma({s'\!\!}_n+1)}\\\nonumber
&\sum_{k_1=0}^{s_1-2}\ldots \sum_{k_n = 0}^{s_n-2}{s_1\choose k_1}{s_1\choose k_1+2}\ldots {s_n\choose k_n}{s_n\choose k_n+2}\\\nonumber
&\sum_{{k'\!\!}_1=0}^{{s'\!\!}_1-2}\ldots \sum_{{k'\!\!}_n = 0}^{{s'\!\!}_n-2}{{s'\!\!}_1\choose {k'\!\!}_1}{{s'\!\!}_1\choose {k'\!\!}_1+2}\ldots {{s'\!\!}_n\choose {k'\!\!}_n}{{s'\!\!}_n\choose {k'\!\!}_n+2}\\\nonumber
&\frac{2^{n-1}}{n}\sum_{\sigma\in P_n}\sum_{\rho\in P_n}
(s_{\sigma(1)}-k_{\sigma(1)}+{k'\!\!}_{\rho(1)})!({s'\!\!}_{\rho(1)}-{k'\!\!}_{\rho(1)}+k_{\sigma(2)})!\\\nonumber
&\ldots(s_{\sigma(n)}-k_{\sigma(n)}+{k'\!\!}_{\rho(n)})!({s'\!\!}_{\rho(n)}-{k'\!\!}_{\rho(n)}+k_{\sigma(1)})!\\\nonumber
&\frac{(x_{\sigma(1)}-y_{\rho(1)})_{z}^{s_{\sigma(1)}-k_{\sigma(1)}+{k'\!\!}_{\rho(1)}}}{\left(( x_{\sigma(1)}-y_{\rho(1)})^2\right)^{s_{\sigma(1)}-k_{\sigma(1)}+{k'\!\!}_{\rho(1)}+1}}\frac{(y_{\rho(1)}-x_{\sigma(2)})_{z}^{{s'\!\!}_{\rho(1)}-{k'\!\!}_{\rho(1)}+k_{\sigma(2)}}}{\left(( y_{\rho(1)}-x_{\sigma(2)})^2\right)^{{s'\!\!}_{\rho(1)}-{k'\!\!}_{\rho(1)}+k_{\sigma(2)}+1}}\\
&\ldots\frac{(x_{\sigma(n)}-y_{\rho(n)})_{z}^{s_{\sigma(n)}-k_{\sigma(n)}+{k'\!\!}_{\rho(n)}}}{\left(( x_{\sigma(n)}-y_{\rho(n)})^2\right)^{s_{\sigma(n)}-k_{\sigma(n)}+{k'\!\!}_{\rho(n)}+1}}
\frac{(y_{\rho(n)}-x_{\sigma(1)})_{z}^{{s'\!\!}_{\rho(n)}-{k'\!\!}_{\rho(n)}+k_{\sigma(1)}}}{\left(( y_{\rho(n)}-x_{\sigma(1)})^2\right)^{{s'\!\!}_{\rho(n)}-{k'\!\!}_{\rho(n)}+k_{\sigma(1)}+1}}
\end{align}
 
 \subsubsection{Extended basis}
 
 We obtain in the extended basis:
  \begin{align}
 \nonumber
\mathcal{A}^E_{s}(x,y)
  = &\frac{1}{(4\pi^2)^2} \frac{N^2-1}{4} 2^{2s}
 \frac{(x-y)_{z}^{2s}}{(( x-y)^2)^{2s+2}}\\
 &\sum_{k_1 = 0}^{s}\sum_{k_2 = 0}^{s}{s\choose k_1}{s\choose k_1}{s\choose k_2}{s\choose k_2}(-1)^{s-k_2+k_1}
 (s-k_1+k_2)!(s+k_1-k_2)!
 \end{align}
 which is equivalent to:
 \begin{align}
 \mathcal{A}^E_{s}(x,y)=&\frac{1}{(4\pi^2)^2} \frac{N^2-1}{4} 2^{2s} (2s)! 
 \frac{x_{{z}}^{2s}}{(x^2)^{2s+2}}
 \end{align}
 For the nonvanishing $3$-point correlators, we get:
 \begin{align}
\nonumber
 \mathcal{A}^E_{s_1s_2s_3}(x,y,z) =&\frac{1}{(4\pi^2)^3} (-1)^{s_1+s_2+s_3}(1+(-1)^{s_1+s_2+s_3})   \frac{N^2-1}{8}2^{s_1+s_2+s_3}\\\nonumber
 &\sum_{k_1 = 0}^{s_1}\sum_{k_2 = 0}^{s_2}\sum_{k_3 = 0}^{s_3}{s_1\choose k_1}{s_1\choose k_1}{s_2\choose k_2}{s_2\choose k_2}{s_3\choose k_3}{s_3\choose k_3}\\\nonumber
 &(s_1-k_1+k_2)!(s_2-k_2+k_3)!(s_3-k_3+k_1)!  \\\nonumber
 &\frac{(x-y)^{s_1-k_1+k_2}_{z}}{(( x-y)^2)^{s_1+1-k_1+k_2}}\frac{(y-z)^{s_2-k_2+k_3}_{z}}{(( y-z)^2)^{s_2+1-k_2+k_3}}\frac{(z-x)^{s_3-k_3+k_1}_{z}}{(( z-x)^2)^{s_3+1-k_3+k_1}} \\
\end{align}
 Moreover, for the $n$-point correlators in the balanced sector, we obtain:
 \begin{align} \label{A0E}
 \nonumber
&\langle \mathbb{A}^E_{s_1}(x_1)\ldots \mathbb{A}^E_{s_n}(x_n)\rangle_{conn} =\frac{1}{(4\pi^2)^n}\frac{N^2-1}{2^n}2^{\sum_{l=1}^n s_l}(-1)^{\sum_{l=1}^ns_l}\\\nonumber
&\sum_{k_1=0}^{s_1}\ldots \sum_{k_n = 0}^{s_n}{s_1\choose k_1}^2\ldots {s_n\choose k_n}^2\frac{1}{n}\sum_{\sigma\in P_n}(s_{\sigma(1)}-k_{\sigma(1)}+k_{\sigma(2)})!\ldots(s_{\sigma(n)}-k_{\sigma(n)}+k_{\sigma(1)})!\\
 &\frac{(x_{\sigma(1)}-x_{\sigma(2)})_{{z}}^{s_{\sigma(1)}-k_{\sigma(1)}+k_{\sigma(2)}}}{\left((  x_{\sigma(1)}-x_{\sigma(2)})^2\right)^{s_{\sigma(1)}-k_{\sigma(1)}+k_{\sigma(2)}+1}}\ldots\frac{(x_{\sigma(n)}-x_{\sigma(1)})_{{z}}^{s_{\sigma(n)}-k_{\sigma(n)}+k_{\sigma(1)}}}{\left(( x_{\sigma(n)}-x_{\sigma(1)})^2\right)^{s_{\sigma(n)}-k_{\sigma(n)}+k_{\sigma(1)}+1}}
 \end{align}
 The very same formula holds for an even number of $\tilde{\mathbb{A}}^E_s$ operators, otherwise the correlators vanish. The nonvanishing correlators in the balanced sector are:
\begin{align} \label{AE}
\nonumber
&\langle \mathbb{A}^E_{s_1}(x_1)\ldots \mathbb{A}^E_{s_n}(x_n)\tilde{\mathbb{A}}^E_{s_{n+1}}(x_{n+1})\ldots \tilde{\mathbb{A}}^E_{s_{n+2m}}(x_{n+2m})\rangle_{conn} \\\nonumber
&=\frac{1}{(4\pi^2)^{n+2m}}\frac{N^2-1}{2^{n+2m}}2^{\sum_{l=1}^{n+2m} s_l}(-1)^{\sum_{l=1}^{n+2m}s_l}\sum_{k_1=0}^{s_1}\ldots \sum_{k_{n+2m} = 0}^{s_{n+2m}}{s_1\choose k_1}^2\ldots{s_{n+2m}\choose k_{n+2m}}^2\\\nonumber
&\frac{1}{n+2m}\sum_{\sigma\in P_{n+2m}}(s_{\sigma(1)}-k_{\sigma(1)}+k_{\sigma(2)})!\ldots(s_{\sigma(n+2m)}-k_{\sigma(n+2m)}+k_{\sigma(1)})!\\
&\frac{(x_{\sigma(1)}-x_{\sigma(2)})_{z}^{s_{\sigma(1)}-k_{\sigma(1)}+k_{\sigma(2)}}}{\left(( x_{\sigma(1)}-x_{\sigma(2)})^2\right)^{s_{\sigma(1)}-k_{\sigma(1)}+k_{\sigma(2)}+1}}\ldots\frac{(x_{\sigma(n+2m)}-x_{\sigma(1)})_{z}^{s_{\sigma(n+2m)}-k_{\sigma(n+2m)}+k_{\sigma(1)}}}{\left(( x_{\sigma(n+2m)}-x_{\sigma(1)})^2\right)^{s_{\sigma(n+2m)}-k_{\sigma(n+2m)}+k_{\sigma(1)}+1}}
\end{align} 
In the unbalanced sector, we get:
 \begin{align} \label{BE}
 \nonumber
 &\langle \mathbb{B}^E_{s_1}(x_1)\ldots \mathbb{B}^E_{s_n}(x_n)\bar{\mathbb{B}}^E_{{s'\!\!}_1}(y_1)\ldots \bar{\mathbb{B}}^E_{{s'\!\!}_n}(y_n)\rangle=\frac{1}{(4\pi^2)^{2n}}\frac{N^2-1}{2^{2n}}2^{\sum_{l=1}^n s_l+{s'\!\!}_l}(-1)^{\sum_{l=1}^ns_l+{s'\!\!}_l}\\\nonumber
 &\sum_{k_1=0}^{s_1}\ldots \sum_{k_n = 0}^{s_n}\sum_{{k'\!\!}_1=0}^{{s'\!\!}_1-2}\ldots \sum_{{k'\!\!}_n = 0}^{{s'\!\!}_n}{s_1\choose k_1}^2\ldots {s_n\choose k_n}^2{{s'\!\!}_1\choose {k'\!\!}_1}^2\ldots {{s'\!\!}_n\choose {k'\!\!}_n}^2\\\nonumber
&\frac{2^{n-1}}{n}\sum_{\sigma\in P_n}\sum_{\rho\in P_n}
(s_{\sigma(1)}-k_{\sigma(1)}+{k'\!\!}_{\rho(1)})!({s'\!\!}_{\rho(1)}-{k'\!\!}_{\rho(1)}+k_{\sigma(2)})!\\\nonumber
&\ldots(s_{\sigma(n)}-k_{\sigma(n)}+{k'\!\!}_{\rho(n)})!({s'\!\!}_{\rho(n)}-{k'\!\!}_{\rho(n)}+k_{\sigma(1)})!\\\nonumber
&\frac{(x_{\sigma(1)}-y_{\rho(1)})_{z}^{s_{\sigma(1)}-k_{\sigma(1)}+{k'\!\!}_{\rho(1)}}}{\left(( x_{\sigma(1)}-y_{\rho(1)})^2\right)^{s_{\sigma(1)}-k_{\sigma(1)}+{k'\!\!}_{\rho(1)}+1}}\frac{(y_{\rho(1)}-x_{\sigma(2)})_{z}^{{s'\!\!}_{\rho(1)}-{k'\!\!}_{\rho(1)}+k_{\sigma(2)}}}{\left(( y_{\rho(1)}-x_{\sigma(2)})^2\right)^{{s'\!\!}_{\rho(1)}-{k'\!\!}_{\rho(1)}+k_{\sigma(2)}+1}}\\
&\ldots\frac{(x_{\sigma(n)}-y_{\rho(n)})_{z}^{s_{\sigma(n)}-k_{\sigma(n)}+{k'\!\!}_{\rho(n)}}}{\left(( x_{\sigma(n)}-y_{\rho(n)})^2\right)^{s_{\sigma(n)}-k_{\sigma(n)}+{k'\!\!}_{\rho(n)}+1}}
\frac{(y_{\rho(n)}-x_{\sigma(1)})_{z}^{{s'\!\!}_{\rho(n)}-{k'\!\!}_{\rho(n)}+k_{\sigma(1)}}}{\left(( y_{\rho(n)}-x_{\sigma(1)})^2\right)^{{s'\!\!}_{\rho(n)}-{k'\!\!}_{\rho(n)}+k_{\sigma(1)}+1}}
 \end{align}

 \subsection{Generating functional of $n$-point correlators in the coordinate representation}
 
Remarkably, we find the following structure for the generating functional of correlators of balanced operators with $\mathcal T=2$ to the lowest order:
\begin{equation} \label{K}
\Gamma_{conf}[\mathcal{O}] \sim \log \Det \left(\mathbb{I}+\mathcal{D}^{-1}\mathcal{O}\right)
\end{equation}
that, in the coordinate representation, is a compact notation for:
\begin{equation}
 \Gamma_{conf}[\mathcal{O}] \sim \log \Det \left(\delta_{s_1k_1,s_2k_2}\delta^{(4)}(x-y)+\mathcal{D}^{-1}_{s_1k_1,s_2k_2}(x-y)\mathcal{O}_{s_2k_2}(y)\right)
\end{equation}
where the source fields $\mathcal{O}_{sk}(x)$ occur in the expansion of the source $\mathcal{O}_s(x)$, in analogy with eq. \eqref{confexp}:
 \begin{equation}
 \mathcal{O}_s(x) = \sum_{k = 0}^{l}\mathcal{O}_{sk}(x)
 \end{equation}
and, by a slight abuse of notation, we have employed for the source fields the very same notation as for the corresponding operators, with $l=s-2$ for the standard basis and $l=s$ for the extended basis.  \par
Though the source fields are originally defined either for even or odd $s$ respectively, to keep the notation simple in the following formulas, we extend their definition to all values of $s$, in such a way that they are $0$ for either odd or even $s$ respectively.\par
 The argument of the determinant is the kernel:
 \begin{equation}
 K_{s_1k_1,s_2k_2}(x,y)=\delta_{s_1k_1,s_2k_2}\delta^{(4)}(x-y)+\mathcal{D}^{-1}_{s_1k_1,s_2k_2}(x-y)\mathcal{O}_{s_2k_2}(y)
\end{equation}
of the integral operator:
 \begin{equation}
\psi_{s_1k_1}(x) =\sum_{s_2k_2} \int K_{s_1k_1,s_2k_2}(x,y) \phi_{s_2k_2}(y) d^4y
\end{equation}
formally of Fredholm type.\par
$\mathcal{D}^{-1}_{s_1k_1,s_2k_2}(x-y)$ -- defined in the following -- is the effective propagator associated to the source fields $\mathcal{O}_{sk}(x)$. \par
Then, the $n$-point correlators are computed by the functional derivatives:
\begin{align}
\langle \mathcal{O}_{s_1}(x_1)\ldots \mathcal{O}_{s_n}(x_n)\rangle_{conn} &=\frac{\delta}{\delta\mathcal{O}_{s_1}(x_1)}\ldots\frac{\delta}{\delta\mathcal{O}_{s_n }(x_n)}\Gamma_{conf}[\mathcal{O}] \nonumber \\
&=\sum_{k_1=0}^{l_1}\ldots\sum_{k_n=0}^{l_n}\frac{\delta}{\delta\mathcal{O}_{s_1 k_1}(x_1)}\ldots\frac{\delta}{\delta\mathcal{O}_{s_n k_n}(x_n)}\Gamma_{conf}[\mathcal{O}]
\end{align}
The generating functional -- reported below -- of correlators of the unbalanced operators with collinear twist $\tau=2$ has a similar structure of the logarithm of a functional determinant.\par
In both cases, we verify that our ansatz reproduces the corresponding correlators.

 \subsubsection{Minkowskian standard basis}
 
 Specifically, we verify by direct computation (section \ref{444}) that:
 \begin{align}
 \nonumber\
 \Gamma_{conf}[\mathbb{O}]=&-(N^2-1)\log \Det\left(\mathbb{I}+\mathcal{D}^{-1}\mathbb{O}\right)\\\nonumber
 \Gamma_{conf}[\tilde{\mathbb{O}}]=& -\frac{N^2-1}{2}\log \Det\left(\mathbb{I}-\mathcal{D}^{-1}\tilde{\mathbb{O}}\mathcal{D}^{-1}\tilde{\mathbb{O}}\right)\\\nonumber
 \Gamma_{conf}[\mathbb{O},\tilde{\mathbb{O}}] =& -\frac{N^2-1}{2}\log \Det\left(\mathbb{I}+\mathcal{D}^{-1}\mathbb{O}+\mathcal{D}^{-1}\tilde{\mathbb{O}}\right)\\\nonumber
 &-\frac{N^2-1}{2}\log \Det\left(\mathbb{I}+\mathcal{D}^{-1}\mathbb{O}-\mathcal{D}^{-1}\tilde{\mathbb{O}}\right)\\\nonumber
 =& -\frac{N^2-1}{2}\log \Det\left(\left(\mathbb{I}+\mathcal{D}^{-1}\mathbb{O}\right)^2-\mathcal{D}^{-1}\tilde{\mathbb{O}}\mathcal{D}^{-1}\tilde{\mathbb{O}}\right)\nonumber\\\
 \Gamma_{conf}[\mathbb{S},\bar{\mathbb{S}}]=& -\frac{N^2-1}{2}\log \Det\left(\mathbb{I}-2\mathcal{D}^{-1}\bar{\mathbb{S}}\mathcal{D}^{-1}\mathbb{S}\right)
 \end{align}
 in Minkowskian space-time.\par
 By making explicit the continuous and discrete indices, the above equations read in the coordinate representation:
 \begin{align}
 \nonumber\
\Gamma_{conf}[\mathbb{O},\tilde{\mathbb{O}}]=&-\frac{N^2-1}{2}\log \Det\Big(\delta_{s_1k_1,s_2k_2}\delta^{(4)}(x-y)\\\nonumber
&+\mathcal{D}^{-1}_{s_1k_1,s_2k_2}(x-y) (\mathbb{O}_{s_2k_2}(y)+\tilde{\mathbb{O}}_{s_2k_2}(y)) \Big)\\\nonumber
&-\frac{N^2-1}{2}\log \Det\Big(\delta_{s_1k_1,s_2k_2}\delta^{(4)}(x-y)\\\nonumber 
&+\mathcal{D}^{-1}_{s_1k_1,s_2k_2}(x-y) (\mathbb{O}_{s_2k_2}(y)-\tilde{\mathbb{O}}_{s_2k_2}(y)) \Big)\\\nonumber
\Gamma_{conf}[\mathbb{S},\bar{\mathbb{S}}] =&-\frac{N^2-1}{2}\log \Det\Big(\delta_{s_1k_1,s_2k_2}\delta^{(4)}(x-y)\\
 \qquad\qquad\qquad&-2\int d^4z\sum_{sk}\mathcal{D}^{-1}_{s_1k_1,sk}(x-z)\bar{\mathbb{S}}_{sk}(z)\mathcal{D}^{-1}_{sk,s_2k_2}(z-y)\mathbb{S}_{s_2k_2}(y)\Big)
\end{align}
 with:
 \begin{align}
 \mathcal{D}^{-1}_{s_1k_1,s_2k_2}(x-y) &=\frac{i^{s_1+1}}{2}\frac{\Gamma(3)\Gamma(s_1+3)}{\Gamma(5)\Gamma(s_1+1)}{s_1\choose k_1}{s_2\choose k_2+2}(-\partial_{+})^{s_1-k_1+k_2}\square^{-1}(x-y)\nonumber\\ &=\frac{i^{s_1}}{8\pi^2}\frac{\Gamma(3)\Gamma(s_1+3)}{\Gamma(5)\Gamma(s_1+1)}{s_1\choose k_1}{s_2\choose k_2+2}(-\partial_{+})^{s_1-k_1+k_2}\frac{1}{\rvert x-y\rvert^2-i\epsilon}
 \end{align}

 \subsubsection{Minkowskian extended basis}

We also verify by direct computation (section \ref{444}) that:
 \begin{align}
 \nonumber\
 \Gamma_{conf}[\mathbb{A}]=&-(N^2-1)\log \Det\left(\mathbb{I}+\mathcal{D}^{-1}\mathbb{A}\right)\\\nonumber
 \Gamma_{conf}[\tilde{\mathbb{A}}]=& -\frac{N^2-1}{2}\log \Det\left(\mathbb{I}-\mathcal{D}^{-1}\tilde{\mathbb{A}}\mathcal{D}^{-1}\tilde{\mathbb{A}}\right)\\\nonumber
 \Gamma_{conf}[\mathbb{A},\tilde{\mathbb{A}}] =& -\frac{N^2-1}{2}\log \Det\left(\mathbb{I}+\mathcal{D}^{-1}\mathbb{A}+\mathcal{D}^{-1}\tilde{\mathbb{A}}\right)\\\nonumber
 &-\frac{N^2-1}{2}\log \Det\left(\mathbb{I}+\mathcal{D}^{-1}\mathbb{A}-\mathcal{D}^{-1}\tilde{\mathbb{A}}\right)\\\nonumber
 =& -\frac{N^2-1}{2}\log \Det\left(\left(\mathbb{I}+\mathcal{D}^{-1}\mathbb{A}\right)^2-\mathcal{D}^{-1}\tilde{\mathbb{A}}\mathcal{D}^{-1}\tilde{\mathbb{A}}\right)\nonumber\\\
 \Gamma_{conf}[\mathbb{B},\bar{\mathbb{B}}]=& -\frac{N^2-1}{2}\log \Det\left(\mathbb{I}-2\mathcal{D}^{-1}\bar{\mathbb{B}}\mathcal{D}^{-1}\mathbb{B}\right)
 \end{align}
  in Minkowskian space-time.\par
 By making explicit the continuous and discrete indices, the above equations read in the coordinate representation:
 \begin{align}
 \nonumber
\Gamma_{conf}[\mathbb{A},\tilde{\mathbb{A}}]=&  -\frac{N^2-1}{2}\log \Det\Big(\delta_{s_1k_1,s_2k_2}\delta^{(4)}(x-y) \\\nonumber
&+\mathcal{D}^{-1}_{s_1k_1,s_2k_2}(x-y) (\mathbb{A}_{s_2k_2}(y)+\tilde{\mathbb{A}}_{s_2k_2}(y)) \Big)\\\nonumber
&-\frac{N^2-1}{2}\log \Det\Big(\delta_{s_1k_1,s_2k_2}\delta^{(4)}(x-y) \\\nonumber
&+\mathcal{D}^{-1}_{s_1k_1,s_2k_2}(x-y) (\mathbb{A}_{s_2k_2}(y)-\tilde{\mathbb{A}}_{s_2k_2}(y)) \Big)\\\nonumber
 \Gamma_{conf}[\mathbb{B},\bar{\mathbb{B}}]=& -\frac{N^2-1}{2}\log \Det\Big(\delta_{s_1k_1,s_2k_2}\delta^{(4)}(x-y)\\
 \qquad\qquad\qquad&-2
\int d^4z\sum_{sk}\mathcal{D}^{-1}_{s_1k_1,sk}(x-z)\bar{\mathbb{B}}_{sk}(z)\mathcal{D}^{-1}_{sk,s_2k_2}(z-y)\mathbb{B}_{s_2k_2}(y)\Big)\\\nonumber
 \end{align}
 with:
 \begin{align}
 \mathcal{D}^{-1}_{s_1k_1,s_2k_2}(x-y) &=\frac{i^{s_1+1}}{2}{s_1\choose k_1}{s_2\choose k_2}(-\partial_{+})^{s_1-k_1+k_2}\square^{-1}(x-y) \nonumber \\
 &= \frac{i^{s_1}}{8\pi^2}{s_1\choose k_1}{s_2\choose k_2}(-\partial_{+})^{s_1-k_1+k_2}\frac{1}{\rvert x-y\rvert^2-i\epsilon } 
 \end{align}

  \subsubsection{Euclidean standard basis}

Similarly:
\begin{align}
 \nonumber\
 \Gamma_{conf}[\mathbb{O}^E]=&-(N^2-1)\log \Det\left(\mathbb{I}+\mathcal{D}^{-1}_E\mathbb{O}^E\right)\\\nonumber
 \Gamma_{conf}[\tilde{\mathbb{O}}^E]=& -\frac{N^2-1}{2}\log \Det\left(\mathbb{I}-\mathcal{D}^{-1}_E\tilde{\mathbb{O}}^E\mathcal{D}^{-1}_E\tilde{\mathbb{O}}^E\right)\\\nonumber
 \Gamma_{conf}[\mathbb{O}^E,\tilde{\mathbb{O}}^E] =& -\frac{N^2-1}{2}\log \Det\left(\mathbb{I}+\mathcal{D}_E^{-1}\mathbb{O}^E+\mathcal{D}^{-1}\tilde{\mathbb{O}}^E\right)\\\nonumber
 &-\frac{N^2-1}{2}\log \Det\left(\mathbb{I}+\mathcal{D}_E^{-1}\mathbb{O}^E-\mathcal{D}_E^{-1}\tilde{\mathbb{O}}^E\right)\\\nonumber
 =& -\frac{N^2-1}{2}\log \Det\left(\left(\mathbb{I}+\mathcal{D}_E^{-1}\mathbb{O}^E\right)^2-\mathcal{D}^{-1}\tilde{\mathbb{O}}^E\mathcal{D}_E^{-1}\tilde{\mathbb{O}}^E\right)\nonumber\\\
 \Gamma_{conf}[\mathbb{S}^E,\bar{\mathbb{S}}^E]=& -\frac{N^2-1}{2}\log \Det\left(\mathbb{I}-2\mathcal{D}_E^{-1}\bar{\mathbb{S}}^E\mathcal{D}_E^{-1}\mathbb{S}^E\right)
 \end{align}
 in Euclidean space-time, with:
 \begin{align}
 \nonumber
 {\mathcal{D}^{-1}_E}_{s_1k_1,s_2k_2}(x-y)&=\frac{(-i)^{-k_1+k_2}}{2}\frac{\Gamma(3)\Gamma(s_1+3)}{\Gamma(5)\Gamma(s_1+1)}{s_1\choose k_1}{s_2\choose k_2+2}\partial_{z}^{s_1-k_1+k_2}\Laplace^{-1}(x-y)\\
 &= -\frac{(-i)^{-k_1+k_2}}{8\pi^2}\frac{\Gamma(3)\Gamma(s_1+3)}{\Gamma(5)\Gamma(s_1+1)}{s_1\choose k_1}{s_2\choose k_2+2}\partial_{z}^{s_1-k_1+k_2}\frac{1}{(x-y)^2}
 \end{align}

\subsubsection{Euclidean extended basis}

 Analogously:
 \begin{align}
 \nonumber\
 \Gamma_{conf}[\mathbb{A}^E]=&-(N^2-1)\log \Det\left(\mathbb{I}+\mathcal{D}^{-1}_E\mathbb{A}^E\right)\\\nonumber
 \Gamma_{conf}[\tilde{\mathbb{A}}^E]=& -\frac{N^2-1}{2}\log \Det\left(\mathbb{I}-\mathcal{D}^{-1}_E\tilde{\mathbb{A}}^E\mathcal{D}^{-1}_E\tilde{\mathbb{A}}^E\right)\\\nonumber
 \Gamma_{conf}[\mathbb{A}^E,\tilde{\mathbb{A}}^E] =& -\frac{N^2-1}{2}\log \Det\left(\mathbb{I}+\mathcal{D}_E^{-1}\mathbb{A}^E+\mathcal{D}^{-1}\tilde{\mathbb{A}}^E\right)\\\nonumber
 &-\frac{N^2-1}{2}\log \Det\left(\mathbb{I}+\mathcal{D}_E^{-1}\mathbb{A}^E-\mathcal{D}_E^{-1}\tilde{\mathbb{A}}^E\right)\\\nonumber
 =& -\frac{N^2-1}{2}\log \Det\left(\left(\mathbb{I}+\mathcal{D}_E^{-1}\mathbb{A}^E\right)^2-\mathcal{D}^{-1}\tilde{\mathbb{A}}^E\mathcal{D}_E^{-1}\tilde{\mathbb{A}}^E\right)\nonumber\\\
 \Gamma_{conf}[\mathbb{B}^E,\bar{\mathbb{B}}^E]=& -\frac{N^2-1}{2}\log \Det\left(\mathbb{I}-2\mathcal{D}_E^{-1}\bar{\mathbb{B}}^E\mathcal{D}_E^{-1}\mathbb{B}^E\right)
 \end{align}
 in Euclidean space-time, with: 
 \begin{align}
 {\mathcal{D}^{-1}_E}_{s_1k_1,s_2k_2}(x-y) &= \frac{(-i)^{-k_1+k_2}}{2}{s_1\choose k_1}{s_2\choose k_2}\partial_{z}^{s_1-k_1+k_2}\Laplace^{-1}(x-y)\nonumber\\
 &= -\frac{(-i)^{-k_1+k_2}}{8\pi^2}{s_1\choose k_1}{s_2\choose k_2}\partial_{z}^{s_1-k_1+k_2}\frac{1}{(x-y)^2}
 \end{align}

 \subsection{Generating functional and $n$-point correlators in the momentum representation}
 
In the momentum representation the generating functionals are exactly the same, but for the fact that the corresponding source fields and effective propagators are Fourier transformed:
\begin{align}
\mathcal{O}_{s}(p) = \int  \mathcal{O}_{s}(x) \, e^{-ip\cdot x} d^4x
\end{align}
In the momentum representation the generating functional in eq. \eqref{K} is (section \ref{55}):
\begin{equation}
\Gamma_{conf}[\mathcal{O}] \sim \log \Det \left(\delta_{s_1k_1,s_2k_2}(2\pi)^4\delta^{(4)}(q_1-q_2)+\mathcal{D}^{-1}_{s_1k_1,s_2k_2}(q_1)\mathcal{O}_{s_2k_2}(q_1-q_2)\right)
\end{equation}
The argument of the determinant is the kernel:
\begin{equation}
K_{s_1k_1,s_2k_2}(q_1,q_2)=\delta_{s_1k_1,s_2k_2}(2\pi)^4\delta^{(4)}(q_1-q_2)+\mathcal{D}^{-1}_{s_1k_1,s_2k_2}(q_1)\mathcal{O}_{s_2k_2}(q_1-q_2)
\end{equation}
of the integral operator:
 \begin{equation}
\psi_{s_1k_1}(q_1) =\sum_{s_2k_2} \int K_{s_1k_1,s_2k_2}(q_1,q_2) \phi_{s_2k_2}(q_2) \frac{d^4q_2}{(2\pi)^4}
\end{equation}
Accordingly, the $n$-point correlators in the momentum representation are computed by the functional derivatives:
\begin{align}
\langle \mathcal{O}_{s_1}(p_1)\ldots \mathcal{O}_{s_n}(p_n)\rangle_{conn} &=\frac{\delta}{\delta\mathcal{O}_{s_1}(p_1)}\ldots\frac{\delta}{\delta\mathcal{O}_{s_n }(p_n)}\Gamma_{conf}[\mathcal{O}] \nonumber \\
&=\sum_{k_1=0}^{l_1}\ldots\sum_{k_n=0}^{l_n}\frac{\delta}{\delta\mathcal{O}_{s_1 k_1}(p_1)}\ldots\frac{\delta}{\delta\mathcal{O}_{s_n k_n}(p_n)}\Gamma_{conf}[\mathcal{O}]
\end{align}

 \subsubsection{Minkowskian standard basis}
 
In the momentum representation, we get in Minkowskian space-time (section \ref{55}):
 \begin{align}
 \nonumber
\Gamma_{conf}[\mathbb{O},\tilde{\mathbb{O}}]=& - \frac{N^2-1}{2}\log \Det \Big(\delta_{s_1k_1,s_2k_2}(2\pi)^4 \delta^{(4)}(q_1-q_2)\\\nonumber
&+\mathcal{D}^{-1}_{s_1k_1,s_2k_2}(q_1)(\mathbb{O}_{s_2k_2}(q_1-q_2)+ \tilde{\mathbb{O}}_{s_2k_2}(q_1-q_2)) \Big) \\\nonumber
& - \frac{N^2-1}{2}\log \Det \Big(\delta_{s_1k_1,s_2k_2} (2\pi)^4\delta^{(4)}(q_1-q_2)\\\nonumber
& +\mathcal{D}^{-1}_{s_1k_1,s_2k_2}(q_1)(\mathbb{O}_{s_2k_2}(q_1-q_2)- \tilde{\mathbb{O}}_{s_2k_2}(q_1-q_2)) \Big)\\\nonumber
\Gamma_{conf}[\mathbb{S},\bar{\mathbb{S}}]= &-\frac{N^2-1}{2}\log \det\Big(\delta_{s_1k_1,s_2k_2}(2\pi)^4 \delta^{(4)}(q_1-q_2)\\
 \qquad\qquad\qquad&-2\int \frac{d^4q}{(2\pi)^4}\sum_{sk}\mathcal{D}^{-1}_{s_1k_1,sk}(q_1)\bar{\mathbb{S}}_{sk}(q_1-q)\mathcal{D}^{-1}_{sk,s_2k_2}(q)\mathbb{S}_{s_2k_2}(q-q_2)\Big)\\\nonumber
 \end{align}
 with:
 \begin{align}
 \mathcal{D}^{-1}_{s_1k_1,s_2k_2}(p) &=\frac{i^{s_1}}{2}\frac{\Gamma(3)\Gamma(s_1+3)}{\Gamma(5)\Gamma(s_1+1)}{s_1\choose k_1}{s_2\choose k_2+2}(-ip_{+})^{s_1-k_1+k_2}\frac{-i}{\rvert p\rvert^2+i\epsilon} 
 \end{align} 
 Explicitly, we obtain by direct computation the nonvanishing correlators in the balanced sector from their generating functional in the momentum representation:
  \begin{align}
 \nonumber
 &\left(\frac{\Gamma(5)\Gamma(s_1+1)}{\Gamma(3)\Gamma(s_1+3)}\right)\ldots\left( \frac{\Gamma(5)\Gamma(s_{n+2m}+1)}{\Gamma(3)\Gamma(s_{n+2m}+3)}\right) \nonumber\\\nonumber
 &\langle \mathbb{O}_{s_1}(p_1)\ldots \mathbb{O}_{s_n}(p_n)\tilde{\mathbb{O}}_{s_{n+1}}(p_{n+1})\ldots \tilde{\mathbb{O}}_{s_{n+2m}}(p_{n+2m})\rangle_{conn} \\\nonumber
 &=\frac{N^2-1}{2^{n+2m}}(2\pi)^4 i^{n+2m}\delta^{(4)}(p_1+\ldots+p_{n+2m})\\\nonumber
 &\sum_{k_1=0}^{s_1-2}\ldots \sum_{k_{n+2m} = 0}^{s_{n+2m}-2}{s_1\choose k_1}{s_1\choose k_1+2}\ldots {s_{n+2m}\choose k_{n+2m}}{s_{n+2m}\choose k_{n+2m}+2}\frac{1}{n+2m}\sum_{\sigma\in P_{n+2m}}\\\nonumber
 &\int \frac{d^4 q}{(2\pi)^4}\frac{(p_{\sigma(1)}+q)_{+}^{s_{\sigma(1)}-k_{\sigma(1)}+k_{\sigma(2)}}}{\rvert p_{\sigma(1)}+q\rvert^2}\frac{(p_{\sigma(1)}+p_{\sigma(2)}+q)_{+}^{s_{\sigma(2)}-k_{\sigma(2)}+k_{\sigma(3)}}}{\rvert p_{\sigma(1)}+p_{\sigma(2)}+q\rvert^2}\\
 &\ldots\frac{(\sum_{l=1}^{n+2m-1} p_{\sigma(l)}+q)_{+}^{s_{\sigma(n+2m-1)}-k_{\sigma(n+2m-1)}+k_{\sigma(n+2m)}}}{\rvert\sum_{l=1}^{n+2m-1} p_{\sigma(l)}+q\rvert^2}\frac{(q)_{+}^{s_{\sigma(n+2m)}-k_{\sigma(n+2m)}+k_{\sigma(1)}}}{\rvert q\rvert^2} 
 \end{align}
In the unbalanced sector, we get:
 \begin{align}
 \nonumber
 &\left(\frac{\Gamma(5)\Gamma(s_1+1)}{\Gamma(3)\Gamma(s_1+3)}\right)\ldots\left(\frac{\Gamma(5)\Gamma(s_n+1)}{\Gamma(3)\Gamma(s_n+3)}\right)\left(\frac{8\pi}{N}\frac{\Gamma(5)\Gamma(s'_1+1)}{\Gamma(3)\Gamma(s'_1+3)}\right)\ldots\left( \frac{\Gamma(5)\Gamma(s'_{n}+1)}{\Gamma(3)\Gamma(s'_{n}+3)}\right)\\\nonumber
 &\langle \mathbb{S}_{s_1}(p_1)\ldots \mathbb{S}_{s_n}(p_n)\bar{\mathbb{S}}_{s'_1}(p'_1)\ldots \bar{\mathbb{S}}_{s'_n}(p'_n)\rangle\\\nonumber
 &=\frac{N^2-1}{2^{2n}}(2\pi)^4 i^{2n}\delta^{(4)}\left(\sum_{l=1}^n p_l+p'_l\right)\\\nonumber
 &\sum_{k_1=0}^{s_1-2}\ldots \sum_{k_n = 0}^{s_n-2}\sum_{k'_1=0}^{s'_1-2}\ldots \sum_{k'_n = 0}^{s'_n-2}{s_1\choose k_1}{s_1\choose k_1+2}\ldots {s_n\choose k_n}{s_n\choose k_n+2}\\\nonumber
&{s'_1\choose k'_1}{s'_1\choose k'_1+2}\ldots {s'_n\choose k'_n}{s'_n\choose k'_n+2}\\\nonumber
 &\frac{2^{n-1}}{n}\sum_{\sigma\in P_n}\sum_{\rho\in P_n}
 \int \frac{d^4 q}{(2\pi)^4}\frac{(p_{\sigma(1)}+q)_{+}^{s_{\sigma(1)}-k_{\sigma(1)}+k'_{\rho(1)}}}{\rvert p_{\sigma(1)}+q\rvert^2}\frac{(p_{\sigma(1)}+p'_{\rho(1)}+q)_{+}^{s'_{\rho(1)}-k'_{\rho(1)}+k_{\sigma(2)}}}{\rvert p_{\sigma(1)}+p'_{\rho(1)}+q\rvert^2}\\\nonumber
 &\frac{(p_{\sigma(1)}+p_{\sigma(2)}+p'_{\rho(1)}+q)_{+}^{s_{\sigma(2)}-k_{\sigma(2)}+k'_{\rho(2)}}}{\rvert p_{\sigma(1)}+p_{\sigma(2)}+p'_{\rho(1)}+q\rvert^2}\frac{(p_{\sigma(1)}+p_{\sigma(2)}+p'_{\rho(1)}+p'_{\rho(2)}+q)_{+}^{s'_{\rho(2)}-k'_{\rho(2)}+k_{\sigma(3)}}}{\rvert p_{\sigma(1)}+p_{\sigma(2)}+p'_{\rho(1)}+p'_{\rho(2)}+q\rvert^2}\\
 &\ldots\frac{(\sum_{l=1}^{n-1} p_{\sigma(l)}+\sum_{l=1}^{n-2} p'_{\rho(l)} +q)_{+}^{s_{\sigma(n-1)}-k_{\sigma(n-1)}+k'_{\rho(n-1)}}}{\rvert\sum_{l=1}^{n-1} p_{\sigma(l)}+\sum_{l=1}^{n-2} p'_{\rho(l)}+q\rvert^2}\nonumber\\
 &\frac{(\sum_{l=1}^{n-1} p_{\sigma(l)}+\sum_{l=1}^{n-1} p'_{\rho(l)}+q)_{+}^{s'_{\rho(n-1)}-k'_{\rho(n-1)}+k_{\sigma(n)}}}{\rvert \sum_{l=1}^{n-1} p_{\sigma(l)}+\sum_{l=1}^{n-1} p'_{\rho(l)}+q\rvert^2}\nonumber\\
 &\frac{(\sum_{l=1}^n p_{\sigma(l)}+\sum_{l=1}^{n-1} p'_{\rho(l)} +q)_{+}^{s_{\sigma(n)}-k_{\sigma(n)}+k'_{\rho(n)}}}{\rvert\sum_{l=1}^n p_{\sigma(l)}+\sum_{l=1}^{n-1} p'_{\rho(l)}+q\rvert^2}\frac{(q)_{+}^{s'_{\rho(n)}-k'_{\rho(n)}+k_{\sigma(1)}}}{\rvert q\rvert^2}
 \end{align}
 
 \subsubsection{Minkowskian extended basis}
 
Similarly:
\begin{align}
\nonumber
\Gamma_{conf}[\mathbb{A},\tilde{\mathbb{A}}]=& - \frac{N^2-1}{2}\log \Det\Big(\delta_{s_1k_1,s_2k_2} (2\pi)^4\delta^{(4)}(q_1-q_2) \\\nonumber
&+\mathcal{D}^{-1}_{s_1k_1,s_2k_2}(q_1)(\mathbb{A}_{s_2k_2}(q_1-q_2)+ \tilde{\mathbb{A}}_{s_2k_2}(q_1-q_2)\Big)\\\nonumber
& - \frac{N^2-1}{2}\log \Det \Big(\delta_{s_1k_1,s_2k_2}(2\pi)^4 \delta^{(4)}(q_1-q_2) \\\nonumber
&+\mathcal{D}^{-1}_{s_1k_1,s_2k_2}(q_1)(\mathbb{A}_{s_2k_2}(q_1-q_2)- \tilde{\mathbb{A}}_{s_2k_2}(q_1-q_2)\Big)\\\nonumber
\Gamma_{conf}[\mathbb{B},\bar{\mathbb{B}}]=& -\frac{N^2-1}{2}\log \Det\Big(\delta_{s_1k_1,s_2k_2}(2\pi)^4\delta^{(4)}(q_1-q_2)\\
\qquad\qquad\qquad&-2\int \frac{d^4q}{(2\pi)^4}\sum_{sk}\mathcal{D}^{-1}_{s_1k_1,sk}(q_1)\bar{\mathbb{B}}_{sk}(q_1-q)\mathcal{D}^{-1}_{sk,s_2k_2}(q)\mathbb{B}_{s_2k_2}(q-q_2)\Big)\\\nonumber
\end{align}
with:
 \begin{align}
 \mathcal{D}^{-1}_{s_1k_1,s_2k_2}(p) =\frac{i^{s_1}}{2}{s_1\choose k_1}{s_2\choose k_2}(-ip_{+})^{s_1-k_1+k_2}\frac{-i}{\rvert p\rvert^2+i\epsilon} 
 \end{align}
 Explicitly, for the nonvanishing correlators of the balanced operators in the momentum representation, we obtain:
\begin{align}
\nonumber
&\langle \mathbb{A}_{s_1}(p_1)\ldots \mathbb{A}_{s_n}(p_n)\tilde{\mathbb{A}}_{s_{n+1}}(p_{n+1})\ldots \tilde{\mathbb{A}}_{s_{n+2m}}(p_{n+2m})\rangle_{conn} \\\nonumber
&=\frac{N^2-1}{2^{n+2m}}(2\pi)^4 i^{n+2m}\delta^{(4)}(p_1+\ldots+p_{n+2m})\\\nonumber
&\sum_{k_1=0}^{s_1}\ldots \sum_{k_{n+2m} = 0}^{s_{n+2m}}{s_1\choose k_1}{s_1\choose k_1}\ldots {s_{n+2m}\choose k_{n+2m}}{s_{n+2m}\choose k_{n+2m}}\frac{1}{n+2m}\sum_{\sigma\in P_{n+2m}}\\\nonumber
&\int \frac{d^4 q}{(2\pi)^4}\frac{(p_{\sigma(1)}+q)_{+}^{s_{\sigma(1)}-k_{\sigma(1)}+k_{\sigma(2)}}}{\rvert p_{\sigma(1)}+q\rvert^2}\frac{(p_{\sigma(1)}+p_{\sigma(2)}+q)_{+}^{s_{\sigma(2)}-k_{\sigma(2)}+k_{\sigma(3)}}}{\rvert p_{\sigma(1)}+p_{\sigma(2)}+q\rvert^2}\\
&\ldots\frac{(\sum_{l=1}^{n+2m-1} p_{\sigma(l)}+q)_{+}^{s_{\sigma(n+2m-1)}-k_{\sigma(n+2m-1)}+k_{\sigma(n+2m)}}}{\rvert\sum_{l=1}^{n+2m-1} p_{\sigma(l)}+q\rvert^2}\frac{(q)_{+}^{s_{\sigma(n+2m)}-k_{\sigma(n+2m)}+k_{\sigma(1)}}}{\rvert q\rvert^2}
\end{align}
In the unbalanced sector, we get:
\begin{align}
\nonumber
&\langle \mathbb{B}_{s_1}(p_1)\ldots \mathbb{B}_{s_n}(p_n)\bar{\mathbb{B}}_{s'_1}(p'_1)\ldots \bar{\mathbb{B}}_{s'_n}(p'_n)\rangle\\\nonumber
&=\frac{N^2-1}{2^{2n}}(2\pi)^4 i^{2n}\delta^{(4)}\left(\sum_{l=1}^n p_l+p'_l\right)\\\nonumber
&\sum_{k_1=0}^{s_1}\ldots \sum_{k_n = 0}^{s_n}\sum_{k'_1=0}^{s'_1}\ldots \sum_{k'_n = 0}^{s'_n}{s_1\choose k_1}{s_1\choose k_1}\ldots {s_n\choose k_n}{s_n\choose k_n}{s'_1\choose k'_1}{s'_1\choose k'_1}\ldots {s'_n\choose k'_n}{s'_n\choose k'_n}\\\nonumber
&\frac{2^{n-1}}{n}\sum_{\sigma\in P_n}\sum_{\rho\in P_n}
\int \frac{d^4 q}{(2\pi)^4}\frac{(p_{\sigma(1)}+q)_{+}^{s_{\sigma(1)}-k_{\sigma(1)}+k'_{\rho(1)}}}{\rvert p_{\sigma(1)}+q\rvert^2}\frac{(p_{\sigma(1)}+p'_{\rho(1)}+q)_{+}^{s'_{\rho(1)}-k'_{\rho(1)}+k_{\sigma(2)}}}{\rvert p_{\sigma(1)}+p'_{\rho(1)}+q\rvert^2}\\\nonumber
&\frac{(p_{\sigma(1)}+p_{\sigma(2)}+p'_{\rho(1)}+q)_{+}^{s_{\sigma(2)}-k_{\sigma(2)}+k'_{\rho(2)}}}{\rvert p_{\sigma(1)}+p_{\sigma(2)}+p'_{\rho(1)}+q\rvert^2}\frac{(p_{\sigma(1)}+p_{\sigma(2)}+p'_{\rho(1)}+p'_{\rho(2)}+q)_{+}^{s'_{\rho(2)}-k'_{\rho(2)}+k_{\sigma(3)}}}{\rvert p_{\sigma(1)}+p_{\sigma(2)}+p'_{\rho(1)}+p'_{\rho(2)}+q\rvert^2}\\
 &\ldots\frac{(\sum_{l=1}^{n-1} p_{\sigma(l)}+\sum_{l=1}^{n-2} p'_{\rho(l)} +q)_{+}^{s_{\sigma(n-1)}-k_{\sigma(n-1)}+k'_{\rho(n-1)}}}{\rvert\sum_{l=1}^{n-1} p_{\sigma(l)}+\sum_{l=1}^{n-2} p'_{\rho(l)}+q\rvert^2}\nonumber\\
 &\frac{(\sum_{l=1}^{n-1} p_{\sigma(l)}+\sum_{l=1}^{n-1} p'_{\rho(l)}+q)_{+}^{s'_{\rho(n-1)}-k'_{\rho(n-1)}+k_{\sigma(n)}}}{\rvert \sum_{l=1}^{n-1} p_{\sigma(l)}+\sum_{l=1}^{n-1} p'_{\rho(l)}+q\rvert^2}\nonumber\\
&\frac{(\sum_{l=1}^n p_{\sigma(l)}+\sum_{l=1}^{n-1} p'_{\rho(l)} +q)_{+}^{s_{\sigma(n)}-k_{\sigma(n)}+k'_{\rho(n)}}}{\rvert\sum_{l=1}^n p_{\sigma(l)}+\sum_{l=1}^{n-1} p'_{\rho(l)}+q\rvert^2}\frac{(q)_{+}^{s'_{\rho(n)}-k'_{\rho(n)}+k_{\sigma(1)}}}{\rvert q\rvert^2}
\end{align}

\subsubsection{Euclidean standard basis}

The Euclidean generating functionals read in the momentum representation:
\begin{align}
\nonumber
\Gamma^E_{conf}[\mathbb{O}^E,\tilde{\mathbb{O}}^E]
=&- \frac{N^2-1}{2}\log \Det\Big(\delta_{s_1k_1,s_2k_2}(2\pi)^4 \delta^{(4)}(q_1-q_2) \\\nonumber
&+{\mathcal{D}_E^{-1}}_{s_1k_1,s_2k_2}(q_1)(\mathbb{O}^E_{s_2k_2}(q_1-q_2)+ \tilde{\mathbb{O}}^E_{s_2k_2}(q_1-q_2)\Big)\\\nonumber
& - \frac{N^2-1}{2}\log \Det\Big(\delta_{s_1k_1,s_2k_2}(2\pi)^4 \delta^{(4)}(q_1-q_2) \\\nonumber
&+{\mathcal{D}_E^{-1}}_{s_1k_1,s_2k_2}(q_1)(\mathbb{O}^E_{s_2k_2}(q_1-q_2)- \tilde{\mathbb{O}}^E_{s_2k_2}(q_1-q_2)\Big)\\\nonumber
\Gamma^E_{conf}[\mathbb{S}^E,\bar{\mathbb{S}}^E]=&- \frac{N^2-1}{2}\log \Det\Big(\delta_{s_1k_1,s_2k_2} (2\pi)^4 \delta^{(4)}(q_1-q_2)\\
\qquad\qquad\qquad&-2\int \frac{d^4q}{(2\pi)^4}\sum_{sk}{\mathcal{D}_E^{-1}}_{s_1k_1,sk}(q_1)\bar{\mathbb{S}}^E_{sk}(q_1-q){\mathcal{D}_E^{-1}}_{sk,s_2k_2}(q)\mathbb{S}^E_{s_2k_2}(q-q_2)\Big)\\\nonumber
\end{align}
with:
\begin{align}
{\mathcal{D}^{-1}_E}_{s_1k_1,s_2k_2}(p) &=\frac{i^{s_1}}{2}\frac{\Gamma(3)\Gamma(s_1+3)}{\Gamma(5)\Gamma(s_1+1)}{s_1\choose k_1}{s_2\choose k_2+2}p_{z}^{s_1-k_1+k_2}\frac{1}{p^2} 
\end{align}
Explicitly, for the nonvanishing correlators of the balanced operators in the momentum representation, we obtain:
\begin{align}
\nonumber
&\left(\frac{\Gamma(5)\Gamma(s_1+1)}{\Gamma(3)\Gamma(s_1+3)}\right)\ldots\left( \frac{\Gamma(5)\Gamma(s_{n+2m}+1)}{\Gamma(3)\Gamma(s_{n+2m}+3)}\right)\\\nonumber
&\langle \mathbb{O}^E_{s_1}(p_1)\ldots \mathbb{O}^E_{s_n}(p_n)\tilde{\mathbb{O}}^E_{s_{n+1}}(p_{n+1})\ldots \tilde{\mathbb{O}}^E_{s_{n+2m}}(p_{n+2m})\rangle_{conn} \\\nonumber
&=\frac{N^2-1}{2^{n+2m}}(2\pi)^4 i^{\sum_{l=1}^{n+2m} s_l} \delta^{(4)}(p_1+\ldots+p_{n+2m})\\\nonumber
&\sum_{k_1=0}^{s_1-2}\ldots \sum_{k_{n+2m} = 0}^{s_{n+2m}-2}{s_1\choose k_1}{s_1\choose k_1+2}\ldots {s_{n+2m}\choose k_{n+2m}}{s_{n+2m}\choose k_{n+2m}+2}\frac{1}{n+2m}\sum_{\sigma\in P_{n+2m}}\\\nonumber
&\int \frac{d^4 q}{(2\pi)^4}\frac{(p_{\sigma(1)}+q)_{z}^{s_{\sigma(1)}-k_{\sigma(1)}+k_{\sigma(2)}}}{( p_{\sigma(1)}+q)^2}\frac{(p_{\sigma(1)}+p_{\sigma(2)}+q)_{z}^{s_{\sigma(2)}-k_{\sigma(2)}+k_{\sigma(3)}}}{( p_{\sigma(1)}+p_{\sigma(2)}+q)^2}\\
&\ldots\frac{(\sum_{l=1}^{n+2m-1} p_{\sigma(l)}+q)_{z}^{s_{\sigma(n+2m-1)}-k_{\sigma(n+2m-1)}+k_{\sigma(n+2m)}}}{(\sum_{l=1}^{n+2m-1} p_{\sigma(l)}+q)^2}\frac{(q)_{z}^{s_{\sigma(n+2m)}-k_{\sigma(n+2m)}+k_{\sigma(1)}}}{(q)^2}
\end{align}
In the unbalanced sector, we get:
\begin{align}
\nonumber
&\left(\frac{\Gamma(5)\Gamma(s_1+1)}{\Gamma(3)\Gamma(s_1+3)}\right)\ldots\left(\frac{\Gamma(5)\Gamma(s_n+1)}{\Gamma(3)\Gamma(s_n+3)}\right)\left(\frac{\Gamma(5)\Gamma(s'_1+1)}{\Gamma(3)\Gamma(s'_1+3)}\right)\ldots\left( \frac{\Gamma(5)\Gamma(s'_{n}+1)}{\Gamma(3)\Gamma(s'_{n}+3)}\right)\\\nonumber
&\langle \mathbb{S}^E_{s_1}(p_1)\ldots \mathbb{S}^E_{s_n}(p_n)\bar{\mathbb{S}}^E_{s'_1}(p'_1)\ldots \bar{\mathbb{S}}^E_{s'_n}(p'_n)\rangle\\\nonumber
&=\frac{N^2-1}{2^{2n}}(2\pi)^4 i^{\sum_{l=1}^{n} s_l+s'_l}\delta^{(4)}\left(\sum_{l=1}^n p_l+p'_l\right)\\\nonumber
&\sum_{k_1=0}^{s_1-2}\ldots \sum_{k_n = 0}^{s_n-2}\sum_{k'_1=0}^{s'_1-2}\ldots \sum_{k'_n = 0}^{s'_n-2}{s_1\choose k_1}{s_1\choose k_1+2}\ldots {s_n\choose k_n}{s_n\choose k_n+2}\\\nonumber
&{s'_1\choose k'_1}{s'_1\choose k'_1+2}\ldots {s'_n\choose k'_n}{s'_n\choose k'_n+2}\\\nonumber
&\frac{2^{n-1}}{n}\sum_{\sigma\in P_n}\sum_{\rho\in P_n}
\int \frac{d^4 q}{(2\pi)^4}\frac{(p_{\sigma(1)}+q)_{z}^{s_{\sigma(1)}-k_{\sigma(1)}+k'_{\rho(1)}}}{( p_{\sigma(1)}+q)^2}\frac{(p_{\sigma(1)}+p'_{\rho(1)}+q)_{z}^{s'_{\rho(1)}-k'_{\rho(1)}+k_{\sigma(2)}}}{( p_{\sigma(1)}+p'_{\rho(1)}+q)^2}\\\nonumber
&\frac{(p_{\sigma(1)}+p_{\sigma(2)}+p'_{\rho(1)}+q)_{z}^{s_{\sigma(2)}-k_{\sigma(2)}+k'_{\rho(2)}}}{( p_{\sigma(1)}+p_{\sigma(2)}+p'_{\rho(1)}+q)^2}\frac{(p_{\sigma(1)}+p_{\sigma(2)}+p'_{\rho(1)}+p'_{\rho(2)}+q)_{z}^{s'_{\rho(2)}-k'_{\rho(2)}+k_{\sigma(3)}}}{( p_{\sigma(1)}+p_{\sigma(2)}+p'_{\rho(1)}+p'_{\rho(2)}+q)^2}\\
 &\ldots\frac{(\sum_{l=1}^{n-1} p_{\sigma(l)}+\sum_{l=1}^{n-2} p'_{\rho(l)} +q)_{z}^{s_{\sigma(n-1)}-k_{\sigma(n-1)}+k'_{\rho(n-1)}}}{(\sum_{l=1}^{n-1} p_{\sigma(l)}+\sum_{l=1}^{n-2} p'_{\rho(l)}+q)^2}\nonumber\\
 &\frac{(\sum_{l=1}^{n-1} p_{\sigma(l)}+\sum_{l=1}^{n-1} p'_{\rho(l)}+q)_{z}^{s'_{\rho(n-1)}-k'_{\rho(n-1)}+k_{\sigma(n)}}}{( \sum_{l=1}^{n-1} p_{\sigma(l)}+\sum_{l=1}^{n-1} p'_{\rho(l)}+q)^2}\nonumber\\
&\frac{(\sum_{l=1}^n p_{\sigma(l)}+\sum_{l=1}^{n-1} p'_{\rho(l)} +q)_{z}^{s_{\sigma(n)}-k_{\sigma(n)}+k'_{\rho(n)}}}{(\sum_{l=1}^n p_{\sigma(l)}+\sum_{l=1}^{n-1} p'_{\rho(l)}+q)^2}\frac{(q)_{+}^{s'_{\rho(n)}-k'_{\rho(n)}+k_{\sigma(1)}}}{( q)^2}
\end{align}

\subsubsection{Euclidean extended basis}

Analogously:
\begin{align}
\nonumber
\Gamma^E_{conf}[\mathbb{A}^E,\tilde{\mathbb{A}}^E]
=&- \frac{N^2-1}{2}\log \Det\Big(\delta_{s_1k_1,s_2k_2} (2\pi)^4\delta^{(4)}(q_1-q_2) \\\nonumber
&+{\mathcal{D}_E^{-1}}_{s_1k_1,s_2k_2}(q_1)(\mathbb{A}^E_{s_2k_2}(q_1-q_2)+ \tilde{\mathbb{A}}^E_{s_2k_2}(q_1-q_2)\Big)\\\nonumber
& - \frac{N^2-1}{2}\log \Det\Big(\delta_{s_1k_1,s_2k_2} (2\pi)^4\delta^{(4)}(q_1-q_2) \\\nonumber
&+{\mathcal{D}_E^{-1}}_{s_1k_1,s_2k_2}(q_1)(\mathbb{O}^E_{s_2k_2}(q_1-q_2)- \tilde{\mathbb{A}}^E_{s_2k_2}(q_1-q_2)\Big)\\\nonumber
\Gamma^E_{conf}[\mathbb{B}^E,\bar{\mathbb{B}}^E]=&- \frac{N^2-1}{2}\log \Det\Big(\delta_{s_1k_1,s_2k_2} (2\pi)^4 \delta^{(4)}(q_1-q_2)\\
\qquad\qquad\qquad&-2\int \frac{d^4q}{(2\pi)^4}\sum_{sk}{\mathcal{D}_E^{-1}}_{s_1k_1,sk}(q_1)\bar{\mathbb{B}}^E_{sk}(q_1-q){\mathcal{D}_E^{-1}}_{sk,s_2k_2}(q)\mathbb{B}^E_{s_2k_2}(q-q_2)\Big)\\\nonumber
\end{align}
with:
\begin{align}
{\mathcal{D}^{-1}_E}_{s_1k_1,s_2k_2}(p) &=\frac{i^{s_1}}{2}{s_1\choose k_1}{s_2\choose k_2}p_{z}^{s_1-k_1+k_2}\frac{1}{p^2} 
\end{align}
Explicitly, for the nonvanishing correlators in the balanced sector, we obtain:
\begin{align}
\nonumber
&\langle \mathbb{A}^E_{s_1}(p_1)\ldots \mathbb{A}^E_{s_n}(p_n)\tilde{\mathbb{A}}^E_{s_{n+1}}(p_{n+1})\ldots \tilde{\mathbb{A}}^E_{s_{n+2m}}(p_{n+2m})\rangle_{conn} \\\nonumber
&=\frac{N^2-1}{2^{n+2m}}(2\pi)^4 i^{\sum_{l=1}^{n+2m} s_l} \delta^{(4)}(p_1+\ldots+p_{n+2m})\\\nonumber
&\sum_{k_1=0}^{s_1}\ldots \sum_{k_{n+2m} = 0}^{s_{n+2m}}{s_1\choose k_1}{s_1\choose k_1}\ldots {s_{n+2m}\choose k_{n+2m}}{s_{n+2m}\choose k_{n+2m}}\frac{1}{n+2m}\sum_{\sigma\in P_{n+2m}}\\\nonumber
&\int \frac{d^4 q}{(2\pi)^4}\frac{(p_{\sigma(1)}+q)_{z}^{s_{\sigma(1)}-k_{\sigma(1)}+k_{\sigma(2)}}}{( p_{\sigma(1)}+q)^2}\frac{(p_{\sigma(1)}+p_{\sigma(2)}+q)_{z}^{s_{\sigma(2)}-k_{\sigma(2)}+k_{\sigma(3)}}}{( p_{\sigma(1)}+p_{\sigma(2)}+q)^2}\\
&\ldots\frac{(\sum_{l=1}^{n+2m-1} p_{\sigma(l)}+q)_{z}^{s_{\sigma(n+2m-1)}-k_{\sigma(n+2m-1)}+k_{\sigma(n+2m)}}}{(\sum_{l=1}^{n+2m-1} p_{\sigma(l)}+q)^2}\frac{(q)_{z}^{s_{\sigma(n+2m)}-k_{\sigma(n+2m)}+k_{\sigma(1)}}}{(q)^2}
\end{align}
In the unbalanced sector, we get:
\begin{align}
\nonumber
&\langle \mathbb{B}^E_{s_1}(p_1)\ldots \mathbb{B}^E_{s_n}(p_n)\bar{\mathbb{B}}^E_{s'_1}(p'_1)\ldots \bar{\mathbb{B}}^E_{s'_n}(p'_n)\rangle\\\nonumber
&=\frac{N^2-1}{2^{2n}}(2\pi)^4 i^{\sum_{l=1}^{n} s_l+s'_l}\delta^{(4)}\left(\sum_{l=1}^n p_l+p'_l\right)\\\nonumber
&\sum_{k_1=0}^{s_1}\ldots \sum_{k_n = 0}^{s_n}\sum_{k'_1=0}^{s'_1}\ldots \sum_{k'_n = 0}^{s'_n}{s_1\choose k_1}{s_1\choose k_1}\ldots {s_n\choose k_n}{s_n\choose k_n}{s'_1\choose k'_1}{s'_1\choose k'_1}\ldots {s'_n\choose k'_n}{s'_n\choose k'_n}\\\nonumber
&\frac{2^{n-1}}{n}\sum_{\sigma\in P_n}\sum_{\rho\in P_n}
\int \frac{d^4 q}{(2\pi)^4}\frac{(p_{\sigma(1)}+q)_{z}^{s_{\sigma(1)}-k_{\sigma(1)}+k'_{\rho(1)}}}{( p_{\sigma(1)}+q)^2}\frac{(p_{\sigma(1)}+p'_{\rho(1)}+q)_{z}^{s'_{\rho(1)}-k'_{\rho(1)}+k_{\sigma(2)}}}{( p_{\sigma(1)}+p'_{\rho(1)}+q)^2}\\\nonumber
&\frac{(p_{\sigma(1)}+p_{\sigma(2)}+p'_{\rho(1)}+q)_{z}^{s_{\sigma(2)}-k_{\sigma(2)}+k'_{\rho(2)}}}{( p_{\sigma(1)}+p_{\sigma(2)}+p'_{\rho(1)}+q)^2}\frac{(p_{\sigma(1)}+p_{\sigma(2)}+p'_{\rho(1)}+p'_{\rho(2)}+q)_{z}^{s'_{\rho(2)}-k'_{\rho(2)}+k_{\sigma(3)}}}{( p_{\sigma(1)}+p_{\sigma(2)}+p'_{\rho(1)}+p'_{\rho(2)}+q)^2}\\
 &\ldots\frac{(\sum_{l=1}^{n-1} p_{\sigma(l)}+\sum_{l=1}^{n-2} p'_{\rho(l)} +q)_{z}^{s_{\sigma(n-1)}-k_{\sigma(n-1)}+k'_{\rho(n-1)}}}{(\sum_{l=1}^{n-1} p_{\sigma(l)}+\sum_{l=1}^{n-2} p'_{\rho(l)}+q)^2}\nonumber\\
 &\frac{(\sum_{l=1}^{n-1} p_{\sigma(l)}+\sum_{l=1}^{n-1} p'_{\rho(l)}+q)_{z}^{s'_{\rho(n-1)}-k'_{\rho(n-1)}+k_{\sigma(n)}}}{( \sum_{l=1}^{n-1} p_{\sigma(l)}+\sum_{l=1}^{n-1} p'_{\rho(l)}+q)^2}\nonumber\\
&\frac{(\sum_{l=1}^n p_{\sigma(l)}+\sum_{l=1}^{n-1} p'_{\rho(l)} +q)_{z}^{s_{\sigma(n)}-k_{\sigma(n)}+k'_{\rho(n)}}}{(\sum_{l=1}^n p_{\sigma(l)}+\sum_{l=1}^{n-1} p'_{\rho(l)}+q)^2}\frac{(q)_{+}^{s'_{\rho(n)}-k'_{\rho(n)}+k_{\sigma(1)}}}{( q)^2}
\end{align}
 
 \section{Plan of the paper}
 
In section \ref{0} we outline our main results and physics motivations. \par
In section \ref{00}, after recalling some basic concepts about the employment of conformal symmetry in massless QCD-like theories, we display our results for the correlators and their generating functionals to the lowest perturbative order both in Minkowskian and Euclidean space-time. \par
In section \ref{operators} we review the classification and construction of the gluonic operators with collinear twist $2$ in Minkowskian space-time both in the standard and extended basis.\par
In section \ref{coordinates} we compute the $2$-point correlators in Minkowskian space-time both in the standard and extended basis.\par
In section \ref{threepoint} we compute the $3$-point correlators in Minkowskian space-time both in the standard and extended basis.\par
In section \ref{npoint} we compute the $n$-point correlators in Minkowskian space-time both in the standard and extended basis in the balanced and unbalanced sectors separately. \par
In section \ref{8} we compute the $n$-point correlators in Euclidean space-time either by analytic continuation or by employing the corresponding Euclidean operators.\par
In section \ref{444} we work out an ansatz for the generating functionals of the $n$-point correlators in the coordinate representation as the logarithm of a functional determinant, and we verify by direct computation that it actually reproduces the correlators in the coordinate representation.\par
In section \ref{55} we work out the corresponding generating functionals in the momentum representation, and we employ them to actually compute the $n$-point correlators in the momentum representation. \par
In the appendices we fix the notation and provide ancillary computations.\par
In particular, in appendix \ref{appcheck} we verify that our results for the $2$- and $3$-point correlators of balanced operators with even collinear spin in Minkowskian space-time coincide with \cite{Kazakov:2012ar}. 

\section{Twist-$2$ gluonic operators in Minkowskian space-time \label{operators}}

We review the construction of the standard and extended conformal bases for gauge-invariant gluonic operators with $\tau=2$. We also work out the dictionary between the spinorial, vectorial and complex bases (appendices \ref{appA2} and \ref{appC}).

\subsection{Standard basis \label{standardb}}

To construct gauge-invariant gluonic operators with $\tau=2$ that are primary (section \ref{00}) for the collinear conformal group, we should find -- according to eq. (\ref{maximum2}) -- suitable gauge-covariant elementary conformal operators. \par
The local gauge-covariant operator with lowest canonical dimension, $d=2$, in YM theory is the field-strength tensor, $F_{\mu\nu}=\partial_{\mu}A_\nu-\partial_{\nu}A_\nu+ig\left[A_\mu,A_\nu\right]$, where $A_{ \mu}=A_\mu^a T^a$ is a traceless Hermitian matrix, with $T^{a}$ the generators of the $SU(N)$ Lie algebra:
\begin{equation}
[T^a,T^b]=if^{abc}T^c
\end{equation}
normalized in the standard way:
\begin{equation}
\label{colour_trace}
\Tr (T^aT^b)=\frac{1}{2}\delta^{ab}
\end{equation}
It is convenient to write $F_{\mu\nu}$ in the spinorial representation \cite{Braun:2008ia} (appendix \ref{appA2}):
\begin{equation}
F_{a\dot{a}b\dot{b}} = \sigma_{a\dot{a}}^\mu\sigma_{b\dot{b}}^\nu F_{\mu\nu}
\end{equation}
It turns out \cite{Braun:2008ia} that:
\begin{equation}
F_{a\dot{a}b\dot{b}} = 2\left(f_{ab}\epsilon_{\dot{a}\dot{b}}- \epsilon_{ab} f_{\dot{a}\dot{b}}\right)
\end{equation}
decomposes into the sum of two chiral representations, $f_{ab} \in (1,0)$ and $f_{\dot{a}\dot{b}}\in (0,1)$, of spin $S=1$ (appendix \ref{appA2}). 
In Minkowskian space-time $f_{\dot{a}\dot{b}} = \bar{f}_{ab}$.
$f_{11}$ and $f_{\dot{1}\dot{1}}$ have maximal collinear spin (appendix \ref{appA2}), $s = 1$, along the $p_+$ direction. Therefore, they have $\tau = d-s = 1$ and $j = s+\frac{\tau}{2} = \frac{3}{2}$.
Hence, they are well suited (section \ref{00}) to build $2$-gluon twist-$2$ primary conformal operators \cite{Ohrndorf:1981qv,conformalops}.
Taking the tensor product of the above representations, we get:
\begin{align}
&(1,0)\otimes(0,1) = (1,1)_+\oplus(1,1)_-\\\nonumber
&(1,0)\otimes(1,0) = (2,0)\oplus\ldots\\\nonumber
&(0,1)\otimes(0,1) = (0,2)\oplus\ldots
\end{align}
where $+$ and $-$ label the parity, and the dots denote terms that do not contribute to the components with maximal collinear spin. Hence, there are four operators with maximal $s$ that can be constructed by means of the corresponding bilinear operators:
\begin{align} \label{44}
&O(x_1,x_2) = f_{11}(x_1) f_{\dot{1}\dot{1}}(x_2) + f_{11}(x_2) f_{\dot{1}\dot{1}}(x_1)\\\nonumber
&O(x_1,x_2) = f_{11}(x_1) f_{\dot{1}\dot{1}}(x_2)- f_{11}(x_2) f_{\dot{1}\dot{1}}(x_1)\\\nonumber
&S(x_1,x_2)= \frac{1}{\sqrt{2}}f_{11}(x_1)f_{11}(x_2)\\\nonumber
&\bar{S}(x_1,x_2)= \frac{1}{\sqrt{2}}f_{\dot{1}\dot{1}}(x_2) f_{\dot{1}\dot{1}}(x_1)
\end{align}
Following \cite{Ohrndorf:1981qv, conformalops, Braun:2003rp}, we build (section \ref{00}) conformal operators with $\tau=2$ and higher collinear spin inserting in eq. \eqref{44} the Gegenbauer polynomials (appendix \ref{appB}), $C^{\alpha}_l(v)$, in the derivatives and afterwards taking the local limit $x_1=x_2$. $l$ is the order of the polynomial, and its relation to the collinear conformal spin is $j =l+j_1+j_2$, where $j_1$ and $j_2$ are the collinear conformal spins of the elementary operators, and $\alpha = 2j_1-\frac{1}{2}$. The Gegenbauer polynomials are either symmetric or antisymmetric for the substitution $v \rightarrow-v$ for $l$ even or odd respectively (appendix \ref{appB}). 
The corresponding primary conformal operators match precisely the ones in \cite{Belitsky:2003sh, Beisert:2004fv, conformalops} up to perhaps the overall normalization:
\begin{align}
\label{basis}
\nonumber
&\mathbb{O}_{s} = \Tr f_{11}(x)(i\overrightarrow{D}_++i\overleftarrow{D}_+)^{s-2}C^{\frac{5}{2}}_{s-2}\left(\frac{\overrightarrow{D}_+-\overleftarrow{D}_+}{\overrightarrow{D}_++\overleftarrow{D}_+}\right) f_{\dot{1}\dot{1}}(x) \qquad s = 2,4,6,\ldots\\\nonumber
&\tilde{\mathbb{O}}_{s} = \Tr f_{11}(x)(i\overrightarrow{D}_++i\overleftarrow{D}_+)^{s-2}C^{\frac{5}{2}}_{s-2}\left(\frac{\overrightarrow{D}_+-\overleftarrow{D}_+}{\overrightarrow{D}_++\overleftarrow{D}_+}\right) f_{\dot{1}\dot{1}}(x) \qquad s = 3,5,7,\ldots\\\nonumber
&\mathbb{S}_{s} =\frac{1}{\sqrt{2}}\Tr f_{11}(x)(i\overrightarrow{D}_++i\overleftarrow{D}_+)^{s-2}C^{\frac{5}{2}}_{s-2}\left(\frac{\overrightarrow{D}_+-\overleftarrow{D}_+}{\overrightarrow{D}_++\overleftarrow{D}_+}\right)f_{11}(x) \qquad s = 2,4,6,\ldots\\
&\bar{\mathbb{S}}_{s} =\frac{1}{\sqrt{2}}\Tr {f}_{\dot{1}\dot{1}}(x)(i\overrightarrow{D}_++i\overleftarrow{D}_+)^{s-2}C^{\frac{5}{2}}_{s-2}\left(\frac{\overrightarrow{D}_+-\overleftarrow{D}_+}{\overrightarrow{D}_++\overleftarrow{D}_+}\right){f}_{\dot{1}\dot{1}}(x)\qquad s = 2,4,6,\ldots
\end{align}
with $j = s+1$, $l =s-2$ and $\alpha = \frac{5}{2}$.
For brevity, we define as in \cite{Kazakov:2012ar}:
\begin{equation}
\label{eq46}
\mathcal{G}_l^\alpha(D_{x^+_1},D_{x^+_2}) = i^l (\overrightarrow{D}_{x^+_2}+\overleftarrow{D}_{x^+_1})^l C^{\alpha}_l\left(\frac{\overrightarrow{D}_{x^+_2}-\overleftarrow{D}_{x^+_1}}{\overrightarrow{D}_{x^+_2}+\overleftarrow{D}_{x^+_1}}\right)
\end{equation}
For $l =s-2$ and $\alpha = \frac{5}{2}$, we get in the light-cone gauge by means eq. (\ref{physicalgegen}):
\begin{align}
\label{defg52}
\mathcal{G}^{\frac{5}{2}}_{s-2}(\partial_{x^+_1},\partial_{x^+_2}) &=i^{s-2}\frac{\Gamma(s+3)\Gamma(3)}{\Gamma(5)\Gamma(s+1)}\sum_{k=0}^{s-2} {s\choose k}{s\choose k+2}(-1)^{s-k} \overleftarrow{\partial}_{x^+_1}^{s-k-2} \overrightarrow{\partial}_{x^+_2}^k\nonumber\\
 &=\frac{2i^{s-2}(s+1)(s+2)}{4!}\sum_{k=0}^{s-2} {s\choose k}{s\choose k+2}(-1)^{s-k} \overleftarrow{\partial}_{x^+_1}^{s-k-2} \overrightarrow{\partial}_{x^+_2}^k
\end{align}

\subsection{Extended basis \label{nonstandardb}}

There exists another choice of the basis for primary conformal operators with $\tau=2$ involving the elementary operators $D_{+}^{-1}f_{11}$ and $D_{+}^{-1} f_{\dot{1}\dot{1}}$, which are nonlocal in general, but local in the light-cone gauge. Yet, gauge invariance ensures that the corresponding gauge-invariant correlators are local. Indeed, in the light-cone gauge (appendix \ref{appC}):
\begin{align}
\partial_{+}^{-1}f_{11} = -\bar{A}\nonumber \\
 \partial_{+}^{-1} f_{\dot{1}\dot{1}} =  -{A} 
\end{align}
where $A$ has $d = 1$, $s = 0$, $ j =\frac{1}{2}$ and $\tau = 1$. The corresponding operators with $\tau=2$ read:
\begin{align}
\nonumber
\label{basisSquasilocal}
&\mathbb{A}_{s} = \Tr D_{+}^{-1}f_{11}(x)\,(i\overrightarrow{D}_++i\overleftarrow{D}_+)^{s}C^{\frac{1}{2}}_{s}\left(\frac{\overrightarrow{D}_+-\overleftarrow{D}_+}{\overrightarrow{D}_++\overleftarrow{D}_+}\right)D_{+}^{-1} f_{\dot{1}\dot{1}}(x)  \qquad s = 0,2,4,\ldots\\\nonumber
&\tilde{\mathbb{A}}_{s} = \Tr D_{+}^{-1}f_{11}(x)\,(i\overrightarrow{D}_++i\overleftarrow{D}_+)^{s}C^{\frac{1}{2}}_{s}\left(\frac{\overrightarrow{D}_+-\overleftarrow{D}_+}{\overrightarrow{D}_++\overleftarrow{D}_+}\right) D_{+}^{-1} f_{\dot{1}\dot{1}}(x)  \qquad s = 1,3,5,\ldots\\\nonumber
&\mathbb{B}_{s} =\frac{1}{\sqrt{2}}\Tr D_{+}^{-1}f_{11}(x)\,(i\overrightarrow{D}_++i\overleftarrow{D}_+)^{s}C^{\frac{1}{2}}_{s}\left(\frac{\overrightarrow{D}_+-\overleftarrow{D}_+}{\overrightarrow{D}_++\overleftarrow{D}_+}\right)D_{+}^{-1}f_{11}(x)\qquad s = 0,2,4,\ldots\\
&\bar{\mathbb{B}}_{s} =\frac{1}{\sqrt{2}}\Tr D_{+}^{-1} f_{\dot{1}\dot{1}}(x)\,(i\overrightarrow{D}_++i\overleftarrow{D}_+)^{s}C^{\frac{1}{2}}_{s}\left(\frac{\overrightarrow{D}_+-\overleftarrow{D}_+}{\overrightarrow{D}_++\overleftarrow{D}_+}\right)D_{+}^{-1} f_{\dot{1}\dot{1}}(x) \qquad s = 0,2,4,\ldots
\end{align}
with $j = s+1$, $l =s$ and $\alpha = \frac{1}{2}$. This basis naturally arises in SUSY calculations \cite{Belitsky:2004sc}, and also includes (nonlocal) operators with $s=0,1$.
We obtain in the light-cone gauge by means of eq. (\ref{physicalgegen2}):
\begin{equation} \label{AA}
\mathcal{G}^{\frac{1}{2}}_{s}(\partial_{x^+_1},\partial_{x^+_2}) = i^s\sum_{k = 0}^{s}{s\choose k}{s\choose k}(-1)^{s-k}\overleftarrow{\partial}_{x^+_1}^{s-k}\overrightarrow{\partial}_{x^+_2}^k
\end{equation}

\section{$2$-point correlators of twist-$2$ gluonic operators\label{coordinates}}

We compute to the lowest perturbative order the $2$-point correlators of the operators in both bases.

\subsection{Standard basis}

In the light-cone gauge, the $2$-point correlators of the balanced operators with even $s$ are given by:
\begin{align}
\langle \mathbb{O}_{s_1}(x) \mathbb{O}_{s_2}(y)\rangle =&  \mathcal{G}_{s_1-2}^{\frac{5}{2}}(\partial_{x_1^+},\partial_{x_2^+})\mathcal{G}_{s_2-2}^{\frac{5}{2}}(\partial_{y_1^+},\partial_{y_2^+}) \nonumber \\
&\langle\Tr f_{11}(x_1) f_{\dot{1}\dot{1}}(x_2)\,\Tr f_{11}(y_1) f_{\dot{1}\dot{1}}(y_2)\rangle\Big\rvert_{x_1=x_2=x}^{y_1=y_2=y}
\end{align}
There is only one Wick contraction:
\begin{align}
\nonumber
\langle \mathbb{O}_{s_1}(x) \mathbb{O}_{s_2}(y)\rangle =&  \mathcal{G}_{s_1-2}^{\frac{5}{2}}(\partial_{x_1^+},\partial_{x_2^+})\mathcal{G}_{s_2-2}^{\frac{5}{2}}(\partial_{y_1^+},\partial_{y_2^+})\\
&\langle\Tr f_{11}(x_1) f_{\dot{1}\dot{1}}(y_2)\rangle\langle\Tr f_{11}(y_1) f_{\dot{1}\dot{1}}(x_2)\rangle\Big\rvert_{x_1=x_2=x}^{y_1=y_2=y}
\end{align}
By means of eq. \eqref{axialprop2}, we get:
\begin{align}
\label{scalarcorr}
\nonumber
\langle \mathbb{O}_{s_1}(x) \mathbb{O}_{s_2}(y)\rangle =& \frac{1}{(4\pi^2)^2} \frac{N^2-1}{4}\mathcal{G}_{s_1-2}^{\frac{5}{2}}(\partial_{x_1^+},\partial_{x_2^+})\mathcal{G}_{s_2-2}^{\frac{5}{2}}(\partial_{y_1^+},\partial_{y_2^+})\\\nonumber
&\partial^+_{x_1}\partial^+_{x_2}\partial^+_{y_1}\partial^+_{y_2}\frac{1}{\rvert x_1-y_2\rvert^2}\frac{1}{\rvert y_1-x_2\rvert^2}\Big\rvert_{x_1=x_2=x}^{y_1=y_2=y}\\
\end{align}
Now we substitute eq. (\ref{defg52}) into the above equation:
\begin{align}
\nonumber
\langle \mathbb{O}_{s_1}(x) \mathbb{O}_{s_2}(y)\rangle = &\frac{1}{(4\pi^2)^2} \frac{N^2-1}{4} \frac{2^2i^{s_1+s_2-4}}{(4!)^2}(s_1+1)(s_1+2)(s_2+1)(s_2+2)\\\nonumber
&\sum_{k_1 = 0}^{s_1-2}\sum_{k_2 = 0}^{s_2-2}{s_1\choose k_1}{s_1\choose k_1+2}{s_2\choose k_2}{s_2\choose k_2+2}\\
&(-\partial_{x_1^+})^{s_1-k_1-1}\partial^{k_1+1}_{x_2^+}(-\partial_{y_1^+})^{s_2-k_2-1}\partial^{k_2+1}_{y_2^+}\frac{1}{\rvert x_1-y_2\rvert^2}\frac{1}{\rvert y_1-x_2\rvert^2}\Big\rvert_{x_1=x_2=x}^{y_1=y_2=y}
\end{align}
We compute the derivatives:
\begin{align}
\nonumber
\label{doubleder}
&\partial_{x^+}^{i}\partial_{y^+}^{j}\frac{1}{\rvert x-y\rvert^2} =\partial_{x^+}^{i}\partial_{y^+}^{j}\frac{1}{2 (x-y)_+(x-y)_--(x-y)^2_\perp} \\\nonumber
&=(-1)^{j}\partial_{x^+}^{i+j}\frac{1}{2 (x-y)_+(x-y)_--(x-y)^2_\perp} \\
&=(-1)^{i} (i+j)!\,2^{i+j} \frac{(x-y)_+^{i+j}}{(\rvert x-y\rvert^2)^{i+j+1}}
\end{align}
by induction on the index $i$:
\begin{align}
\nonumber
&\partial_{x^+}^{i+1}\partial_{y^+}^{j}\frac{1}{\rvert x-y\rvert^2} =(-1)^{i} (i+j)!\,2^{i+j} \partial_{x^+} \frac{(x-y)_+^{i+j}}{(\rvert x-y\rvert^2)^{i+j+1}}\\\nonumber
&=(-1)^{i} (i+j)!\,2^{i+j} \partial_{x_-}\frac{(x-y)_+^{i+j}}{(2 (x-y)_+(x-y)_--(x-y)^2_\perp)^{i+j+1}}\\\nonumber
&=(-1)^{i} (i+j)!\,2^{i+j} (x-y)_+^{i+j} \frac{-2(i+j+1)(x-y)_+ }{(2 (x-y)_+(x-y)_--(x-y)^2_\perp)^{i+j+2}}\\
&=(-1)^{i+1} (i+j+1)!\,2^{i+j+1} \frac{(x-y)_+^{i+j+1}}{(\rvert x-y\rvert^2)^{i+j+2}}
\end{align}
We obtain:
\begin{align}
\nonumber
\langle \mathbb{O}_{s_1}(x) &\mathbb{O}_{s_2}(y)\rangle = \frac{1}{(4\pi^2)^2} \frac{N^2-1}{4} \frac{2^2i^{s_1+s_2-4}}{(4!)^2}(s_1+1)(s_1+2)(s_2+1)(s_2+2)\\\nonumber
&\sum_{k_1 = 0}^{s_1-2}\sum_{k_2 = 0}^{s_2-2}{s_1\choose k_1}{s_1\choose k_1+2}{s_2\choose k_2}{s_2\choose k_2+2}\\\nonumber
&(-1)^{s_1-k_1-1}\, (-1)^{s_1-k_1-1}2^{s_1-k_1+k_2}(s_1-k_1+k_2)!\frac{(x_1-y_2)_+^{s_1-k_1+k_2}}{(\rvert x_1-y_2\rvert^2)^{s_1-k_1+k_2+1}}\\
&(-1)^{s_2-k_2-1}\,(-1)^{s_2-k_2-1}2^{s_2+k_1-k_2}(s_2+k_1-k_2)!\frac{(y_1-x_2)_+^{s_2+k_1-k_2}}{(\rvert y_1-x_2\rvert^2)^{s_2+k_1-k_2+1}}\Big\rvert_{x_1=x_2=x}^{y_1=y_2=y}
\end{align}
that becomes:
\begin{align}
\nonumber
\langle \mathbb{O}_{s_1}(x) \mathbb{O}_{s_2}(y)\rangle = &\frac{1}{(4\pi^2)^2} \frac{N^2-1}{4} \frac{2^{s_1+s_2+2}i^{s_1+s_2-4}}{(4!)^2}(s_1+1)(s_1+2)(s_2+1)(s_2+2)\\\nonumber
&\sum_{k_1 = 0}^{s_1-2}\sum_{k_2 = 0}^{s_2-2}{s_1\choose k_1}{s_1\choose k_1+2}{s_2\choose k_2}{s_2\choose k_2+2}\\
&(s_1-k_1+k_2)!(s_2+k_1-k_2)!\frac{(x-y)_+^{s_1-k_1+k_2}}{(\rvert x-y\rvert^2)^{s_1-k_1+k_2+1}}\frac{(y-x)_+^{s_2+k_1-k_2}}{(\rvert y-x\rvert^2)^{s_2+k_1-k_2+1}}
\end{align}
and simplifies as follows:
\begin{align}
\label{59}
\nonumber
\langle \mathbb{O}_{s_1}(x) \mathbb{O}_{s_2}(y)\rangle =& \frac{1}{(4\pi^2)^2} \frac{N^2-1}{4} \frac{2^{s_1+s_2+2}i^{s_1+s_2-4}}{(4!)^2}\\\nonumber
&(s_1+1)(s_1+2)(s_2+1)(s_2+2)\frac{(x-y)_+^{s_1+s_2}}{(\rvert x-y\rvert^2)^{s_1+s_2+2}}\\\nonumber
&\sum_{k_1 = 0}^{s_1-2}\sum_{k_2 = 0}^{s_2-2}{s_1\choose k_1}{s_1\choose k_1+2}{s_2\choose k_2}{s_2\choose k_2+2}\\\nonumber
&(-1)^{s_2-k_2+k_1}(s_1-k_1+k_2)!(s_2+k_1-k_2)!  \nonumber \\
=& \mathcal{C}_{s_1}(x,y)\delta_{s_1s_2}
\end{align}
since the correlator is zero for $s_1\neq s_2$ (appendix \ref{appA3}).
By setting $s=s_1=s_2$, we get:
\begin{align}  \label{CC}
\nonumber
 \mathcal{C}_{s}(x,y) = &\frac{1}{(4\pi^2)^2} \frac{N^2-1}{4} \frac{2^{2s+2}i^{2s-4}}{(4!)^2}(s+1)^2(s+2)^2
\frac{(x-y)_+^{2s}}{(\rvert x-y\rvert^2)^{2s+2}}\\\nonumber
&\sum_{k_1 = 0}^{s-2}\sum_{k_2 = 0}^{s-2}{s\choose k_1}{s\choose k_1+2}{s\choose k_2}{s\choose k_2+2}(-1)^{s-k_2+k_1}\\
&(s-k_1+k_2)!(s+k_1-k_2)! 
\end{align}
Moreover, by the substitution in eq. (\ref{59}):
\begin{equation}
{k'\!\!}_2=s_2-2-k_2
\end{equation}
we obtain:
\begin{align}
\nonumber
\langle \mathbb{O}_{s_1}(x) \mathbb{O}_{s_2}(y)\rangle = &\frac{1}{(4\pi^2)^2} \frac{N^2-1}{4} \frac{2^{s_1+s_2+2}i^{s_1+s_2-4}}{(4!)^2}\\\nonumber
&(s_1+1)(s_1+2)(s_2+1)(s_2+2)
\frac{(x-y)_+^{s_1+s_2}}{(\rvert x-y\rvert^2)^{s_1+s_2+2}}\\\nonumber
&\sum_{k_1 = 0}^{s_1-2}\sum_{{k'\!\!}_2 = 0}^{s_2-2}{s_1\choose k_1}{s_1\choose k_1+2}{s_2\choose s_2-2-{k'\!\!}_2}{s_2\choose s_2-{k'\!\!}_2}(-1)^{s_2-s_2+2+{k'\!\!}_2+k_1}\\
&(s_1-k_1+s_2-2-{k'\!\!}_2)!(s_2+k_1-s_2+2+{k'\!\!}_2)!
\end{align}
that becomes:
\begin{align}
\label{qesum}
\nonumber
\langle \mathbb{O}_{s_1}(x) \mathbb{O}_{s_2}(y)\rangle = &\frac{1}{(4\pi^2)^2} \frac{N^2-1}{4} \frac{2^{s_1+s_2+2}i^{s_1+s_2-4}}{(4!)^2}\\\nonumber
&(s_1+1)(s_1+2)(s_2+1)(s_2+2)
\frac{(x-y)_+^{s_1+s_2}}{(\rvert x-y\rvert^2)^{s_1+s_2+2}}\\\nonumber
&\sum_{k_1 = 0}^{s_1-2}\sum_{k_2 = 0}^{s_2-2}{s_1\choose k_1}{s_1\choose k_1+2}{s_2\choose k_2+2}{s_2\choose k_2}(-1)^{k_2+k_1}\\\nonumber
&(s_1+s_2-k_1-k_2-2)!(k_1+k_2+2)!\\\nonumber
=&\frac{1}{(4\pi^2)^2} \frac{N^2-1}{4} \frac{2^{s_1+s_2+2}i^{s_1+s_2-4}}{(4!)^2}\\\nonumber
&(s_1+1)(s_1+2)(s_2+1)(s_2+2)
(s_1+s_2)!\frac{(x-y)_+^{s_1+s_2}}{(\rvert x-y\rvert^2)^{s_1+s_2+2}}\\
&\sum_{k_1 = 0}^{s_1-2}\sum_{k_2 = 0}^{s_2-2}{s_1\choose k_1}{s_1\choose k_1+2}{s_2\choose k_2+2}{s_2\choose k_2}(-1)^{k_2+k_1} \frac{1}{{s_1+s_2\choose k_1+k_2+2}}
\end{align}
Besides, according to the trick in \cite{Kazakov:2012ar} (appendix \ref{appA3}), we get:
 \begin{align} \label{qesum1}
 \langle \mathbb{O}_{s_1}(x) \mathbb{O}_{s_2}(y)\rangle =& \delta_{s_1s_2}\frac{1}{(4\pi^2)^2} \frac{N^2-1}{4} \frac{2^{2{s_1}+2}}{(4!)^2}(-1)^{s_1}\nonumber\\
 &({s_1}-1){s_1}({s_1}+1)({s_1}+2)(2{s_1})!
 \frac{(x-y)_+^{2{s_1}}}{(\rvert x-y\rvert^2)^{2{s_1}+2}}
 \end{align}
Hence, comparing eqs. \eqref{qesum} and \eqref{qesum1}, we can virtually perform the sums in eq. \eqref{qesum} to obtain the identity: 
\begin{align}
\label{id2}
\delta_{s_1s_2}\frac{s_1(s_1-1)}{(s_1+1)(s_1+2)}= &
\sum_{k_1 = 0}^{s_1-2}\sum_{k_2 = 0}^{s_2-2}{s_1\choose k_1}{s_1\choose k_1+2}{s_2\choose k_2}{s_2\choose k_2+2}(-1)^{k_2+k_1}\frac{1}{{s_1+s_2\choose k_1+k_2+2}}
 \end{align}
Similarly, for the balanced operators with odd $s$, we get as well:
\begin{equation}
\langle \tilde{\mathbb{O}}_{s_1}(x) \tilde{\mathbb{O}}_{s_2}(y)\rangle =\mathcal{C}_{s_1}(x,y) \delta_{s_1s_2}
\end{equation}
where the definition of $\mathcal{C}_{s}(x,y)$ in eq. \eqref{CC} has been extended to odd $s$. Correspondingly, eq. \eqref{id2} extends to odd $s_1,s_2$ as well. \par
Now we compute the only correlators of two unbalanced operators that are nonzero:
\begin{align}
\nonumber
\langle\mathbb{S}_{s_1}(x)\bar{\mathbb{S}}_{s_2}(y)\rangle =&\frac{1}{2}\mathcal{G}_{s_1-2}^{\frac{5}{2}}(\partial_{x_1^+},\partial_{x_2^+})\mathcal{G}_{s_2-2}^{\frac{5}{2}}(\partial_{y_1^+},\partial_{y_2^+})\\
&\langle\Tr f_{11}(x_1)f_{11}(x_2)\,\Tr f_{\dot{1}\dot{1}}(y_1) f_{\dot{1}\dot{1}}(y_2)\rangle\Big\rvert_{x_1=x_2=x}^{y_1=y_2=y}
\end{align}
There are two Wick contractions but an extra factor of $\frac{1}{2}$ in the normalization of the operators, in such a way that the result is the same as for the correlators of balanced operators with even $s$:
\begin{align}
\langle\mathbb{S}_{s_1}(x)\bar{\mathbb{S}}_{s_2}(y)\rangle =& \frac{1}{(4\pi^2)^2} \frac{N^2-1}{4}\mathcal{G}_{s_1-2}^{\frac{5}{2}}(\partial_{x_1^+},\partial_{x_2^+})\mathcal{G}_{s_2-2}^{\frac{5}{2}}(\partial_{y_1^+},\partial_{y_2^+})\\\nonumber
&\partial^+_{x_1}\partial^+_{x_2}\partial^+_{y_1}\partial^+_{y_2}\frac{1}{\rvert x_1-y_2\rvert^2}\frac{1}{\rvert y_1-x_2\rvert^2}\Big\rvert_{x_1=x_2=x}^{y_1=y_2=y}\\\nonumber
=&\mathcal{C}_{s_1}(x,y)\delta_{s_1s_2}
\end{align}
All the remaining $2$-point correlators vanish.

\subsection{Extended basis}

Similarly, employing eq. \eqref{AA}, we obtain the correlators in the extended basis:
\begin{align}
\label{semilocal}
\nonumber
\langle \mathbb{A}_{s}(x) \mathbb{A}_{s}(y)\rangle = &\frac{1}{(4\pi^2)^2} \frac{N^2-1}{4} 2^{2s}i^{2s}
\frac{(x-y)_+^{2s}}{(\rvert x-y\rvert^2)^{2s+2}}\\
&\sum_{k_1 = 0}^{s}\sum_{k_2 = 0}^{s}{s\choose k_1}{s\choose k_1}{s\choose k_2}{s\choose k_2}(-1)^{s-k_2+k_1}
(s-k_1+k_2)!(s+k_1-k_2)! \nonumber\\
= & \mathcal{A}_{s}(x,y)
\end{align}
We extend the above definition of $\mathcal{A}_{s}(x,y)$ to odd $s$.
We get for even $s$:
\begin{align}
\langle\mathbb{B}_{s_1}(x)\bar{\mathbb{B}}_{s_2}(y)\rangle 
= &\mathcal{A}_{s_1}(x,y)  \delta_{s_1s_2} 
\nonumber \\
\end{align}
and for odd $s$:
\begin{align}
\langle \tilde{\mathbb{A}}_{s_1}(x) \tilde{\mathbb{A}}_{s_2}(y)\rangle
=&\mathcal{A}_{s_1}(x,y) \delta_{s_1s_2} 
\end{align}
We obtain (appendix \ref{appA3}):
\begin{align}
\mathcal{A}_{s}(x,y) = &\frac{1}{(4\pi^2)^2} (-1)^s\frac{N^2-1}{4} 2^{2s}(2s)!
\frac{(x-y)_+^{2s}}{(\rvert x-y\rvert^2)^{2s+2}}
\end{align}
Similarly, it follows the identity:
\begin{align}
\label{id1}
\delta_{s_1s_2}= &
\sum_{k_1 = 0}^{s_1}\sum_{k_2 = 0}^{s_2}{s_1\choose k_1}{s_1\choose k_1}{s_2\choose k_2}{s_2\choose k_2}(-1)^{k_2+k_1}\frac{1}{{s_1+s_2\choose k_1+k_2}}
\end{align}
All the remaining $2$-point correlators vanish.

\section{$3$-point correlators of twist-$2$ gluonic operators \label{threepoint}}

We compute to the lowest perturbative order the $3$-point correlators of the operators in both bases.

\subsection{Standard basis}

In the light-cone gauge, for the $3$-point correlators of the balanced operators with even $s$, we obtain:
\begin{align}
\nonumber
\langle {\mathbb{O}}_{s_1}(x){\mathbb{O}}_{s_2}(y){\mathbb{O}}_{s_3}&(z)\rangle =\mathcal{G}_{s_1-2}^{\frac{5}{2}}(\partial_{x_1^+},\partial_{x_2^+})\mathcal{G}_{s_2-2}^{\frac{5}{2}}(\partial_{y_1^+},\partial_{y_2^+})\mathcal{G}_{s_3-2}^{\frac{5}{2}}(\partial_{z_1^+},\partial_{z_2^+})\\
&\langle\Tr f_{11}(x_1) f_{\dot{1}\dot{1}}(x_2)\,\Tr f_{11}(y_1) f_{\dot{1}\dot{1}}(y_2)\,\Tr f_{11}(z_1) f_{\dot{1}\dot{1}}(z_2)\rangle\Big\rvert_{x_1=x_2=x}^{y_1=y_2=y,z_1=z_2=z}
\end{align}
Since there are two Wick contractions, employing eq. \eqref{axialprop2}, we get:
\begin{align}
\label{62}
\nonumber
\langle {\mathbb{O}}_{s_1}(x)&{\mathbb{O}}_{s_2}(y){\mathbb{O}}_{s_3}(z)\rangle =\frac{1}{(4\pi^2)^3}\frac{N^2-1}{8}\Bigg(\mathcal{G}_{s_1-2}^{\frac{5}{2}}(\partial_{x_1^+},\partial_{x_2^+})\mathcal{G}_{s_2-2}^{\frac{5}{2}}(\partial_{y_1^+},\partial_{y_2^+})\mathcal{G}_{s_3-2}^{\frac{5}{2}}(\partial_{z_1^+},\partial_{z_2^+})\\\nonumber
&(-\partial_{x_1^+})\partial_{x_2^+}(-\partial_{y_1^+})\partial_{y_2^+}(-\partial_{z_1^+})\partial_{z_2^+}\frac{1}{\rvert x_1-y_2\rvert^2}\frac{1}{\rvert y_1-z_2\rvert^2}\frac{1}{\rvert z_1-x_2\rvert^2}\\\nonumber
&+\mathcal{G}_{s_1-2}^{\frac{5}{2}}(\partial_{x_1^+},\partial_{x_2^+})\mathcal{G}_{s_2-2}^{\frac{5}{2}}(\partial_{y_1^+},\partial_{y_2^+})\mathcal{G}_{s_3-2}^{\frac{5}{2}}(\partial_{z_1^+},\partial_{z_2^+})\\
&(-\partial_{x_1^+})\partial_{x_2^+}(-\partial_{y_1^+})\partial_{y_2^+}(-\partial_{z_1^+})\partial_{z_2^+}\frac{1}{\rvert x_1-z_2\rvert^2}\frac{1}{\rvert z_1-y_2\rvert^2}\frac{1}{\rvert y_1-x_2\rvert^2}\Bigg)\Bigg\rvert_{x_1=x_2=x}^{y_1=y_2=y,z_1=z_2=z}
\end{align}
In the second term of eq. (\ref{62}) we may conveniently relabel the coordinates, $x_1\rightarrow x_2$,  $y_1\rightarrow y_2$,  $z_1\rightarrow z_2$, and vice versa, because they coincide in the local limit:
\begin{align}
\nonumber
\langle {\mathbb{O}}_{s_1}(x)&{\mathbb{O}}_{s_2}(y){\mathbb{O}}_{s_3}(z)\rangle =\frac{1}{(4\pi^2)^3}\frac{N^2-1}{8}\Bigg(\mathcal{G}_{s_1-2}^{\frac{5}{2}}(\partial_{x_1^+},\partial_{x_2^+})\mathcal{G}_{s_2-2}^{\frac{5}{2}}(\partial_{y_1^+},\partial_{y_2^+})\mathcal{G}_{s_3-2}^{\frac{5}{2}}(\partial_{z_1^+},\partial_{z_2^+})\\\nonumber
&(-\partial_{x_1^+})\partial_{x_2^+}(-\partial_{y_1^+})\partial_{y_2^+}(-\partial_{z_1^+})\partial_{z_2^+}\frac{1}{\rvert x_1-y_2\rvert^2}\frac{1}{\rvert y_1-z_2\rvert^2}\frac{1}{\rvert z_1-x_2\rvert^2}\\\nonumber
&+\mathcal{G}_{s_1-2}^{\frac{5}{2}}(\partial_{x_2^+},\partial_{x_1^+})\mathcal{G}_{s_2-2}^{\frac{5}{2}}(\partial_{y_2^+},\partial_{y_1^+})\mathcal{G}_{s_3-2}^{\frac{5}{2}}(\partial_{z_2^+},\partial_{z_1^+})\\
&(-\partial_{x_2^+})\partial_{x_1^+}(-\partial_{y_2^+})\partial_{y_1^+}(-\partial_{z_2^+})\partial_{z_1^+}\frac{1}{\rvert x_2-z_1\rvert^2}\frac{1}{\rvert z_2-y_1\rvert^2}\frac{1}{\rvert y_2-x_1\rvert^2}\Bigg)\Bigg\rvert_{x_1=x_2=x}^{y_1=y_2=y,z_1=z_2=z}
\end{align}
In the second term above, we employ the property of the Gegenbauer polynomials (appendix \ref{appB}):
\begin{equation} \label{GG}
\mathcal{G}_{l}^{\alpha}(\partial_{x^+_1},\partial_{x^+_2})=(-1)^l\mathcal{G}_{l}^{\alpha}(\partial_{x^+_2},\partial_{x^+_1})
\end{equation}
to get:
\begin{align}
\nonumber
\langle {\mathbb{O}}_{s_1}(x)&{\mathbb{O}}_{s_2}(y){\mathbb{O}}_{s_3}(z)\rangle =\frac{1}{(4\pi^2)^3}\frac{N^2-1}{8}\Bigg(\mathcal{G}_{s_1-2}^{\frac{5}{2}}(\partial_{x_1^+},\partial_{x_2^+})\mathcal{G}_{s_2-2}^{\frac{5}{2}}(\partial_{y_1^+},\partial_{y_2^+})\mathcal{G}_{s_3-2}^{\frac{5}{2}}(\partial_{z_1^+},\partial_{z_2^+})\\\nonumber
&(-\partial_{x_1^+})\partial_{x_2^+}(-\partial_{y_1^+})\partial_{y_2^+}(-\partial_{z_1^+})\partial_{z_2^+}\frac{1}{\rvert x_1-y_2\rvert^2}\frac{1}{\rvert y_1-z_2\rvert^2}\frac{1}{\rvert z_1-x_2\rvert^2}\\\nonumber
&+(-1)^{s_1+s_2+s_3}\mathcal{G}_{s_1-2}^{\frac{5}{2}}(\partial_{x_1^+},\partial_{x_2^+})\mathcal{G}_{s_2-2}^{\frac{5}{2}}(\partial_{y_1^+},\partial_{y_2^+})\mathcal{G}_{s_3-2}^{\frac{5}{2}}(\partial_{z_1^+},\partial_{z_2^+})\\
&(-\partial_{x_1^+})\partial_{x_2^+}(-\partial_{y_1^+})\partial_{y_2^+}(-\partial_{z_1^+})\partial_{z_2^+}\frac{1}{\rvert x_1-y_2\rvert^2}\frac{1}{\rvert y_1-z_2\rvert^2}\frac{1}{\rvert z_1-x_2\rvert^2}\Bigg)\Bigg\rvert_{x_1=x_2=x}^{y_1=y_2=y,z_1=z_2=z}
\end{align}
Therefore:
\begin{align}
\label{3point}
\nonumber
\langle {\mathbb{O}}_{s_1}(x)&{\mathbb{O}}_{s_2}(y){\mathbb{O}}_{s_3}(z)\rangle =\frac{1}{(4\pi^2)^3}(1+(-1)^{s_1+s_2+s_3})\frac{N^2-1}{8}\\\nonumber
&\mathcal{G}_{s_1-2}^{\frac{5}{2}}(\partial_{x_1^+},\partial_{x_2^+})\mathcal{G}_{s_2-2}^{\frac{5}{2}}(\partial_{y_1^+},\partial_{y_2^+})\mathcal{G}_{s_3-2}^{\frac{5}{2}}(\partial_{z_1^+},\partial_{z_2^+})\\
&(-\partial_{x_1^+})\partial_{x_2^+}(-\partial_{y_1^+})\partial_{y_2^+}(-\partial_{z_1^+})\partial_{z_2^+}\frac{1}{\rvert x_1-y_2\rvert^2}\frac{1}{\rvert y_1-z_2\rvert^2}\frac{1}{\rvert z_1-x_2\rvert^2}\Bigg\rvert_{x_1=x_2=x}^{y_1=y_2=y,z_1=z_2=z}
\end{align}
Since the collinear spins are all even, $1+(-1)^{s_1+s_2+s_3}=2$, so that:
\begin{align}
\label{3point2}
\nonumber
\langle {\mathbb{O}}_{s_1}(x)&{\mathbb{O}}_{s_2}(y){\mathbb{O}}_{s_3}(z)\rangle =\frac{1}{(4\pi^2)^3}2\frac{N^2-1}{8}\mathcal{G}_{s_1-2}^{\frac{5}{2}}(\partial_{x_1^+},\partial_{x_2^+})\mathcal{G}_{s_2-2}^{\frac{5}{2}}(\partial_{y_1^+},\partial_{y_2^+})\mathcal{G}_{s_3-2}^{\frac{5}{2}}(\partial_{z_1^+},\partial_{z_2^+})\\
&(-\partial_{x_1^+})\partial_{x_2^+}(-\partial_{y_1^+})\partial_{y_2^+}(-\partial_{z_1^+})\partial_{z_2^+}\frac{1}{\rvert x_1-y_2\rvert^2}\frac{1}{\rvert y_1-z_2\rvert^2}\frac{1}{\rvert z_1-x_2\rvert^2}\Bigg\rvert_{x_1=x_2=x}^{y_1=y_2=y,z_1=z_2=z}
\end{align}
Eq. (\ref{3point}) also holds for the $3$-point correlators of $\mathbb{O}_s$, $\tilde{\mathbb{O}}_s$, $\mathbb{S}_s$ and $\bar{\mathbb{S}}_s$ below, with the factor of $ 1+(-1)^{s_1+s_2+s_3}$ selecting which of these correlators yield a nonzero result. 
Therefore, by defining:
\begin{align}
\label{3point1}
\nonumber
& \mathcal{C}_{s_1s_2s_3}(x,y,z) =\frac{1}{(4\pi^2)^3}(1+(-1)^{s_1+s_2+s_3})\frac{N^2-1}{8}\\\nonumber
&\mathcal{G}_{s_1-2}^{\frac{5}{2}}(\partial_{x_1^+},\partial_{x_2^+})\mathcal{G}_{s_2-2}^{\frac{5}{2}}(\partial_{y_1^+},\partial_{y_2^+})\mathcal{G}_{s_3-2}^{\frac{5}{2}}(\partial_{z_1^+},\partial_{z_2^+})\\
&(-\partial_{x_1^+})\partial_{x_2^+}(-\partial_{y_1^+})\partial_{y_2^+}(-\partial_{z_1^+})\partial_{z_2^+}\frac{1}{\rvert x_1-y_2\rvert^2}\frac{1}{\rvert y_1-z_2\rvert^2}\frac{1}{\rvert z_1-x_2\rvert^2}\Bigg\rvert_{x_1=x_2=x}^{y_1=y_2=y,z_1=z_2=z}
\end{align}
the nonvanishing correlators are:
\begin{align}
\label{defc3}
\langle {\mathbb{O}}_{s_1}(x){\mathbb{O}}_{s_2}(y){\mathbb{O}}_{s_3}(z)\rangle=\langle {\mathbb{O}}_{s_1}(x){\mathbb{S}}_{s_2}(y)\bar{\mathbb{S}}_{s_3}(z)\rangle = \mathcal{C}_{s_1s_2s_3}(x,y,z)
\end{align}
and:
\begin{align}
\label{defc3}
 \langle {\mathbb{O}}_{s_1}(x)\tilde{\mathbb{O}}_{s_2}(y) \tilde{\mathbb{O}}_{s_3}(z)\rangle = \mathcal{C}_{s_1s_2s_3}(x,y,z)
\end{align}
After substituting the definition of the Gegenbauer polynomials in eq. \eqref{defg52}, we obtain:
\begin{align}
\nonumber
 \mathcal{C}_{s_1s_2s_3}(x,y,z)
 =&\frac{1}{(4\pi^2)^3} (1+(-1)^{s_1+s_2+s_3})   \left(\frac{2}{4!}\right)^3\frac{N^2-1}{8}i^{s_1+s_2+s_3-6}\\\nonumber
& (s_1+1)(s_1+2)(s_2+1)(s_2+2)(s_3+1)(s_3+2)\\\nonumber
&\sum_{k_1 = 0}^{s_1-2}\sum_{k_2 = 0}^{s_2-2}\sum_{k_3 = 0}^{s_3-2}{s_1\choose k_1}{s_1\choose k_1+2}{s_2\choose k_2}{s_2\choose k_2+2}{s_3\choose k_3}{s_3\choose k_3+2}\\\nonumber
&(-\partial_{x_1^+})^{s_1-k_1-1}\partial^{k_1+1}_{x_2^+}(-\partial_{y_1^+})^{s_2-k_2-1}\partial^{k_2+1}_{y_2^+}(-\partial_{z_1^+})^{s_3-k_3-1}\partial^{k_3+1}_{z_2^+}\\
&\frac{1}{\rvert x_1-y_2\rvert^2}\frac{1}{\rvert y_1-z_2\rvert^2}\frac{1}{\rvert z_1-x_2\rvert^2}\Bigg\rvert_{x_1=x_2=x}^{y_1=y_2=y,z_1=z_2=z}
\end{align}
Employing eq. \eqref{doubleder}, we get:
\begin{align}
\label{3pointO}
 \mathcal{C}_{s_1s_2s_3}(x,y,z)
=&-\frac{1}{(4\pi^2)^3} (1+(-1)^{s_1+s_2+s_3})   \left(\frac{2}{4!}\right)^3\frac{N^2-1}{8}i^{s_1+s_2+s_3}2^{s_1+s_2+s_3}\\\nonumber
& (s_1+1)(s_1+2)(s_2+1)(s_2+2)(s_3+1)(s_3+2)\\\nonumber
&\sum_{k_1 = 0}^{s_1-2}\sum_{k_2 = 0}^{s_2-2}\sum_{k_3 = 0}^{s_3-2}{s_1\choose k_1}{s_1\choose k_1+2}{s_2\choose k_2}{s_2\choose k_2+2}{s_3\choose k_3}{s_3\choose k_3+2}\\\nonumber
&(s_1-k_1+k_2)!(s_2-k_2+k_3)!(s_3-k_3+k_1)!  \\\nonumber
&\frac{(x-y)^{s_1-k_1+k_2}_+}{(\rvert x-y\rvert^2)^{s_1+1-k_1+k_2}}\frac{(y-z)^{s_2-k_2+k_3}_+}{(\rvert y-z\rvert^2)^{s_2+1-k_2+k_3}}\frac{(z-x)^{s_3-k_3+k_1}_+}{(\rvert z-x\rvert^2)^{s_3+1-k_3+k_1}}
\end{align}
All the remaining $3$-point correlators vanish. 

\subsection{Extended basis}

The $3$-point correlators in the extended basis are computed analogously. \par
The nonvanishing correlators are:
\begin{align}
\label{defc3}
\langle {\mathbb{A}}_{s_1}(x){\mathbb{A}}_{s_2}(y){\mathbb{A}}_{s_3}(z)\rangle=\langle {\mathbb{A}}_{s_1}(x){\mathbb{B}}_{s_2}(y)\bar{\mathbb{B}}_{s_3}(z)\rangle= \mathcal{A}_{s_1s_2s_3}(x,y,z)
\end{align}
and:
\begin{align}
\label{defc3}
\langle {\mathbb{A}}_{s_1}(x)\tilde{\mathbb{A}}_{s_2}(y)\tilde{\mathbb{A}}_{s_3}(z)\rangle  = \mathcal{A}_{s_1s_2s_3}(x,y,z)
\end{align}
with:
\begin{align}
\nonumber
 \mathcal{A}_{s_1s_2s_3}(x,y,z)
 =&-\frac{1}{(4\pi^2)^3} (1+(-1)^{s_1+s_2+s_3})   \frac{N^2-1}{8}i^{s_1+s_2+s_3}2^{s_1+s_2+s_3}\\\nonumber
&\sum_{k_1 = 0}^{s_1}\sum_{k_2 = 0}^{s_2}\sum_{k_3 = 0}^{s_3}{s_1\choose k_1}{s_1\choose k_1}{s_2\choose k_2}{s_2\choose k_2}{s_3\choose k_3}{s_3\choose k_3}\\\nonumber
&(s_1-k_1+k_2)!(s_2-k_2+k_3)!(s_3-k_3+k_1)!  \\\nonumber
&\frac{(x-y)^{s_1-k_1+k_2}_+}{(\rvert x-y\rvert^2)^{s_1+1-k_1+k_2}}\frac{(y-z)^{s_2-k_2+k_3}_+}{(\rvert y-z\rvert^2)^{s_2+1-k_2+k_3}}\frac{(z-x)^{s_3-k_3+k_1}_+}{(\rvert z-x\rvert^2)^{s_3+1-k_3+k_1}}\nonumber \\
\end{align}
All the remaining $3$-point correlators vanish.

\section{$n$-point correlators of twist-$2$ gluonic operators \label{npoint}}

We compute to the lowest perturbative order $n$-point correlators of the operators in both bases. \par

\subsection{Standard basis}

Given the bilocal operators:
\begin{equation} \label{tr}
O(x_i^A,x_i^B) = \Tr f_{11}(x_i^A) f_{\dot{1}\dot{1}}(x_i^B) = \frac{1}{2} f^a_{11}(x_i^A) f^a_{\dot{1}\dot{1}}(x_i^B)
\end{equation} 
in the light-cone gauge, the corresponding $n$-point correlator that is connected in the local limit, $x^A_i=x^B_i=x_i$, takes the form:
\begin{align}
\label{perm0}
\nonumber
\langle O(x_1^A,x_1^B)\ldots O(x_n^A,x_n^B) \rangle=&\frac{1}{n}\frac{1}{2^n} \sum_{i_1 = 1}^n\ldots \sum_{i_{n}=1}^n\Bigg\rvert_{i_1\neq i_2\neq\ldots\neq i_{n}}\hspace{-1.5cm}\langle f^{a_{i_1}}_{11}(x_{i_1}^A) f^{a_{i_2}}_{\dot{1}\dot{1}}(x^B_{i_2})\rangle\\
&\langle f^{a_{i_2}}_{11}(x_{i_2}^A) f^{a_{i_3}}_{\dot{1}\dot{1}}(x^B_{i_3})\rangle\ldots \langle f^{a_{i_n}}_{11}(x_{i_{n}}^A) f^{a_{i_1}}_{\dot{1}\dot{1}}(x^B_{i_1})\rangle
\end{align}
The factor of $\frac{1}{n}$ arises because, if the first index -- for example $i_1=1$ -- is kept fixed, there are only $(n-1)!$ Wick contractions that contribute to the connected correlator.
A nicer -- but completely equivalent -- formula is written in terms of permutations. If we denote by $P_n$ the set of permutations of $1 \ldots n$, it follows identically: 
\begin{align}
\nonumber
\langle O(x_1^A,x_1^B)\ldots O(x_n^A,x_n^B)\rangle =& \frac{1}{n}\frac{1}{2^n}\sum_{\sigma\in P_n}\langle f^{a_{\sigma(1)}}_{11}(x_{\sigma(1)}^A) f^{a_{\sigma(2)}}_{\dot{1}\dot{1}}(x^B_{\sigma(2)})\rangle\\
&\langle  f^{a_{\sigma(2)}}_{11}(x_{\sigma(2)}^A) f^{a_{\sigma(3)}}_{\dot{1}\dot{1}}(x^B_{\sigma(3)})\rangle\ldots \langle f^{a_{\sigma(n)}}_{11}(x_{\sigma(n)}^A) f^{a_{\sigma(1)}}_{\dot{1}\dot{1}}(x^B_{\sigma(1)})\rangle
\end{align}
Besides, eq. \eqref{axialprop2} reads:
\begin{equation}
\langle f^a_{11}(x_i) f^b_{\dot{1}\dot{1}}(x_j)\rangle =-\partial_{x_i^+}\partial_{x_j^+}\frac{\delta^{ab}}{4\pi^2\rvert x_i-x_j\rvert^2}
\end{equation}
Hence, for the balanced operators with even $s$, we get in the light-cone gauge:
\begin{align}
\label{bilinear1}
\nonumber
\langle \mathbb{O}_{s_1}(x_1)\ldots \mathbb{O}_{s_n}(x_n)\rangle=&\frac{1}{2^n}\mathcal{G}^{\frac{5}{2}}_{s_1-2}(\partial_{x_1^{A\,+}},\partial_{x_1^{B\,+}})\ldots\mathcal{G}^{\frac{5}{2}}_{s_n-2}(\partial_{x_n^{A\,+}},\partial_{x_n^{B\,+}})\\
&\langle f^{a_1}_{11}(x_1^A)f^{a_1}_{\dot{1}\dot{1}}(x_1^B)\ldots f^{a_n}_{11}(x_n^A){f}^{a_n}_{\dot{1}\dot{1}}(x_n^B)\rangle\Big\rvert_{A=B}
\end{align}
where:
\begin{align} \label{g}
\mathcal{G}^{\frac{5}{2}}_{s-2}(\partial_{x_i^+},\partial_{x_j^+}) &=\frac{i^{s-2}\Gamma(3)\Gamma(s+3)}{\Gamma(5)\Gamma(s+1)}\sum_{k=0}^{s-2} {s\choose k}{s\choose k+2}(-1)^{s-k} \overleftarrow{\partial}_{x^+_i}^{s-k-2} \overrightarrow{\partial}_{x^+_j}^{k}\nonumber\\ &=\frac{2i^{s-2}(s+1)(s+2)}{4!}\sum_{k=0}^{s-2} {s\choose k}{s\choose k+2}(-1)^{s-k} \overleftarrow{\partial}_{x^+_i}^{s-k-2} \overrightarrow{\partial}_{x^+_j}^{k}
\end{align}
It follows from eq. (\ref{doubleder}) that, correspondingly, the $n$-point correlator contains factors of the form:
\begin{align} \label{no}
\nonumber
&-\partial_{x_i^+}^{s_i-k_i-2}\partial_{x_j^+}^{k_j}\partial_{x_i^+}\partial_{x_j^+}\frac{1}{4\pi^2\rvert x_i-x_j\rvert^2} \\
&=\frac{1}{4\pi^2}
 (-1)^{s_i-k_i} (s_i-k_i+k_j)!\,2^{s_i-k_i+k_j} \frac{(x_i-x_j)_+^{s_i-k_i+k_j}}{(\rvert x_i-x_j\rvert^2)^{s_i-k_i+k_j+1}}
\end{align}
Therefore, we get:
\begin{align} \label{000}
\nonumber
&\langle \mathbb{O}_{s_1}(x_1)\ldots \mathbb{O}_{s_n}(x_n)\rangle_{conn} =\frac{(-1)^n}{(4\pi^2)^n}\frac{N^2-1}{2^n}i^{\sum_{l=1}^n s_l}\frac{\Gamma(3)\Gamma(s_1+3)}{\Gamma(5)\Gamma(s_1+1)}\ldots \frac{\Gamma(3)\Gamma(s_n+3)}{\Gamma(5)\Gamma(s_n+1)}\frac{1}{n}\sum_{\sigma\in P_n}\\\nonumber
 &\sum_{k_{\sigma(1)}=0}^{s_{\sigma(1)}-2}\ldots\sum_{k_{\sigma(n)} = 0}^{s_{\sigma(n)}-2}{s_{\sigma(1)}\choose k_{\sigma(1)}}{s_{\sigma(1)}\choose k_{\sigma(1)}+2}(-1)^{s_{\sigma(1)}-k_{\sigma(1)}}\ldots {s_{\sigma(n)}\choose k_{\sigma(n)}}{s_{\sigma(n)}\choose k_{\sigma(n)}+2}(-1)^{s_{\sigma(n)}-k_{\sigma(n)}}\\\nonumber
 &2^{s_{\sigma(1)}-k_{\sigma(1)}+k_{\sigma(2)}}(-1)^{s_{\sigma(1)}-k_{\sigma(1)}}(s_{\sigma(1)}-k_{\sigma(1)}+k_{\sigma(2)})!\frac{(x_{\sigma(1)}-x_{\sigma(2)})_+^{s_{\sigma(1)}-k_{\sigma(1)}+k_{\sigma(2)}}}{\left(\rvert x_{\sigma(1)}-x_{\sigma(2)}\rvert^2\right)^{s_{\sigma(1)}-k_{\sigma(1)}+k_{\sigma(2)}+1}}\\
 &\ldots 2^{s_{\sigma(n)}-k_{\sigma(n)}+k_{\sigma(1)}}(-1)^{s_{\sigma(n)}-k_{\sigma(n)}}(s_{\sigma(n)}-k_{\sigma(n)}+k_{\sigma(1)})!\frac{(x_{\sigma(n)}-x_{\sigma(1)})_+^{s_{\sigma(n)}-k_{\sigma(n)}+k_{\sigma(1)}}}{\left(\rvert x_{\sigma(n)}-x_{\sigma(1)}\rvert^2\right)^{s_{\sigma(n)}-k_{\sigma(n)}+k_{\sigma(1)}+1}}
\end{align}
where we have set $x^A_i=x^B_i=x_i$ in order to implement the local limit of the bilocal operators. The color factor comes from the contraction of the $n$ Kronecker delta:
\begin{equation}
N^2-1 = \delta^{a_{\sigma(1)}a_{\sigma(2)}}\delta^{a_{\sigma(2)}a_{\sigma(3)}}\ldots \delta^{a_{\sigma(n)}a_{\sigma(1)}}
\end{equation}
The overall factor of $(-1)^n$ occurs because of the factor of $i^{-2}$, which is a partial factor of $i^{s-2}$ in eq. \eqref{g}.\par
After cancelling between themselves the pairs of factors of the kind $(-1)^{s_a-k_a}$ in eq. \eqref{000}, and moving out of the sum over the permutations the product of the binomial coefficients, since it is independent of the permutations, we obtain:
\begin{align}
\label{standardn}
\nonumber
&\langle \mathbb{O}_{s_1}(x_1)\ldots \mathbb{O}_{s_n}(x_n)\rangle_{conn} =\frac{1}{(4\pi^2)^n}\frac{N^2-1}{2^n}2^{\sum_{l=1}^n s_l}i^{\sum_{l=1}^n s_l}\\\nonumber
&\frac{\Gamma(3)\Gamma(s_1+3)}{\Gamma(5)\Gamma(s_1+1)}\ldots \frac{\Gamma(3)\Gamma(s_n+3)}{\Gamma(5)\Gamma(s_n+1)}\sum_{k_1=0}^{s_1-2}\ldots \sum_{k_n = 0}^{s_n-2}{s_1\choose k_1}{s_1\choose k_1+2}\ldots {s_n\choose k_n}{s_n\choose k_n+2}\\\nonumber
&\frac{(-1)^n}{n}\sum_{\sigma\in P_n}(s_{\sigma(1)}-k_{\sigma(1)}+k_{\sigma(2)})!\ldots(s_{\sigma(n)}-k_{\sigma(n)}+k_{\sigma(1)})!\\
&\frac{(x_{\sigma(1)}-x_{\sigma(2)})_+^{s_{\sigma(1)}-k_{\sigma(1)}+k_{\sigma(2)}}}{\left(\rvert x_{\sigma(1)}-x_{\sigma(2)}\rvert^2\right)^{s_{\sigma(1)}-k_{\sigma(1)}+k_{\sigma(2)}+1}}\ldots\frac{(x_{\sigma(n)}-x_{\sigma(1)})_+^{s_{\sigma(n)}-k_{\sigma(n)}+k_{\sigma(1)}}}{\left(\rvert x_{\sigma(n)}-x_{\sigma(1)}\rvert^2\right)^{s_{\sigma(n)}-k_{\sigma(n)}+k_{\sigma(1)}+1}}
\end{align}
Actually, if $n$ is even, eq. \eqref{standardn} also holds for the $n$-point correlator of the operators $\tilde{\mathbb{O}}_s$, with the only difference that their collinear spin is odd. \par
Otherwise, if $n$ is odd, the correlator vanishes. To verify it, it suffices to notice that, in the sum over the permutations, for every permutation the inverse permutation also occurs with the opposite sign. For example, for $3$-point correlators we get pairs of terms of the kind:
\begin{align}
\label{sommaperm}
\nonumber
&\sum_{k_1= 0}^{s_1-2}\sum_{k_2= 0}^{s_2-2}\sum_{k_3= 0}^{s_3-2}   \ldots (s_{1}-k_{1}+k_{2})!(s_{2}-k_{2}+k_{3})!(s_{3}-k_{3}+k_{1})!\\\nonumber
&(x_1-x_2)^{s_1-k_1+k_2}(x_2-x_3)^{s_2-k_2+k_3}(x_3-x_1)^{s_3-k_3+k_1} \\\nonumber
&+\sum_{k_1= 0}^{s_1-2}\sum_{k_2= 0}^{s_2-2}\sum_{k_3= 0}^{s_3-2}    \ldots  (s_{2}-k_{2}+k_{1})!(s_{3}-k_{3}+k_{2})!(s_{1}-k_{1}+k_{3})!\\
&(x_2-x_1)^{s_2-k_2+k_1}(x_3-x_2)^{s_3-k_3+k_2}(x_1-x_3)^{s_1-k_1+k_3} 
\end{align}
Employing the substitution ${k'\!\!}_{i} = s_{i}-2-k_{i}$, we obtain for the last term above:
\begin{align}
&\sum_{k_1= 0}^{s_1-2}\sum_{k_2= 0}^{s_2-2}\sum_{k_3= 0}^{s_3-2}   \ldots (-1)^{s_1+s_2+s_3}(s_{1}-k_{1}+k_{2})!(s_{2}-k_{2}+k_{3})!(s_{3}-k_{3}+k_{1})!\nonumber \\
&(x_1-x_2)^{s_1-k_1+k_2}(x_2-x_3)^{s_2-k_2+k_3}(x_3-x_1)^{s_3-k_3+k_1}
\end{align}
that cancels the first term in eq. (\ref{sommaperm}).\par
The same reasoning applies to the $n+2m+1$-point correlators of balanced operators:
\begin{equation}
\langle \mathbb{O}_{s_1}(x_1)\ldots \mathbb{O}_{s_{n}}(x_{n})\tilde{\mathbb{O}}_{s_{n+1}}(x_{n+1})\ldots \tilde{\mathbb{O}}_{s_{n+2m+1}}(x_{n+2m+1})\rangle_{conn} = 0
\end{equation}
Otherwise, we get:
\begin{align}
\label{standardn2}
\nonumber
&\langle \mathbb{O}_{s_1}(x_1)\ldots \mathbb{O}_{s_n}(x_n)\tilde{\mathbb{O}}_{s_{n+1}}(x_{n+1})\ldots \tilde{\mathbb{O}}_{s_{n+2m}}(x_{n+2m})\rangle_{conn}\\\nonumber
& =\frac{1}{(4\pi^2)^{n+2m}}\frac{N^2-1}{2^{n+2m}}2^{\sum_{l=1}^{n+2m} s_l}i^{\sum_{l=1}^{n+2m} s_l}\frac{\Gamma(3)\Gamma(s_1+3)}{\Gamma(5)\Gamma(s_1+1)} \ldots\frac{\Gamma(3)\Gamma(s_{n+2m}+3)}{\Gamma(5)\Gamma(s_{n+2m}+1)}\\\nonumber
&\sum_{k_1=0}^{s_1-2}\ldots \sum_{k_{n+2m} = 0}^{s_{n+2m}-2}{s_1\choose k_1}{s_1\choose k_1+2}\ldots{s_{n+2m}\choose k_{n+2m}}{s_{n+2m}\choose k_{n+2m}+2}\\\nonumber
&\frac{(-1)^{n+2m}}{n+2m}\sum_{\sigma\in P_{n+2m}}(s_{\sigma(1)}-k_{\sigma(1)}+k_{\sigma(2)})!\ldots(s_{\sigma(n+2m)}-k_{\sigma(n+2m)}+k_{\sigma(1)})!\\
&\frac{(x_{\sigma(1)}-x_{\sigma(2)})_+^{s_{\sigma(1)}-k_{\sigma(1)}+k_{\sigma(2)}}}{\left(\rvert x_{\sigma(1)}-x_{\sigma(2)}\rvert^2\right)^{s_{\sigma(1)}-k_{\sigma(1)}+k_{\sigma(2)}+1}}\ldots\frac{(x_{\sigma(n+2m)}-x_{\sigma(1)})_+^{s_{\sigma(n+2m)}-k_{\sigma(n+2m)}+k_{\sigma(1)}}}{\left(\rvert x_{\sigma(n+2m)}-x_{\sigma(1)}\rvert^2\right)^{s_{\sigma(n+2m)}-k_{\sigma(n+2m)}+k_{\sigma(1)}+1}}
\end{align}
For the correlators of the unbalanced operators, we obtain:
\begin{align}
\label{bilinear}
\nonumber
&\langle \mathbb{S}_{s_1}(x_1)\ldots \mathbb{S}_{s_n}(x_n)\bar{\mathbb{S}}_{{s'\!\!}_1}(y_1)\ldots \bar{\mathbb{S}}_{{s'\!\!}_n}(y_n)\rangle\\\nonumber
&=\frac{1}{2^{2n}}\mathcal{G}^{\frac{5}{2}}_{s_1-2}(\partial_{x_1^{A\,+}},\partial_{x_1^{B\,+}})\ldots\mathcal{G}^{\frac{5}{2}}_{s_n-2}(\partial_{x_n^{A\,+}},\partial_{x_n^{B\,+}})\mathcal{G}^{\frac{5}{2}}_{{s'\!\!}_1-2}(\partial_{y_1^{A\,+}},\partial_{y_1^{B\,+}})\ldots\mathcal{G}^{\frac{5}{2}}_{{s'\!\!}_n-2}(\partial_{y_n^{A\,+}},\partial_{y_n^{B\,+}})\\
&\quad\frac{1}{2^{n}} \langle f^{a_1}_{11}(x_1^A)f^{a_1}_{11}(x_1^B)\ldots f^{a_n}_{11}(x_n^A){f}^{a_n}_{11}(x_n^B) f^{b_1}_{\dot{1}\dot{1}}(y_1^A) f^{b_1}_{\dot{1}\dot{1}}(y_1^B)\ldots f^{b_n}_{\dot{1}\dot{1}}(y_n^A) f^{b_n}_{\dot{1}\dot{1}}(y_n^B)\rangle\Big\rvert_{A=B}
\end{align}
The factor of $\frac{1}{2^{2n}}$ arises from the normalization of the color trace in eq. \eqref{tr}, while the factor of $\frac{1}{2^n}$ comes from the normalization of the operators.\par
We get the very same correlator by exchanging $A$ and $B$ in all the couples $(x_i^A,x_i^B)$ and $(y_k^A,y_k^B)$ simultaneously:
\begin{align}
\label{bilinear2}
\nonumber
&\langle \mathbb{S}_{s_1}(x_1)\ldots \mathbb{S}_{s_n}(x_n)\bar{\mathbb{S}}_{{s'\!\!}_1}(y_1)\ldots \bar{\mathbb{S}}_{{s'\!\!}_n}(y_n)\rangle\\\nonumber
&=\frac{1}{2^n}\mathcal{G}^{\frac{5}{2}}_{s_1-2}(\partial_{x_1^{B\,+}},\partial_{x_1^{A\,+}})\ldots\mathcal{G}^{\frac{5}{2}}_{s_n-2}(\partial_{x_n^{B\,+}},\partial_{x_n^{A\,+}})\mathcal{G}^{\frac{5}{2}}_{{s'\!\!}_1-2}(\partial_{y_1^{B\,+}},\partial_{y_1^{A\,+}})\ldots\mathcal{G}^{\frac{5}{2}}_{{s'\!\!}_n-2}(\partial_{y_n^{B\,+}},\partial_{y_n^{A\,+}})\\
&\quad\frac{1}{2^{2n}} \langle f^{a_1}_{11}(x_1^B)f^{a_1}_{11}(x_1^A)\ldots f^{a_n}_{11}(x_n^B){f}^{a_n}_{11}(x_n^A) f^{b_1}_{\dot{1}\dot{1}}(y_1^B) f^{b_1}_{\dot{1}\dot{1}}(y_1^A)\ldots f^{b_n}_{\dot{1}\dot{1}}(y_n^B) f^{b_n}_{\dot{1}\dot{1}}(y_n^A)\rangle\Big\rvert_{A=B}
\end{align}
Indeed, in eq. (\ref{717}) we may conveniently relabel the coordinates, $x_i^A\rightarrow x_i^B$, $y_k^A\rightarrow y_k^B$, and vice versa for each $i$ and $k$, since they coincide in the local limit.
Moreover, according to eq. \eqref{GG}, $\mathcal{G}^{\frac{5}{2}}_{s-2}(\partial_{x^{A\,+}},\partial_{x^{B\,+}})$ in eqs. \eqref{bilinear} and \eqref{bilinear2} is symmetric for the exchange of its arguments, because the collinear spin is even. \par
We evaluate the Wick contractions:
\begin{equation}
\label{717}
\langle f^{a_1}_{11}(x_1^A)\ldots f^{a_n}_{11}(x_n^A) f^{b_1}_{\dot{1}\dot{1}}(y_1^A)\ldots  f^{b_n}_{\dot{1}\dot{1}}(y_n^A)  f^{a_1}_{11}(x_1^B)\ldots{f}^{a_n}_{11}(x_n^B) f^{b_1}_{\dot{1}\dot{1}}(y_1^B)\ldots f^{b_n}_{\dot{1}\dot{1}}(y_n^B)\rangle\Big\rvert_{A=B}
\end{equation}
We exploit the symmetry above: We only perform the Wick contractions involving the pairing of $x_i^A$ with $y_k^A$ and of $x_i^B$ with $y_k^B$ for any $i,k$, since all the remaining contractions provide the very same result due to the symmetry, and can be taken into account by a symmetry factor that we compute momentarily.\par
Besides, since we only are interested in the connected correlator, once $x^A_i$ has been contracted with some $y^A_k$, $x^B_i$ cannot be contracted with $y^B_k$, because the corresponding contribution to the correlator is not connected. \par
Hence, we construct the correlator as follows: We contract all the $x_i^A$ with the $y_k^A$ and all the $x_{i'}^B$ with the $y_{{k'\!\!}}^B$ with $i \neq i'$ if $k = {k'\!\!}$ and $k \neq {k'\!\!}$ if $i = i'$, in such a way that we build a single connected loop. \par
This is realized by summing over two sets of independent permutations arranged in such a way that no disconnected piece may be created: Firstly, we contract $x_{i_1}^A$ with  $y_{k_1}^A$, secondly, we contract $y_{k_1}^B$ with $x_{i_2}^B$ for $i_1\neq i_2$, then, we contract $x_{i_2}^A$ with $y_{k_2}^A$ for $k_2\neq k_1$, afterwards, we contract $y_{k_2}^B$ with $x_{i_3}^B$ for $i_3\neq i_2 \neq i_1$ and so on, until we arrive at $x_{i_1}^B$, which we contract with the last remaining $y_{k_n}^A$ with $k_n\neq k_{n-1}\neq\ldots\neq k_1$, in order to close the loop. We end up with a chain that looks like:
\begin{align}
&\sum_{i_1\neq i_2\neq i_3\ldots\neq i_n}\sum_{k_1\neq k_2\neq k_3\ldots\neq k_n}\langle f^{a_{i_1}}_{11}(x_{i_1}^A)f^{b_{k_1}}_{\dot{1}\dot{1}}(y_{k_1}^A)\rangle\langle f^{b_{k_1}}_{\dot{1}\dot{1}}(y_{k_1}^B)f^{a_{i_2}}_{11}(x_{i_2}^B)\rangle\nonumber\\
&\langle f^{a_{i_2}}_{11}(x_{i_2}^A)f^{b_{k_2}}_{\dot{1}\dot{1}}(y_{k_2}^A)\rangle\langle f^{b_{k_2}}_{\dot{1}\dot{1}}(y_{k_2}^B)f^{a_{i_3}}_{11}(x_{i_3}^B)\rangle\ldots\langle f^{a_{i_n}}_{11}(x_{i_n}^A) f^{b_{k_n}}_{\dot{1}\dot{1}}(y_{k_n}^A)\rangle\langle f^{b_{k_n}}_{\dot{1}\dot{1}}(y_{k_n}^B)f^{a_{i_1}}_{11}(x_{i_1}^B)\rangle 
\end{align}
Yet, now we are creating a redundancy, since we also are summing on the possible $n$ choices of the starting point of the loop. Therefore, we divide the sum by a factor of $n$:
 \begin{align}
 &\frac{1}{n}\sum_{i_1\neq i_2\neq i_3\ldots\neq i_n}\sum_{k_1\neq k_2\neq k_3\ldots\neq k_n}\langle f^{a_{i_1}}_{11}(x_{i_1}^A)f^{b_{k_1}}_{\dot{1}\dot{1}}(y_{k_1}^A)\rangle\langle f^{b_{k_1}}_{\dot{1}\dot{1}}(y_{k_1}^B)f^{a_{i_2}}_{11}(x_{i_2}^B)\rangle\nonumber\\
 &\langle f^{a_{i_2}}_{11}(x_{i_2}^A)f^{b_{k_2}}_{\dot{1}\dot{1}}(y_{k_2}^A)\rangle\langle f^{b_{k_2}}_{\dot{1}\dot{1}}(y_{k_2}^B)f^{a_{i_3}}_{11}(x_{i_3}^B)\rangle\ldots\langle f^{a_{i_n}}_{11}(x_{i_n}^A) f^{b_{k_n}}_{\dot{1}\dot{1}}(y_{k_n}^A)\rangle\langle f^{b_{k_n}}_{\dot{1}\dot{1}}(y_{k_n}^B)f^{a_{i_1}}_{11}(x_{i_1}^B)\rangle 
 \end{align}
A nicer -- but completely equivalent -- formula is written in terms of permutations. 
It follows identically: 
\begin{align}
\frac{1}{n}\sum_{\sigma\in P_n}\sum_{\rho\in P_n} &\langle f^{a_{\sigma_1}}_{11}(x_{\sigma_1}^A)f^{b_{\rho_1}}_{\dot{1}\dot{1}}(y_{\rho_1}^A)\rangle\langle f^{b_{\rho_1}}_{\dot{1}\dot{1}}(y_{\rho_1}^B)f^{a_{\sigma_2}}_{11}(x_{\sigma_2}^B)\rangle\langle f^{a_{\sigma_2}}_{11}(x_{\sigma_2}^A)f^{b_{\rho_2}}_{\dot{1}\dot{1}}(y_{\rho_2}^A)\rangle\langle f^{b_{\rho_2}}_{\dot{1}\dot{1}}(y_{\rho_2}^B)f^{a_{\sigma_3}}_{11}(x_{\sigma_3}^B)\rangle\nonumber\\
&\ldots \langle f^{a_{\sigma_n}}_{11}(x_{\sigma_n}^A) f^{b_{\rho_n}}_{\dot{1}\dot{1}}(y_{\rho_n}^A)\rangle\langle f^{b_{\rho_n}}_{\dot{1}\dot{1}}(y_{\rho_n}^B)f^{a_{\sigma_1}}_{11}(x_{\sigma_1}^B)\rangle 
\end{align}
All the remaining contractions are obtained from this formula by exchanging the coordinates in each couple, $(x_i^A,x_i^B)$ and  $(y_k^A,y_k^B)$, for each $i$ and $k$.
There are $2^{2n}$ of such exchanges. \par
However, the actual degeneration factor is $2^{2n-1}$. Indeed, the extra factor of $\frac{1}{2}$ comes from the fact that the simultaneous exchange of the coordinates in each couple, $(x_i^A,x_i^B)$ and $(y_k^A,y_k^B)$, yields a contraction that has already been counted due to the symmetry of  eqs. \eqref{bilinear} and \eqref{bilinear2} with respect of the simultaneous exchange of $A$ with $B$ in all the coordinate pairs.\par
Hence, by combining the degeneration factor of $2^{2n-1}$ with the factor of $\frac{1}{2^n}$ from the normalization of the operators, the overall factor of $2^{n-1}$ survives. It follows:
\begin{align}
\label{unbalanced}
\nonumber
&\langle \mathbb{S}_{s_1}(x_1)\ldots \mathbb{S}_{s_n}(x_n)\bar{\mathbb{S}}_{{s'\!\!}_1}(y_1)\ldots \bar{\mathbb{S}}_{{s'\!\!}_n}(y_n)\rangle=\frac{1}{(4\pi^2)^{2n}}\frac{N^2-1}{2^{2n}}2^{\sum_{l=1}^n s_l+{s'\!\!}_l}i^{\sum_{l=1}^n s_l+{s'\!\!}_l}\\\nonumber
&\frac{\Gamma(3)\Gamma(s_1+3)}{\Gamma(5)\Gamma(s_1+1)}\ldots \frac{\Gamma(3)\Gamma(s_n+3)}{\Gamma(5)\Gamma(s_n+1)}\frac{\Gamma(3)\Gamma({s'\!\!}_1+3)}{\Gamma(5)\Gamma({s'\!\!}_1+1)}\ldots \frac{\Gamma(3)\Gamma({s'\!\!}_n+3)}{\Gamma(5)\Gamma({s'\!\!}_n+1)}\\\nonumber
&\sum_{k_1=0}^{s_1-2}\ldots \sum_{k_n = 0}^{s_n-2}{s_1\choose k_1}{s_1\choose k_1+2}\ldots {s_n\choose k_n}{s_n\choose k_n+2}\\\nonumber
&\sum_{{k'\!\!}_1=0}^{{s'\!\!}_1-2}\ldots \sum_{{k'\!\!}_n = 0}^{{s'\!\!}_n-2}{{s'\!\!}_1\choose {k'\!\!}_1}{{s'\!\!}_1\choose {k'\!\!}_1+2}\ldots {{s'\!\!}_n\choose {k'\!\!}_n}{{s'\!\!}_n\choose {k'\!\!}_n+2}\\\nonumber
&\frac{2^{n-1}}{n}\sum_{\sigma\in P_n}\sum_{\rho\in P_n}
(s_{\sigma(1)}-k_{\sigma(1)}+{k'\!\!}_{\rho(1)})!({s'\!\!}_{\rho(1)}-{k'\!\!}_{\rho(1)}+k_{\sigma(2)})!\\\nonumber
&\ldots(s_{\sigma(n)}-k_{\sigma(n)}+{k'\!\!}_{\rho(n)})!({s'\!\!}_{\rho(n)}-{k'\!\!}_{\rho(n)}+k_{\sigma(1)})!\\\nonumber
&\frac{(x_{\sigma(1)}-y_{\rho(1)})_+^{s_{\sigma(1)}-k_{\sigma(1)}+{k'\!\!}_{\rho(1)}}}{\left(\rvert x_{\sigma(1)}-y_{\rho(1)}\rvert^2\right)^{s_{\sigma(1)}-k_{\sigma(1)}+{k'\!\!}_{\rho(1)}+1}}\frac{(y_{\rho(1)}-x_{\sigma(2)})_+^{{s'\!\!}_{\rho(1)}-{k'\!\!}_{\rho(1)}+k_{\sigma(2)}}}{\left(\rvert y_{\rho(1)}-x_{\sigma(2)}\rvert^2\right)^{{s'\!\!}_{\rho(1)}-{k'\!\!}_{\rho(1)}+k_{\sigma(2)}+1}}\\
&\ldots\frac{(x_{\sigma(n)}-y_{\rho(n)})_+^{s_{\sigma(n)}-k_{\sigma(n)}+{k'\!\!}_{\rho(n)}}}{\left(\rvert x_{\sigma(n)}-y_{\rho(n)}\rvert^2\right)^{s_{\sigma(n)}-k_{\sigma(n)}+{k'\!\!}_{\rho(n)}+1}}
\frac{(y_{\rho(n)}-x_{\sigma(1)})_+^{{s'\!\!}_{\rho(n)}-{k'\!\!}_{\rho(n)}+k_{\sigma(1)}}}{\left(\rvert y_{\rho(n)}-x_{\sigma(1)}\rvert^2\right)^{{s'\!\!}_{\rho(n)}-{k'\!\!}_{\rho(n)}+k_{\sigma(1)}+1}}
\end{align}

\subsection{Extended basis}

Similarly, in the extended basis, we get:
\begin{align}
\label{bilinear3}
\nonumber
\langle \mathbb{A}_{s_1}(x_1)\ldots \mathbb{A}_{s_n}(x_n)\rangle=&\frac{1}{2^n}\mathcal{G}^{\frac{1}{2}}_{s_1}(\partial_{x_1^{A\,+}},\partial_{x_1^{B\,+}})\ldots\mathcal{G}^{\frac{1}{2}}_{s_n}(\partial_{x_n^{A\,+}},\partial_{x_n^{B\,+}})\\
&\langle \partial_{x_1^{A\,+}}^{-1}f^{a_1}_{11}(x_1^A)\partial_{x_1^{B\,+}}^{-1}f^{a_1}_{\dot{1}\dot{1}}(x_1^B)\ldots \partial_{x_n^{A\,+}}^{-1}f^{a_n}_{11}(x_n^A)\partial_{x_n^{B\,+}}^{-1}{f}^{a_n}_{\dot{1}\dot{1}}(x_n^B)\rangle\Big\rvert_{A=B}
\end{align}
in the light-cone gauge, where:
\begin{align} \label{gext}
\mathcal{G}^{\frac{1}{2}}_{s}(\partial_{x_1^+},\partial_{x_2^+}) &=i^{s}\sum_{k=0}^{s} {s\choose k}{s\choose k}(-1)^{s-k} \overleftarrow{\partial}_{x^+_1}^{s-k} \overrightarrow{\partial}_{x^+_2}^{k}
\end{align}
It follows from eqs. \eqref{axialpropeuc2} and \eqref{doubleder} that, correspondingly, the $n$-point correlator contains factors of the form:
\begin{align} \label{na}
\nonumber
&-\partial_{x_i^+}^{s_i-k_i}\partial_{x_j^+}^{k_j}\frac{1}{4\pi^2\rvert x_i-x_j\rvert^2} \\
&=-\frac{1}{4\pi^2}
(-1)^{s_i-k_i} (s_i-k_i+k_j)!\,2^{s_i-k_i+k_j} \frac{(x_i-x_j)_+^{s_i-k_i+k_j}}{(\rvert x_i-x_j\rvert^2)^{s_i-k_i+k_j+1}}
\end{align}
Therefore:
\begin{align}
\nonumber
&\langle \mathbb{A}_{s_1}(x_1)\ldots \mathbb{A}_{s_n}(x_n)\rangle_{conn} =\frac{1}{(4\pi^2)^n}\frac{N^2-1}{2^n}2^{\sum_{l=1}^n s_l}i^{\sum_{l=1}^n s_l}\\\nonumber
&\sum_{k_1=0}^{s_1}\ldots \sum_{k_n = 0}^{s_n}{s_1\choose k_1}^2\ldots {s_n\choose k_n}^2\frac{(-1)^n}{n}\sum_{\sigma\in P_n}(s_{\sigma(1)}-k_{\sigma(1)}+k_{\sigma(2)})!\ldots(s_{\sigma(n)}-k_{\sigma(n)}+k_{\sigma(1)})!\\
&\frac{(x_{\sigma(1)}-x_{\sigma(2)})_+^{s_{\sigma(1)}-k_{\sigma(1)}+k_{\sigma(2)}}}{\left(\rvert x_{\sigma(1)}-x_{\sigma(2)}\rvert^2\right)^{s_{\sigma(1)}-k_{\sigma(1)}+k_{\sigma(2)}+1}}\ldots\frac{(x_{\sigma(n)}-x_{\sigma(1)})_+^{s_{\sigma(n)}-k_{\sigma(n)}+k_{\sigma(1)}}}{\left(\rvert x_{\sigma(n)}-x_{\sigma(1)}\rvert^2\right)^{s_{\sigma(n)}-k_{\sigma(n)}+k_{\sigma(1)}+1}}
\end{align}
where now the overall factor of $(-1)^n$ occurs because of the extra minus sign in eq. \eqref{na} with respect to eq. \eqref{no}. \par
The very same formula holds for an even number of operators $\tilde{\mathbb{A}}_s$, otherwise the correlators vanish. 
We obtain as well:
\begin{align}
\nonumber
&\langle \mathbb{A}_{s_1}(x_1)\ldots \mathbb{A}_{s_n}(x_n)\tilde{\mathbb{A}}_{s_{n+1}}(x_{n+1})\ldots \tilde{\mathbb{A}}_{s_{n+2m}}(x_{n+2m})\rangle_{conn} \\\nonumber
&=\frac{1}{(4\pi^2)^{n+2m}}\frac{N^2-1}{2^{n+2m}}2^{\sum_{l=1}^{n+2m} s_l}i^{\sum_{l=1}^{n+2m} s_l}\sum_{k_1=0}^{s_1}\ldots \sum_{k_{n+2m} = 0}^{s_{n+2m}}{s_1\choose k_1}^2\ldots{s_{n+2m}\choose k_{n+2m}}^2\\\nonumber
&\frac{(-1)^{n+2m}}{n+2m}\sum_{\sigma\in P_{n+2m}}(s_{\sigma(1)}-k_{\sigma(1)}+k_{\sigma(2)})!\ldots(s_{\sigma(n+2m)}-k_{\sigma(n+2m)}+k_{\sigma(1)})!\\
&\frac{(x_{\sigma(1)}-x_{\sigma(2)})_+^{s_{\sigma(1)}-k_{\sigma(1)}+k_{\sigma(2)}}}{\left(\rvert x_{\sigma(1)}-x_{\sigma(2)}\rvert^2\right)^{s_{\sigma(1)}-k_{\sigma(1)}+k_{\sigma(2)}+1}}\ldots\frac{(x_{\sigma(n+2m)}-x_{\sigma(1)})_+^{s_{\sigma(n+2m)}-k_{\sigma(n+2m)}+k_{\sigma(1)}}}{\left(\rvert x_{\sigma(n+2m)}-x_{\sigma(1)}\rvert^2\right)^{s_{\sigma(n+2m)}-k_{\sigma(n+2m)}+k_{\sigma(1)}+1}}\,
\end{align}
Analogously, for the unbalanced operators in the extended basis, we get:
\begin{align}
\label{unbalancedext}
\nonumber
&\langle \mathbb{B}_{s_1}(x_1)\ldots \mathbb{B}_{s_n}(x_n)\bar{\mathbb{B}}_{{s'\!\!}_1}(y_1)\ldots \bar{\mathbb{B}}_{{s'\!\!}_n}(y_n)\rangle=\frac{1}{(4\pi^2)^{2n}}\frac{N^2-1}{2^{2n}}2^{\sum_{l=1}^n s_l+{s'\!\!}_l}i^{\sum_{l=1}^n s_l+{s'\!\!}_l}\\\nonumber
&\sum_{k_1=0}^{s_1}\ldots \sum_{k_n = 0}^{s_n}\sum_{{k'\!\!}_1=0}^{{s'\!\!}_1-2}\ldots \sum_{{k'\!\!}_n = 0}^{{s'\!\!}_n}{s_1\choose k_1}^2\ldots {s_n\choose k_n}^2{{s'\!\!}_1\choose {k'\!\!}_1}^2\ldots {{s'\!\!}_n\choose {k'\!\!}_n}^2\\\nonumber
&\frac{2^{n-1}}{n}\sum_{\sigma\in P_n}\sum_{\rho\in P_n}
(s_{\sigma(1)}-k_{\sigma(1)}+{k'\!\!}_{\rho(1)})!({s'\!\!}_{\rho(1)}-{k'\!\!}_{\rho(1)}+k_{\sigma(2)})!\\\nonumber
&\ldots(s_{\sigma(n)}-k_{\sigma(n)}+{k'\!\!}_{\rho(n)})!({s'\!\!}_{\rho(n)}-{k'\!\!}_{\rho(n)}+k_{\sigma(1)})!\\\nonumber
&\frac{(x_{\sigma(1)}-y_{\rho(1)})_+^{s_{\sigma(1)}-k_{\sigma(1)}+{k'\!\!}_{\rho(1)}}}{\left(\rvert x_{\sigma(1)}-y_{\rho(1)}\rvert^2\right)^{s_{\sigma(1)}-k_{\sigma(1)}+{k'\!\!}_{\rho(1)}+1}}\frac{(y_{\rho(1)}-x_{\sigma(2)})_+^{{s'\!\!}_{\rho(1)}-{k'\!\!}_{\rho(1)}+k_{\sigma(2)}}}{\left(\rvert y_{\rho(1)}-x_{\sigma(2)}\rvert^2\right)^{{s'\!\!}_{\rho(1)}-{k'\!\!}_{\rho(1)}+k_{\sigma(2)}+1}}\\
&\ldots\frac{(x_{\sigma(n)}-y_{\rho(n)})_+^{s_{\sigma(n)}-k_{\sigma(n)}+{k'\!\!}_{\rho(n)}}}{\left(\rvert x_{\sigma(n)}-y_{\rho(n)}\rvert^2\right)^{s_{\sigma(n)}-k_{\sigma(n)}+{k'\!\!}_{\rho(n)}+1}}
\frac{(y_{\rho(n)}-x_{\sigma(1)})_+^{{s'\!\!}_{\rho(n)}-{k'\!\!}_{\rho(n)}+k_{\sigma(1)}}}{\left(\rvert y_{\rho(n)}-x_{\sigma(1)}\rvert^2\right)^{{s'\!\!}_{\rho(n)}-{k'\!\!}_{\rho(n)}+k_{\sigma(1)}+1}}
\end{align}

\section{$n$-point correlators and twist-$2$ gluonic operators in Euclidean space-time} \label{8}

\subsection{Analytic continuation of $n$-point correlators to Euclidean space-time}

 The Minkowskian $n$-point correlators can be analytically continued to Euclidean space-time by substituting (appendix \ref{appN}):
 \begin{equation} \label{ab}
 x_+\rightarrow -i x_{z}
 \end{equation}
 and:
 \begin{equation} \label{bc}
 \frac{1}{\rvert x \rvert^2}\rightarrow-\frac{1}{x^2}
 \end{equation}
We describe the effect of the analytic continuation about various numerical factors.\par
In the standard basis, for the $(n+2m)$-point correlators of balanced operators, the extra factor of $(-i)^{\sum_{l=1}^{n+2m} s_l}$, which arises from the substitution of eq. \eqref{ab} into the numerators of eq. \eqref{O}, cancels out the factor of $i^{\sum_{l=1}^{n+2m} s_l}$ -- already present in eq. \eqref{O} -- that comes from the definition of the Minkowskian operators, but the extra factor of $(-1)^{\sum_{l=1}^{n+2m} (s_l+1)}=(-1)^{\sum_{l=1}^{n+2m} s_l} (-1)^n$ arises, which comes from the substitution of eq. \eqref{bc} into the denominators of eq. \eqref{O}. 
It combines with the factor of $(-1)^n$ already present in eq. \eqref{O}, in such a way that only the factor of $(-1)^{\sum_{l=1}^{n+2m} s_l}$ survives after the analytic continuation.\par
Similarly, for the $2n$-point correlators of unbalanced operators, the factor of  $(-i)^{\sum_{l=1}^n s_l+{s'\!\!}_l}$, which arises from the substitution of eq. \eqref{ab} into the numerators of eq. \eqref{S}, cancels out the factor of $i^{\sum_{l=1}^n s_l+{s'\!\!}_l}$ -- already present in eq. \eqref{S} -- that comes from the definition of the Minkowskian operators. Thus, only the factor of $(-1)^{\sum_{l=1}^n (s_l+1+{s'\!\!}_l+1)}=(-1)^{\sum_{l=1}^n s_l+{s'\!\!}_l}$, which comes from the substitution of eq. \eqref{bc} into the denominators of eq. \eqref{S}, survives after the analytic continuation.
Exactly the same factors survive in the analytic continuation of the corresponding correlators in the extended basis. \par

\subsection{Twist-$2$ gluonic operators in Euclidean space-time}

Alternatively, the correlators may be computed by first defining the Euclidean operators and afterwards evaluating them in Euclidean space-time.
Of course, the two procedures must furnish identical results, as we verify momentarily.\par
By performing the Wick rotation (appendix \ref{appN}) to Euclidean space-time, the operators get rotated as follows.
The derivative along the $p_+$ direction transforms as:
\begin{equation} \label{par}
\partial_+\rightarrow i\partial_{z}
\end{equation}
Correspondingly, for the elementary operators in the light-cone gauge (appendix \ref{appC}), we get:
\begin{align}
\nonumber
&f_{11}=-\partial_{+} \bar{A}\longrightarrow  f^E_{11}=-i\partial_{z} \bar{A}^E\\
&f_{\dot{1}\dot{1}}=-\partial_{+} {A}\longrightarrow  f_{\dot{1}\dot{1}}^E=-i\partial_{z} {A}^E
\end{align}
and:
\begin{align}
\nonumber
&\partial_+^{-1} f_{11}= -\bar{A}\longrightarrow  -i \partial_z^{-1} f^E_{11}=- \bar{A}^E\\
&\partial_+^{-1} f_{\dot{1}\dot{1}}=  -{A}\longrightarrow   -i \partial_z^{-1} f_{\dot{1}\dot{1}}^E= -{A}^E
\end{align}
We observe that the structure and the sign of the propagators (appendix \ref{appA1}) of the Euclidean elementary operators, $f^E_{11}, f_{\dot{1}\dot{1}}^E$ and $ \partial_z^{-1} f^E_{11}, \partial_z^{-1} f_{\dot{1}\dot{1}}^E$, are the same as for the Minkowskian operators,  $f_{11}, f_{\dot{1}\dot{1}}$ and $ \partial_+^{-1} f_{11}, \partial_+^{-1} f_{\dot{1}\dot{1}}$, respectively. \par
Therefore, the change of the numerical factors in the Euclidean correlators with respect to the Minkowskian correlators may only arise from the change of the numerical factors in the definition of the Euclidean composite operators in terms of the Euclidean elementary operators, $f^E_{11}, f_{\dot{1}\dot{1}}^E$ and $ \partial_z^{-1} f^E_{11}, \partial_z^{-1} f_{\dot{1}\dot{1}}^E$. \par
In the standard basis, we get:
\begin{align}
\label{basisE}
\nonumber
&\mathbb{O}_{s}\rightarrow (-1)^s\Tr f^E_{11}(x)(\overrightarrow{D}_{z}+\overleftarrow{D}_{z})^{s-2}C^{\frac{5}{2}}_{s-2}\left(\frac{\overrightarrow{D}_{z}-\overleftarrow{D}_{z}}{\overrightarrow{D}_{z}+\overleftarrow{D}_{z}}\right) f^E_{\dot{1}\dot{1}}(x)=  \mathbb{O}^E_{s}\\\nonumber
&\tilde{\mathbb{O}}_{s}\rightarrow (-1)^s\Tr f^E_{11}(x)(\overrightarrow{D}_{z}+\overleftarrow{D}_{z})^{s-2}C^{\frac{5}{2}}_{s-2}\left(\frac{\overrightarrow{D}_{z}-\overleftarrow{D}_{z}}{\overrightarrow{D}_{z}+\overleftarrow{D}_{z}}\right) f^E_{\dot{1}\dot{1}}(x)=\tilde{\mathbb{O}}^E_{s}\\\nonumber
&\mathbb{S}_{s} \rightarrow\frac{1}{\sqrt{2}}(-1)^s\Tr f^E_{11}(x)(\overrightarrow{D}_{z}+\overleftarrow{D}_{z})^{s-2}C^{\frac{5}{2}}_{s-2}\left(\frac{\overrightarrow{D}_{z}-\overleftarrow{D}_{z}}{\overrightarrow{D}_{z}+\overleftarrow{D}_{z}}\right)f^E_{11}(x)=  \mathbb{S}_{s}^E\\
&\bar{\mathbb{S}}_{s} \rightarrow\frac{1}{\sqrt{2}}(-1)^s\Tr{f}^E_{\dot{1}\dot{1}}(x) (\overrightarrow{D}_{z}+\overleftarrow{D}_{z})^{s-2}C^{\frac{5}{2}}_{s-2}\left(\frac{\overrightarrow{D}_{z}-\overleftarrow{D}_{z}}{\overrightarrow{D}_{z}+\overleftarrow{D}_{z}}\right) f^E_{\dot{1}\dot{1}}(x)= \bar{\mathbb{S}}^E_{s}
\end{align}
since the factor of $i^{s-2}$, which comes from the substitution of eq. \eqref{par} into eq. \eqref{OO}, combines with the already present factor of $i^{s-2}$ in eq. \eqref{OO} to produce the factor of $(-1)^{s}$. \par
As a consequence, for all the Euclidean $n$-point correlators in the standard basis, a factor of $i^{\sum_{l=1}^n (s_l-2)}= i^{\sum_{l=1}^n s_l}(-1)^n$ disappears with respect to the Minkowskian correlators, because it disappears from the definition of the Euclidean operators, but a factor of $(-1)^{\sum_{l=1}^n s_l}$ takes its place, in such a way that only the latter survives, thus providing the same result as in the preceding discussion about the analytic continuation. \par
Similarly, in the extended basis, we get:
\begin{align}
\label{EbasisE}
\nonumber
&\mathbb{A}_{s}\rightarrow (-1)^{s+1}\Tr D_{z}^{-1}f^E_{11}(x)(\overrightarrow{D}_{z}+\overleftarrow{D}_{z})^{s}C^{\frac{1}{2}}_{s}\left(\frac{\overrightarrow{D}_{z}-\overleftarrow{D}_{z}}{\overrightarrow{D}_{z}+\overleftarrow{D}_{z}}\right)D_{z}^{-1} f^E_{\dot{1}\dot{1}}(x)= \mathbb{A}^E_{s}\\\nonumber
&\tilde{\mathbb{A}}_{s}\rightarrow(-1)^{s+1}\Tr D_{z}^{-1}f^E_{11}(x)(\overrightarrow{D}_{z}+\overleftarrow{D}_{z})^{s}C^{\frac{1}{2}}_{s}\left(\frac{\overrightarrow{D}_{z}-\overleftarrow{D}_{z}}{\overrightarrow{D}_{z}+\overleftarrow{D}_{z}}\right)D_{z}^{-1} f^E_{\dot{1}\dot{1}}(x)= \tilde{\mathbb{A}}^E_{s} \\\nonumber
&\mathbb{B}_{s} \rightarrow\frac{1}{\sqrt{2}}(-1)^{s+1}\Tr D_{z}^{-1}f^E_{11}(x)(\overrightarrow{D}_{z}+\overleftarrow{D}_{z})^{s}C^{\frac{1}{2}}_{s}\left(\frac{\overrightarrow{D}_{z}-\overleftarrow{D}_{z}}{\overrightarrow{D}_{z}+\overleftarrow{D}_{z}}\right)D_{z}^{-1}f^E_{11}(x)= \mathbb{B}_{s}^E\\
&\bar{\mathbb{B}}_{s} \rightarrow\frac{1}{\sqrt{2}}(-1)^{s+1}\Tr D_{z}^{-1} f^E_{\dot{1}\dot{1}}(x) (\overrightarrow{D}_{z}+\overleftarrow{D}_{z})^{s}C^{\frac{1}{2}}_{s}\left(\frac{\overrightarrow{D}_{z}-\overleftarrow{D}_{z}}{\overrightarrow{D}_{z}+\overleftarrow{D}_{z}}\right)D_{z}^{-1} f^E_{\dot{1}\dot{1}}(x)=\bar{\mathbb{B}}^E_{s}
\end{align}
since an extra minus sign with respect to the operators in the standard basis comes from the analytic continuation of the two operators $D_+^{-1}$, which contribute $(-i)^2=-1$.\par
Correspondingly, for all the Euclidean $n$-point correlators in the extended basis, a factor of $i^{\sum_{l=1}^n s_l}$ disappears with respect to the Minkowskian correlators,  because it disappears from the definition of the Euclidean operators, but a factor of $(-1)^{\sum_{l=1}^n s_l}(-1)^n$ takes its place, which combines with the factor of $(-1)^n$ already present in the Minkowskian correlators, in such a way that only the factor of  $(-1)^{\sum_{l=1}^n s_l}$ survives, thus providing the same result as in the preceding discussion about the analytic continuation.\par

 \section{Generating functional of $n$-point correlators in the coordinate representation} \label{444}

We work out an ansatz for the generating functional of the correlators in the coordinate representation and verify by direct computation that it reproduces the $n$-point correlators computed in the previous sections. \par
The basic idea is that a conformal operator $\mathcal{O}_s(x)$ of spin $s$ is actually the sum of -- not necessarily conformal -- operators $\mathcal{O}_{sk}(x)$ according to eq. \eqref{confexp}:
\begin{equation}
\mathcal{O}_s(x) = \sum_{k = 0}^{l}\mathcal{O}_{sk}(x)
\end{equation}
where $l=s-2$ for the standard basis and $l=s$ for the extended basis. \par
Hence, the generating functional should be expressed in terms of the corresponding source fields, which by a slight abuse of notation we label by the very same symbols, $\mathcal{O}_s(x)$
and $\mathcal{O}_{sk}(x)$.\par
Originally, the source fields are defined either for even or odd $s$ respectively but, to keep the notation simple in the following formulas, it is convenient to extend their definition to all values of $s$, in such a way that they are $0$ for either odd or even $s$ respectively.\par

\subsection{Minkowskian standard basis} \label{9.1}

Our ansatz for the generating functional of correlators of balanced operators with even spin  is:
\begin{align}
\nonumber
\Gamma_{conf}[\mathbb{O}]&=-(N^2-1)\log \Det\left(\mathbb{I}+\mathcal{D}^{-1}\mathbb{O}\right)\\
\end{align}
where:
\begin{align}
\label{standardker}
\mathcal{D}^{-1}_{s_1k_1,s_2k_2}(x-y) &=\frac{i^{s_1+1}}{2}\frac{\Gamma(3)\Gamma(s_1+3)}{\Gamma(5)\Gamma(s_1+1)}{s_1\choose k_1}{s_2\choose k_2+2}(-\partial_{+})^{s_1-k_1+k_2}\square^{-1}(x-y)\nonumber\\ &=\frac{i^{s_1}}{8\pi^2}\frac{\Gamma(3)\Gamma(s_1+3)}{\Gamma(5)\Gamma(s_1+1)}{s_1\choose k_1}{s_2\choose k_2+2}(-\partial_{+})^{s_1-k_1+k_2}\frac{1}{\rvert x-y\rvert^2-i\epsilon}
\end{align}
with:
\begin{equation}
\square= g^{\mu\nu}\partial_{\mu}\partial_{\nu}=\partial_0^2-\sum_{i=1}^{3}\partial_i^2
\end{equation}
and:
\begin{equation}
\frac{1}{4\pi^2}\frac{1}{\rvert x-y \rvert^2-i\epsilon} =i \square^{-1}\delta^{(4)}(x-y)
\end{equation}
The corresponding correlators are computed by the functional derivatives:
\begin{align} \label{ocorr}
\langle \mathbb{O}_{s_1}(x_1)\ldots \mathbb{O}_{s_n}(x_n)\rangle_{conn} &=\frac{\delta}{\delta\mathbb{O}_{s_1}(x_1)}\ldots\frac{\delta}{\delta\mathbb{O}_{s_n }(x_n)}\Gamma_{conf}[\mathbb{O}] \nonumber \\
&=\sum_{k_1=0}^{s_1-2}\ldots\sum_{k_n=0}^{s_n-2}\frac{\delta}{\delta\mathbb{O}_{s_1 k_1}(x_1)}\ldots\frac{\delta}{\delta\mathbb{O}_{s_n k_n}(x_n)}\Gamma_{conf}[\mathbb{O}]
\end{align}
Explicitly:
\begin{align}
\Gamma_{conf}[\mathbb{O}]=&-(N^2-1)\log \Det\Bigg(\delta_{s_1k_1,s_2k_2}\delta^{(4)}(x-y)\nonumber\\
&+\frac{i^{s_1}}{8\pi^2}\frac{\Gamma(3)\Gamma(s_1+3)}{\Gamma(5)\Gamma(s_1+1)}{s_1\choose k_1}{s_2\choose k_2+2}(-\partial_{+})^{s_1-k_1+k_2}\frac{1}{\rvert x-y\rvert^2-i\epsilon}\mathbb{O}_{s_2k_2}(y)\Bigg)
\end{align}
From now on, we skip the $i \epsilon$ in the effective propagators.\par
Expanding the logarithm of the determinant:
\begin{align}
\nonumber
\Gamma_{conf}[\mathbb{O}] =&-(N^2-1)\sum_{n=1}^{\infty}\frac{(-1)^{n+1}}{n}\int d^4x_1\ldots d^4x_n\, \sum_{s_1k_1}\ldots\sum_{s_nk_n}\\
&\mathcal{D}^{-1}_{s_1k_1,s_2k_2}(x_1-x_2)\mathbb{O}_{s_2k_2}(x_2)
\ldots \mathcal{D}^{-1}_{s_nk_n,s_1k_1}(x_n-x_1)\mathbb{O}_{s_1k_1}(x_1)
\end{align}
and performing the functional derivatives in eq.\eqref{ocorr},
we obtain:
\begin{align}
\nonumber
&\langle \mathbb{O}_{s_1}(x_1)\ldots \mathbb{O}_{s_n}(x_n)\rangle_{conn} = (N^2-1)\sum_{k_1=0}^{s_1-2}\ldots\sum_{k_n=0}^{s_n-2}\frac{(-1)^{n}}{n}\sum_{\sigma\in P_n}\\
&\mathcal{D}^{-1}_{s_{\sigma(1)}k_{\sigma(1)},s_{\sigma(2)}k_{\sigma(2)}}(x_{\sigma(1)}-x_{\sigma(2)})
\ldots \mathcal{D}^{-1}_{s_{\sigma(n)}k_{\sigma(n)},s_{\sigma(1)}k_{\sigma(1)}}(x_{\sigma(n)}-x_{\sigma(1)})
\end{align}
Employing eqs. \eqref{standardker} and \eqref{doubleder},
we get:
\begin{align}
\nonumber
\langle \mathbb{O}_{s_1}(x_1)\ldots \mathbb{O}_{s_n}&(x_n)\rangle_{conn} =\frac{1}{(4\pi^2)^n}\frac{N^2-1}{2^n}2^{\sum_{l=1}^n s_l}i^{\sum_{l=1}^n s_l}\frac{\Gamma(3)\Gamma(s_1+3)}{\Gamma(5)\Gamma(s_1+1)}\ldots \frac{\Gamma(3)\Gamma(s_n+3)}{\Gamma(5)\Gamma(s_n+1)}\\\nonumber
&\sum_{k_1=0}^{s_1-2}\ldots \sum_{k_n = 0}^{s_n-2}{s_1\choose k_1}{s_1\choose k_1+2}\ldots {s_n\choose k_n}{s_n\choose k_n+2}\frac{(-1)^n}{n}\sum_{\sigma\in P_n}\\\nonumber
&(s_{\sigma(1)}-k_{\sigma(1)}+k_{\sigma(2)})!  \ldots(s_{\sigma(n)}-k_{\sigma(n)}+k_{\sigma(1)})!\\
&\frac{(x_{\sigma(1)}-x_{\sigma(2)})_+^{s_{\sigma(1)}-k_{\sigma(1)}+k_{\sigma(2)}}}{\left(\rvert x_{\sigma(1)}-x_{\sigma(2)}\rvert^2\right)^{s_{\sigma(1)}-k_{\sigma(1)}+k_{\sigma(2)}+1}}\ldots\frac{(x_{\sigma(n)}-x_{\sigma(1)})_+^{s_{\sigma(n)}-k_{\sigma(n)}+k_{\sigma(1)}}}{\left(\rvert x_{\sigma(n)}-x_{\sigma(1)}\rvert^2\right)^{s_{\sigma(n)}-k_{\sigma(n)}+k_{\sigma(1)}+1}}
\end{align}
which coincides exactly with the correlator in the coordinate representation in eq. \eqref{O0}.\par 
Now, in the whole balanced sector, we demonstrate that:
\begin{equation}
\label{gfO}
\Gamma_{conf}[\mathbb{O},\tilde{\mathbb{O}}] =  -\frac{N^2-1}{2}\log \Det\left(\left(\mathbb{I}+\mathcal{D}^{-1}\mathbb{O}\right)^2-\mathcal{D}^{-1}\tilde{\mathbb{O}}\mathcal{D}^{-1}\tilde{\mathbb{O}}\right)
\end{equation}
Of course, setting $\mathbb{O}$ or $\tilde{\mathbb{O}}$ to zero, we recover the generating functionals for the single operators:
\begin{align}
\nonumber
\Gamma_{conf}[\mathbb{O}]&=-(N^2-1)\log \Det\left(\mathbb{I}+\mathcal{D}^{-1}\mathbb{O}\right)\\
\Gamma_{conf}[\tilde{\mathbb{O}}]&= -\frac{N^2-1}{2}\log \Det\left(\mathbb{I}-\mathcal{D}^{-1}\tilde{\mathbb{O}}\mathcal{D}^{-1}\tilde{\mathbb{O}}\right)
\end{align}
We rewrite eq. \eqref{gfO} as:
\begin{equation}
\Gamma_{conf}[\mathbb{O},\tilde{\mathbb{O}}] =  -\frac{N^2-1}{2}\log \Det\left[\left(\mathbb{I}+\mathcal{D}^{-1}\mathbb{O}+\mathcal{D}^{-1}\tilde{\mathbb{O}}\right)\left(\mathbb{I}+\mathcal{D}^{-1}\mathbb{O}-\mathcal{D}^{-1}\tilde{\mathbb{O}}\right)\right]
\end{equation}
Thus:
\begin{align}
\Gamma_{conf}[\mathbb{O},\tilde{\mathbb{O}}] =&  -\frac{N^2-1}{2}\log \Det\left(\mathbb{I}+\mathcal{D}^{-1}\mathbb{O}+\mathcal{D}^{-1}\tilde{\mathbb{O}}\right)\nonumber\\
&-\frac{N^2-1}{2}\log \Det\left(\mathbb{I}+\mathcal{D}^{-1}\mathbb{O}-\mathcal{D}^{-1}\tilde{\mathbb{O}}\right)
\end{align}
By expanding the logarithm of the determinant, it follows that:
\begin{align}
\nonumber
\Gamma_{conf}[\mathbb{O},\tilde{\mathbb{O}}] =&-\frac{N^2-1}{2}\sum_{n=1}^{\infty}\frac{(-1)^{n+1}}{n}\int d^4x_1\ldots d^4x_n\, \sum_{s_1k_1}\ldots\sum_{s_nk_n}\\\nonumber
&\mathcal{D}^{-1}_{s_1k_1,s_2k_2}(x_1-x_2)(\mathbb{O}_{s_2k_2}(x_2)+\tilde{\mathbb{O}}_{s_2k_2}(x_2))\\\nonumber
&\ldots \mathcal{D}^{-1}_{s_nk_n,s_1k_1}(x_n-x_1)(\mathbb{O}_{s_1k_1}(x_1)+\tilde{\mathbb{O}}_{s_1k_1}(x_1))\\\nonumber
&-\frac{N^2-1}{2}\sum_{n=1}^{\infty}\frac{(-1)^{n+1}}{n}\int d^4x_1\ldots d^4x_n\, \sum_{s_1k_1}\ldots\sum_{s_nk_n}\\\nonumber
&\mathcal{D}^{-1}_{s_1k_1,s_2k_2}(x_1-x_2)(\mathbb{O}_{s_2k_2}(x_2)-\tilde{\mathbb{O}}_{s_2k_2}(x_2))\\
&\ldots \mathcal{D}^{-1}_{s_nk_n,s_1k_1}(x_n-x_1)(\mathbb{O}_{s_1k_1}(x_1)-\tilde{\mathbb{O}}_{s_1k_1}(x_1))
\end{align}
The correlators read:
\begin{align}
&\langle \mathbb{O}_{s_1}(x_1)\ldots \mathbb{O}_{s_n}(x_n)\tilde{\mathbb{O}}_{s_{n+1}}(x_{n+1})\ldots \tilde{\mathbb{O}}_{s_{m+n}}(x_{m+n})\rangle_{conn}\nonumber\\
&=\frac{\delta}{\delta\mathbb{O}_{s_1}(x_1)}\ldots\frac{\delta}{\delta\mathbb{O}_{s_n }(x_n)}\frac{\delta}{\delta\tilde{\mathbb{O}}_{s_1}(x_{n+1})}\ldots\frac{\delta}{\delta\tilde{\mathbb{O}}_{s_n }(x_{n+m})}\Gamma_{conf}[\mathbb{O},\tilde{\mathbb{O}}]
\end{align}
Performing the functional derivatives yields:
\begin{align}
\nonumber\
&\langle \mathbb{O}_{s_1}(x_1)\ldots \mathbb{O}_{s_n}(x_n)\tilde{\mathbb{O}}_{s_{n+1}}(x_{n+1})\ldots \tilde{\mathbb{O}}_{s_{m+n}}(x_{m+n})\rangle_{conn}\\\nonumber
=&-\frac{N^2-1}{2}\sum_{k_1=0}^{s_1-2}\ldots\sum_{k_{n+m}=0}^{s_{n+m}-2}\frac{(-1)^{n+m+1}}{n+m}\sum_{\sigma\in P_{n+m}}\\\nonumber
&\mathcal{D}^{-1}_{s_{\sigma(1)}k_{\sigma(1)},s_{\sigma(2)}k_{\sigma(2)}}(x_{\sigma(1)}-x_{\sigma(2)})
\ldots \mathcal{D}^{-1}_{s_{\sigma(n)}k_{\sigma(n+m)},s_{\sigma(1)}k_{\sigma(1)}}(x_{\sigma(n+m)}-x_{\sigma(1)})\\\nonumber
&-(-1)^m\frac{N^2-1}{2}\sum_{k_1=0}^{s_1-2}\ldots\sum_{k_{n+m}=0}^{s_{n+m}-2}\frac{(-1)^{n+m+1}}{n+m}\sum_{\sigma\in P_{n+m}}\\\nonumber
&\mathcal{D}^{-1}_{s_{\sigma(1)}k_{\sigma(1)},s_{\sigma(2)}k_{\sigma(2)}}(x_{\sigma(1)}-x_{\sigma(2)})
\ldots \mathcal{D}^{-1}_{s_{\sigma(n)}k_{\sigma(n+m)},s_{\sigma(1)}k_{\sigma(1)}}(x_{\sigma(n+m)}-x_{\sigma(1)})\\\nonumber
=&-(1+(-1)^m)\frac{N^2-1}{2}\sum_{k_1=0}^{s_1-2}\ldots\sum_{k_{n+m}=0}^{s_{n+m}-2}\frac{(-1)^{n+m+1}}{n+m}\sum_{\sigma\in P_{n+m}}\\
&\mathcal{D}^{-1}_{s_{\sigma(1)}k_{\sigma(1)},s_{\sigma(2)}k_{\sigma(2)}}(x_{\sigma(1)}-x_{\sigma(2)})
\ldots \mathcal{D}^{-1}_{s_{\sigma(n)}k_{\sigma(n+m)},s_{\sigma(1)}k_{\sigma(1)}}(x_{\sigma(n+m)}-x_{\sigma(1)})
\end{align}
Hence, the correlators are nonzero only if $m$ is even, as it should be (section \ref{npoint}). Therefore:
\begin{align}
&\langle \mathbb{O}_{s_1}(x_1)\ldots \mathbb{O}_{s_n}(x_n)\tilde{\mathbb{O}}_{s_{n+1}}(x_{n+1})\ldots \tilde{\mathbb{O}}_{s_{m+n}}(x_{m+n})\rangle_{conn}\nonumber\\\nonumber
=&(1+(-1)^m)\frac{N^2-1}{2}\sum_{k_1=0}^{s_1-2}\ldots\sum_{k_{n+m}=0}^{s_{n+m}-2}\frac{(-1)^{n+m}}{n+m}\sum_{\sigma\in P_{n+m}}\mathcal{D}^{-1}_{s_{\sigma(1)}k_{\sigma(1)},s_{\sigma(2)}k_{\sigma(2)}}(x_{\sigma(1)}-x_{\sigma(2)})\\
&\ldots \mathcal{D}^{-1}_{s_{\sigma(n)}k_{\sigma(n+m)},s_{\sigma(1)}k_{\sigma(1)}}(x_{\sigma(n+m)}-x_{\sigma(1)})
\end{align}
which matches eq. \eqref{O} once everything is made explicit by means of eqs. \eqref{standardker} and \eqref{doubleder}:
\begin{align}
\nonumber
&\langle \mathbb{O}_{s_1}(x_1)\ldots \mathbb{O}_{s_n}(x_n)\tilde{\mathbb{O}}_{s_{n+1}}(x_{n+1})\ldots \tilde{\mathbb{O}}_{s_{m+n}}(x_{m+n})\rangle_{conn} =
\frac{(1+(-1)^m)}{2}
\frac{1}{(4\pi^2)^{n+m}}\\\nonumber
&\frac{N^2-1}{2^{n+m}} 2^{\sum_{l=1}^{n+m} s_l}i^{\sum_{l=1}^{n+m} s_l}\frac{\Gamma(3)\Gamma(s_1+3)}{\Gamma(5)\Gamma(s_1+1)}
\ldots\frac{\Gamma(3)\Gamma(s_{m+n}+3)}{\Gamma(5)\Gamma(s_{m+n}+1)}\\\nonumber
&\sum_{k_1=0}^{s_1-2}\ldots \sum_{k_{m+n} = 0}^{s_{m+n}-2}{s_1\choose k_1}{s_1\choose k_1+2}   \dots {s_{m+n}\choose k_{m+n}}{s_{m+n}\choose k_{m+n}+2}    \\\nonumber
&\frac{(-1)^{n+m}}{n+{m}}\sum_{\sigma\in P_{n+m}}(s_{\sigma(1)}-k_{\sigma(1)}+k_{\sigma(2)})!
\ldots(s_{\sigma(n+{m})}-k_{\sigma(n+{m})}+k_{\sigma(1)})!\\
&\frac{(x_{\sigma(1)}-x_{\sigma(2)})_+^{s_{\sigma(1)}-k_{\sigma(1)}+k_{\sigma(2)}}} {\left(\rvert x_{\sigma(1)}-x_{\sigma(2)}\rvert^2\right)^{s_{\sigma(1)}-k_{\sigma(1)}+k_{\sigma(2)}+1}}\ldots\frac{(x_{\sigma(n+m)}-x_{\sigma(1)})_+^{s_{\sigma(n+{m})}-k_{\sigma(n+{m})}+k_{\sigma(1)}}}{\left(\rvert x_{\sigma(n+{m})}-x_{\sigma(1)}\rvert^2\right)^{s_{\sigma(n+{m})}-k_{\sigma(n+{m})}+k_{\sigma(1)}+1}}
\end{align}
In the unbalanced sector, we demonstrate that:
\begin{equation} \label{gfS}
\Gamma_{conf}[\mathbb{S},\bar{\mathbb{S}}]= -\frac{N^2-1}{2}\log \Det\left(\mathbb{I}-2\mathcal{D}^{-1}\bar{\mathbb{S}}\mathcal{D}^{-1}\mathbb{S}\right)
\end{equation}
which, more explicitly, reads:
\begin{align}
\Gamma_{conf}[\mathbb{S},\bar{\mathbb{S}}]=& -\frac{N^2-1}{2}\log \Det\Big(\delta_{s_1k_1,s_2k_2}\delta^{(4)}(x-y)\nonumber\\
&-2 \int d^4z\, \sum_{sk}\mathcal{D}^{-1}_{s_1k_1,sk}(x-z)\bar{\mathbb{S}}_{sk}(z)\mathcal{D}^{-1}_{sk,s_2k_2}(z-y)\mathbb{S}_{s_2k_2}(y)\Big)
\end{align}
By expanding the logarithm of the determinant, it follows that:
\begin{align} 
\Gamma_{conf}[\mathbb{S},\bar{\mathbb{S}}]=& \frac{N^2-1}{2}\sum_{n=1}^{\infty}\frac{2^n}{n}\int d^4x_1\ldots d^4x_n d^4y_1\ldots d^4y_n\sum_{s_1k_1}\ldots\sum_{s_nk_n}\sum_{s'_1k'_1}\ldots\sum_{s'_nk'_n}\nonumber\\\nonumber
&\mathcal{D}^{-1}_{s_1k_1,s'_1k'_1}(x_1-y_1)\bar{\mathbb{S}}_{s'_1k'_1}(y_1)\mathcal{D}^{-1}_{s'_1k'_1,s_2k_2}(y_1-x_2)\mathbb{S}_{s_2k_2}(x_2)\ldots\nonumber\\ &\ldots\mathcal{D}^{-1}_{s_{n}k_{n},s'_nk'_n}(x_n-y_n)\bar{\mathbb{S}}_{s'_nk'_n}(y_n)\mathcal{D}^{-1}_{s'_nk'_n,s_1k_1}(y_n-x_1)\mathbb{S}_{s_1k_1}(x_1)
\end{align}
As a consequence, we obtain the $2n$-point correlator:
\begin{eqnarray} 
&\langle \mathbb{S}_{s_1}(x_1)\ldots \mathbb{S}_{s_n}(x_n)\bar{\mathbb{S}}_{s'_1}(y_1)\ldots \bar{\mathbb{S}}_{s'_n}(y_n)\rangle= \nonumber\\
&\frac{\delta}{\delta\mathbb{S}_{s_1}(x_1)}\ldots\frac{\delta}{\delta\mathbb{S}_{s_n }(x_n)}\frac{\delta}{\delta\bar{\mathbb{S}}_{s'_1}(y_1)}\ldots\frac{\delta}{\delta\bar{\mathbb{S}}_{s'_n }(y_n)}\Gamma_{conf}[\mathbb{S},\bar{\mathbb{S}}]
\end{eqnarray}
by means of eqs. \eqref{standardker} and \eqref{doubleder}:
\begin{align} 
\nonumber
&\langle \mathbb{S}_{s_1}(x_1)\ldots \mathbb{S}_{s_n}(x_n)\bar{\mathbb{S}}_{{s'\!\!}_1}(y_1)\ldots \bar{\mathbb{S}}_{{s'\!\!}_n}(y_n)\rangle=\frac{1}{(4\pi^2)^{2n}}\frac{N^2-1}{2^{2n}}2^{\sum_{l=1}^n s_l+{s'\!\!}_l}i^{\sum_{l=1}^n s_l+{s'\!\!}_l}\\\nonumber
&\frac{\Gamma(3)\Gamma(s_1+3)}{\Gamma(5)\Gamma(s_1+1)}\ldots \frac{\Gamma(3)\Gamma(s_n+3)}{\Gamma(5)\Gamma(s_n+1)}\frac{\Gamma(3)\Gamma({s'\!\!}_1+3)}{\Gamma(5)\Gamma({s'\!\!}_1+1)}\ldots \frac{\Gamma(3)\Gamma({s'\!\!}_n+3)}{\Gamma(5)\Gamma({s'\!\!}_n+1)}\\\nonumber
&\sum_{k_1=0}^{s_1-2}\ldots \sum_{k_n = 0}^{s_n-2}{s_1\choose k_1}{s_1\choose k_1+2}\ldots {s_n\choose k_n}{s_n\choose k_n+2}\\\nonumber
&\sum_{{k'\!\!}_1=0}^{{s'\!\!}_1-2}\ldots \sum_{{k'\!\!}_n = 0}^{{s'\!\!}_n-2}{{s'\!\!}_1\choose {k'\!\!}_1}{{s'\!\!}_1\choose {k'\!\!}_1+2}\ldots {{s'\!\!}_n\choose {k'\!\!}_n}{{s'\!\!}_n\choose {k'\!\!}_n+2}\\\nonumber
&\frac{2^{n-1}}{n}\sum_{\sigma\in P_n}\sum_{\rho\in P_n}
(s_{\sigma(1)}-k_{\sigma(1)}+{k'\!\!}_{\rho(1)})!({s'\!\!}_{\rho(1)}-{k'\!\!}_{\rho(1)}+k_{\sigma(2)})!\\\nonumber
&\ldots(s_{\sigma(n)}-k_{\sigma(n)}+{k'\!\!}_{\rho(n)})!({s'\!\!}_{\rho(n)}-{k'\!\!}_{\rho(n)}+k_{\sigma(1)})!\\\nonumber
&\frac{(x_{\sigma(1)}-y_{\rho(1)})_+^{s_{\sigma(1)}-k_{\sigma(1)}+{k'\!\!}_{\rho(1)}}}{\left(\rvert x_{\sigma(1)}-y_{\rho(1)}\rvert^2\right)^{s_{\sigma(1)}-k_{\sigma(1)}+{k'\!\!}_{\rho(1)}+1}}\frac{(y_{\rho(1)}-x_{\sigma(2)})_+^{{s'\!\!}_{\rho(1)}-{k'\!\!}_{\rho(1)}+k_{\sigma(2)}}}{\left(\rvert y_{\rho(1)}-x_{\sigma(2)}\rvert^2\right)^{{s'\!\!}_{\rho(1)}-{k'\!\!}_{\rho(1)}+k_{\sigma(2)}+1}}\\
&\ldots\frac{(x_{\sigma(n)}-y_{\rho(n)})_+^{s_{\sigma(n)}-k_{\sigma(n)}+{k'\!\!}_{\rho(n)}}}{\left(\rvert x_{\sigma(n)}-y_{\rho(n)}\rvert^2\right)^{s_{\sigma(n)}-k_{\sigma(n)}+{k'\!\!}_{\rho(n)}+1}}
\frac{(y_{\rho(n)}-x_{\sigma(1)})_+^{{s'\!\!}_{\rho(n)}-{k'\!\!}_{\rho(n)}+k_{\sigma(1)}}}{\left(\rvert y_{\rho(n)}-x_{\sigma(1)}\rvert^2\right)^{{s'\!\!}_{\rho(n)}-{k'\!\!}_{\rho(n)}+k_{\sigma(1)}+1}}
\end{align}
which coincides exactly with the correlator in eq. \eqref{S}.

\subsection{Minkowskian extended basis} \label{9.2}

We demonstrate by direct computation that:
\begin{align}
\nonumber
\Gamma_{conf}[\mathbb{A}] =  -(N^2-1)\log \Det\left(\mathbb{I}+\mathcal{D}^{-1}\mathbb{A}\right)\\
\end{align}
with:
\begin{align}
\label{ker}
\mathcal{D}^{-1}_{s_1k_1,s_2k_2}(x-y) &=\frac{i^{s_1+1}}{2}{s_1\choose k_1}{s_2\choose k_2}(-\partial_{+})^{s_1-k_1+k_2}\square^{-1}(x-y) \nonumber \\
&= \frac{i^{s_1}}{8\pi^2}{s_1\choose k_1}{s_2\choose k_2}(-\partial_{+})^{s_1-k_1+k_2}\frac{1}{\rvert x-y\rvert^2-i\epsilon} 
\end{align}
Expanding the logarithm of the determinant:
\begin{align}
\nonumber
\Gamma_{conf}[\mathbb{A}] =&-(N^2-1)\sum_{n=1}^{\infty}\frac{(-1)^{n+1}}{n}\int d^4x_1\ldots d^4x_n\, \sum_{s_1k_1}\ldots\sum_{s_nk_n}\\
&\mathcal{D}^{-1}_{s_1k_1,s_2k_2}(x_1-x_2)\mathbb{A}_{s_2k_2}(x_2)
\ldots \mathcal{D}^{-1}_{s_nk_n,s_1k_1}(x_n-x_1)\mathbb{A}_{s_1k_1}(x_1)
\end{align}
and performing the functional derivatives, we get:
\begin{align}
\nonumber
&\langle \mathbb{A}_{s_1}(x_1)\ldots \mathbb{A}_{s_n}(x_n)\rangle_{conn} = (1-N^2)\sum_{k_1=0}^{s_1}\ldots\sum_{k_n=0}^{s_n}\frac{(-1)^{n+1}}{n}\sum_{\sigma\in P_n}\\
&\mathcal{D}^{-1}_{s_{\sigma(1)}k_{\sigma(1)},s_{\sigma(2)}k_{\sigma(2)}}(x_{\sigma(1)}-x_{\sigma(2)})
\ldots \mathcal{D}^{-1}_{s_{\sigma(n)}k_{\sigma(n)},s_{\sigma(1)}k_{\sigma(1)}}(x_{\sigma(n)}-x_{\sigma(1)})
\end{align}
Employing eq. \eqref{ker},
we obtain:
\begin{align}
\nonumber
\langle \mathbb{A}_{s_1}(x_1)\ldots& \mathbb{A}_{s_n}(x_n)\rangle_{conn}
=\frac{1}{(4\pi^2)^n}\frac{N^2-1}{2^n}2^{\sum_{l=1}^n s_l}i^{\sum_{l=1}^n s_l}\sum_{k_1=0}^{s_1}\ldots\sum_{k_n=0}^{s_n}{s_1\choose k_1}^2\ldots {s_n\choose k_n}^2\\\nonumber
&\frac{(-1)^n}{n}\sum_{\sigma\in P_n}(s_{\sigma(1)}-k_{\sigma(1)}+k_{\sigma(2)})! \ldots(s_{\sigma(n)}-k_{\sigma(n)}+k_{\sigma(1)})!\\
&\frac{(x_{\sigma(1)}-x_{\sigma(2)})_+^{s_{\sigma(1)}-k_{\sigma(1)}+k_{\sigma(2)}}}{\left(\rvert x_{\sigma(1)}-x_{\sigma(2)}\rvert^2\right)^{s_{\sigma(1)}-k_{\sigma(1)}+k_{\sigma(2)}+1}}\ldots\frac{(x_{\sigma(n)}-x_{\sigma(1)})_+^{s_{\sigma(n)}-k_{\sigma(n)}+k_{\sigma(1)}}}{\left(\rvert x_{\sigma(n)}-x_{\sigma(1)}\rvert^2\right)^{s_{\sigma(n)}-k_{\sigma(n)}+k_{\sigma(1)}+1}}
\end{align}
which coincides exactly with eq. \eqref{A0}.\par
Moreover, in analogy with eqs. \eqref{gfO} and \eqref{gfS}, we get:
\begin{align}
\nonumber
&\Gamma_{conf}[\mathbb{A},\tilde{\mathbb{A}}] =  -\frac{N^2-1}{2}\log \det\left(\left(\mathbb{I}+\mathcal{D}^{-1}\mathbb{A}\right)^2-\mathcal{D}^{-1}\tilde{\mathbb{A}}\mathcal{D}^{-1}\tilde{\mathbb{A}}\right)\\
&\Gamma_{conf}[\mathbb{B},\bar{\mathbb{B}}]= -\frac{N^2-1}{2} \log \det\left(\mathbb{I}-2\mathcal{D}^{-1}\bar{\mathbb{B}}\mathcal{D}^{-1}\mathbb{B}\right)
\end{align}

\subsection{Euclidean standard basis} \label{9.3}

Similarly, the generating functionals of the Euclidean correlators are:
\begin{align}
\nonumber
&\Gamma^E_{conf}[\mathbb{O}^E,\tilde{\mathbb{O}}^E] =  -\frac{N^2-1}{2}\log \Det\left(\left(\mathbb{I}+\mathcal{D}_E^{-1}\mathbb{O}^E\right)^2-\mathcal{D}^{-1}_E\tilde{\mathbb{O}}^E\mathcal{D}^{-1}_E\tilde{\mathbb{O}}^E\right)\\
&\Gamma^E_{conf}[\mathbb{S}^E,\bar{\mathbb{S}}^E]= -\frac{N^2-1}{2}\log \Det\left(\mathbb{I}-2\mathcal{D}_E^{-1}\bar{\mathbb{S}}^E\mathcal{D}^{-1}_E\mathbb{S}^E\right)
\end{align}
where the kernel in eq. \eqref{standardker} is analytically continued to Euclidean space-time:
\begin{align}
\nonumber
{\mathcal{D}^{-1}_E}_{s_1k_1,s_2k_2}(x-y)&=\frac{(-i)^{-k_1+k_2}}{2}\frac{\Gamma(3)\Gamma(s_1+3)}{\Gamma(5)\Gamma(s_1+1)}{s_1\choose k_1}{s_2\choose k_2+2}\partial_{z}^{s_1-k_1+k_2}\Laplace^{-1}(x-y)\\
&= -\frac{(-i)^{-k_1+k_2}}{8\pi^2}\frac{\Gamma(3)\Gamma(s_1+3)}{\Gamma(5)\Gamma(s_1+1)}{s_1\choose k_1}{s_2\choose k_2+2}\partial_{z}^{s_1-k_1+k_2}\frac{1}{(x-y)^2}
\end{align}
with:
\begin{equation}
\Laplace= \delta_{\mu\nu}\partial_{\mu}\partial_{\nu}=\partial_4^2+\sum_{i=1}^{3}\partial_i^2
\end{equation}
and:
\begin{equation}
\frac{1}{4\pi^2}\frac{1}{(x-y)^2} = -\Laplace^{-1}\delta^{(4)}(x-y)
\end{equation}

\subsection{Euclidean extended basis} \label{9.4}

Analogously:
\begin{align}
\nonumber
&\Gamma^E_{conf}[\mathbb{A}^E,\tilde{\mathbb{A}}^E] =  -\frac{N^2-1}{2}\log \Det\left(\left(\mathbb{I}+\mathcal{D}_E^{-1}\mathbb{A}^E\right)^2-\mathcal{D}^{-1}_E\tilde{\mathbb{A}}^E\mathcal{D}^{-1}_E\tilde{\mathbb{A}}^E\right)\\
&\Gamma^E_{conf}[\mathbb{B}^E,\bar{\mathbb{B}}^E]= -\frac{N^2-1}{2}\log \Det\left(\mathbb{I}-2\mathcal{D}_E^{-1}\bar{\mathbb{B}}^E\mathcal{D}^{-1}_E\mathbb{B}^E\right)
\end{align}
where the kernel in eq. \eqref{ker} is analytically continued to Euclidean space-time:
\begin{align}
{\mathcal{D}^{-1}_E}_{s_1k_1,s_2k_2}(x-y) &= \frac{(-i)^{-k_1+k_2}}{2}{s_1\choose k_1}{s_2\choose k_2}\partial_{z}^{s_1-k_1+k_2}\Laplace^{-1}(x-y)\nonumber\\
&= -\frac{(-i)^{-k_1+k_2}}{8\pi^2}{s_1\choose k_1}{s_2\choose k_2}\partial_{z}^{s_1-k_1+k_2}\frac{1}{(x-y)^2}
\end{align}

\section{Generating functional and $n$-point correlators in the momentum representation} \label{55}

The generating functionals of correlators in the momentum representation are obtained from the corresponding generating functionals in the coordinate representation (section \ref{444}) as follows. \par
Given the generating functional in the coordinate representation:
\begin{equation}
\Gamma_{conf}[\mathcal{O}] \sim \log \det \left(\delta_{s_1k_1,s_2k_2}\delta^{(4)}(x-y)+\mathcal{D}^{-1}_{s_1k_1,s_2k_2}(x-y)\mathcal{O}_{s_2k_2}(y)\right)
\end{equation}
the argument of the determinant is the kernel:
\begin{equation} \label{KK}
K_{s_1k_1,s_2k_2}(x,y)=\delta_{s_1k_1,s_2k_2}\delta^{(4)}(x-y)+\mathcal{D}^{-1}_{s_1k_1,s_2k_2}(x-y)\mathcal{O}_{s_2k_2}(y)
\end{equation}
of the integral operator:
\begin{equation}
\psi_{s_1k_1}(x) =\sum_{s_2k_2} \int K_{s_1k_1,s_2k_2}(x,y) \phi_{s_2k_2}(y) d^4y
\end{equation}
In order to obtain the kernel in the momentum representation, we perform the Fourier transform of the lhs:
\begin{equation}
\psi_{s_1k_1}(q)=\int  \psi_{s_1k_1}(x) e^{-iqx} d^4x=\sum_{s_2k_2} \int K_{s_1k_1,s_2k_2}(x,y) \phi_{s_2k_2}(y) e^{-iqx}\,d^4x\,d^4y
\end{equation}
and write the rhs in terms of the Fourier-transformed fields:
\begin{equation}
\psi_{s_1k_1}(q) =\sum_{s_2k_2} \int K_{s_1k_1,s_2k_2}(x,y) \phi_{s_2k_2}(p)e^{ipy} e^{-iqx}\,d^4x\,d^4y\frac{d^4p}{(2\pi)^4}
\end{equation}
By substituting the kernel in eq. \eqref{KK}:
\begin{align}
\psi_{s_1k_1}(q) =&\sum_{s_2k_2} \int \delta_{s_1k_1,s_2k_2}\delta^{(4)}(x-y)\phi_{s_2k_2}(p)e^{ipy} e^{-iqx}\nonumber\\
&+ \mathcal{D}^{-1}_{s_1k_1,s_2k_2}(x-y)\mathcal{O}_{s_2k_2}(y) \phi_{s_2k_2}(p)e^{ipy} e^{-iqx}\,d^4x\,d^4y\frac{d^4p}{(2\pi)^4}
\end{align}
we get:
\begin{align}
\psi_{s_1k_1}(q) =&\sum_{s_2k_2} \int \delta_{s_1k_1,s_2k_2}(2\pi)^4\delta^{(4)}(p-q)\phi_{s_2k_2}(p)\frac{d^4p}{(2\pi)^4}\nonumber\\
&+ \sum_{s_2k_2} \int \mathcal{D}^{-1}_{s_1k_1,s_2k_2}(x-y)\mathcal{O}_{s_2k_2}(y) \phi_{s_2k_2}(p)e^{ipy} e^{-iqx}\,d^4x\,d^4y\frac{d^4p}{(2\pi)^4}
\end{align}
The second line in the above equation becomes:
\begin{align}
\sum_{s_2k_2} \int\mathcal{D}^{-1}_{s_1k_1,s_2k_2}(k_1)\mathcal{O}_{s_2k_2}(k_2) \phi_{s_2k_2}(p)e^{ik_1(x-y)}e^{ik_2y}e^{ipy} e^{-iqx}\,d^4x\,d^4y\frac{d^4p}{(2\pi)^4}\frac{d^4k_1}{(2\pi)^4}\frac{d^4k_2}{(2\pi)^4}
\end{align}
which simplifies to:
\begin{align}
\sum_{s_2k_2} \int\mathcal{D}^{-1}_{s_1k_1,s_2k_2}(q)\mathcal{O}_{s_2k_2}(q-p) \phi_{s_2k_2}(p)\frac{d^4p}{(2\pi)^4}
\end{align}
Therefore, the kernel in the momentum representation reads:
\begin{equation}
K_{s_1k_1,s_2k_2}(q_1,q_2)=\delta_{s_1k_1,s_2k_2}(2\pi)^4\delta^{(4)}(q_1-q_2)+\mathcal{D}^{-1}_{s_1k_1,s_2k_2}(q_1)\mathcal{O}_{s_2k_2}(q_1-q_2)
\end{equation}
which defines the integral operator:
\begin{equation}
\psi_{s_1k_1}(q_1) =\sum_{s_2k_2} \int K_{s_1k_1,s_2k_2}(q_1,q_2) \phi_{s_2k_2}(q_2) \frac{d^4q_2}{(2\pi)^4}
\end{equation}
Equivalently, we may expand the logarithm of the functional determinant in the coordinate representation:
\begin{align}
\nonumber
\Gamma_{conf}[\mathcal{O}] \sim&\sum_{n=1}^{\infty}\frac{(-1)^{n+1}}{n}\int d^4x_1\ldots d^4x_n\, \sum_{s_1k_1}\ldots\sum_{s_nk_n}\mathcal{D}^{-1}_{s_1k_1,s_2k_2}(x_1-x_2)\mathcal{O}_{s_2k_2}(x_2)\\
&\ldots \mathcal{D}^{-1}_{s_nk_n,s_1k_1}(x_n-x_1)\mathcal{O}_{s_1k_1}(x_1)
\end{align}
and express the source fields and effective propagators in terms of their Fourier transforms $\mathcal{O}_{sk}(p)$:
\begin{align}
\mathcal{O}_{sk}(x) = \int \frac{d^4p}{(2\pi)^4} \mathcal{O}_{sk}(p)\, e^{ip\cdot x}
\end{align}
and $\mathcal{D}^{-1}_{s_1k_1,s_2k_2}(q)$;
\begin{align}
\mathcal{D}^{-1}_{s_1k_1,s_2k_2}(x-y) = \int \frac{d^4q}{(2\pi)^4} \mathcal{D}^{-1}_{s_1k_1,s_2k_2}(q)\, e^{iq\cdot(x-y)}
\end{align}
We get:
\begin{align}
\nonumber
\Gamma_{conf}[\mathcal{O}] \sim&\sum_{n=1}^{\infty}\frac{(-1)^{n+1}}{n}\int d^4x_1\ldots d^4x_n\,\frac{d^4q_1}{(2\pi)^4}\ldots \frac{d^4q_n}{(2\pi)^4}\,\frac{d^4p_1}{(2\pi)^4}\ldots \frac{d^4p_n}{(2\pi)^4}\nonumber\\ &\sum_{s_1k_1}\ldots\sum_{s_nk_n}\mathcal{D}^{-1}_{s_1k_1,s_2k_2}(q_1)\mathcal{O}_{s_2k_2}(p_1)\ldots \mathcal{D}^{-1}_{s_nk_n,s_1k_1}(q_n)\mathcal{O}_{s_1k_1}(p_n)\nonumber\\
&e^{iq_1 \cdot(x_1-x_2)}\ldots e^{iq_n \cdot(x_n-x_1)}\,e^{ip_1 \cdot x_2}\ldots e^{ip_n \cdot x_1}
\end{align}
Performing the product of the exponentials above:
\begin{align}
e^{-i x_1 \cdot(q_n-q_1-p_n)}e^{-i x_2 \cdot(q_1-q_2-p_1)}e^{-i x_3 \cdot(q_2-q_3-p_2)}\ldots e^{-i x_n \cdot(q_{n-1}-q_n-p_{n-1})}
\end{align}
and employing:
\begin{equation}
\frac{1}{(2\pi)^4}\int d^4x\, e^{-iq\cdot x} = \delta^{(4)}(q)
\end{equation}
we obtain:
\begin{align}
\nonumber
\Gamma_{conf}[\mathcal{O}] \sim&\sum_{n=1}^{\infty}\frac{(-1)^{n+1}}{n}\int \frac{d^4q_1}{(2\pi)^4}\ldots \frac{d^4q_n}{(2\pi)^4}\sum_{s_1k_1}\ldots\sum_{s_nk_n}\mathcal{D}^{-1}_{s_1k_1,s_2k_2}(q_1)\mathcal{O}_{s_2k_2}(q_1-q_2)\nonumber\\ &\mathcal{D}^{-1}_{s_2k_2,s_3k_3}(q_2)\mathcal{O}_{s_3k_3}(q_2-q_3)\ldots \mathcal{D}^{-1}_{s_nk_n,s_1k_1}(q_n)\mathcal{O}_{s_1k_1}(q_n-q_1)
\end{align}
Either way, the generating functional in the momentum representation reads:
\begin{equation}
\Gamma_{conf}[\mathcal{O}] \sim \log\Det\left(\delta_{s_1k_1,s_2k_2} (2\pi)^4\delta^{(4)}(q_1-q_2)+\mathcal{D}^{-1}_{s_1k_1,s_2k_2}(q_1)\mathcal{O}_{s_2k_2}(q_1-q_2)\right)
\end{equation}
and the corresponding correlators are obtained by the functional derivatives:
\begin{align} \label{ocorrp}
\langle \mathcal{O}_{s_1}(p_1)\ldots  \mathcal{O}_{s_n}(p_n)\rangle_{conn} &=\frac{\delta}{\delta  \mathcal{O}_{s_1}(p_1)}\ldots\frac{\delta}{\delta  \mathcal{O}_{s_n }(p_n)}\Gamma_{conf}[ \mathcal{O}] \nonumber \\
&=\sum_{k_1=0}^{l_1}\ldots\sum_{k_n=0}^{l_n}\frac{\delta}{\delta  \mathcal{O}_{s_1 k_1}(p_1)}\ldots\frac{\delta}{\delta  \mathcal{O}_{s_n k_n}(p_n)}\Gamma_{conf}[ \mathcal{O}]
\end{align}

\subsection{Minkowskian standard basis}

The generating functionals in the coordinate representation in Minkowskian space-time read (section \ref{9.1}):
 \begin{align}
\nonumber
\Gamma_{conf}[\mathbb{O}] =&-(N^2-1)\log \Det\left(\mathbb{I}+\mathcal{D}^{-1}\mathbb{O}\right)\\\nonumber
\Gamma_{conf}[\tilde{\mathbb{O}}] =& -\frac{N^2-1}{2}\log \Det\left(\mathbb{I}-\mathcal{D}^{-1}\tilde{\mathbb{O}}\mathcal{D}^{-1}\tilde{\mathbb{O}}\right)\\\nonumber
\Gamma_{conf}[\mathbb{O},\tilde{\mathbb{O}}] =& -\frac{N^2-1}{2}\log \Det\left(\mathbb{I}+\mathcal{D}^{-1}\mathbb{O}+\mathcal{D}^{-1}\tilde{\mathbb{O}}\right)\nonumber\\
&-\frac{N^2-1}{2}\log \det\left(\mathbb{I}+\mathcal{D}^{-1}\mathbb{O}-\mathcal{D}^{-1}\tilde{\mathbb{O}}\right)\nonumber\\
=& -\frac{N^2-1}{2}\log \det\left(\left(\mathbb{I}+\mathcal{D}^{-1}\mathbb{O}\right)^2-\mathcal{D}^{-1}\tilde{\mathbb{O}}\mathcal{D}^{-1}\tilde{\mathbb{O}}\right)\nonumber\\
\Gamma_{conf}[\mathbb{S},\bar{\mathbb{S}}] =& -\frac{N^2-1}{2}\log \Det\left(\mathbb{I}-2\mathcal{D}^{-1}\bar{\mathbb{S}}\mathcal{D}^{-1}\mathbb{S}\right)
\end{align}
By making explicit the continuous and discrete indices, the above equations read in the momentum  representation:
\begin{align}
\nonumber
\Gamma_{conf}[\mathbb{O}]=&-(N^2-1)\log \Det\left(\delta_{s_1k_1,s_2k_2}(2\pi)^4\delta^{(4)}(q_1-q_2)+\mathcal{D}^{-1}_{s_1k_1,s_2k_2}(q_1)\mathbb{O}_{s_2k_2}(q_1-q_2)\right)\\\nonumber
\Gamma_{conf}[\tilde{\mathbb{O}}]=& -\frac{N^2-1}{2}\log \Det\Big(\delta_{s_1k_1,s_2k_2}(2\pi)^4\delta^{(4)}(q_1-q_2)\\\nonumber
\qquad\qquad\qquad&-\int \frac{d^4q}{(2\pi)^4}\sum_{sk}\mathcal{D}^{-1}_{s_1k_1,sk}(q_1)\tilde{\mathbb{O}}_{sk}(q_1-q)\mathcal{D}^{-1}_{sk,s_2k_2}(q)\tilde{\mathbb{O}}_{s_2k_2}(q-q_2)\Big)\\\nonumber
\Gamma_{conf}[\mathbb{O},\tilde{\mathbb{O}}]= &- \frac{N^2-1}{2}\log \Det\Big(\delta_{s_1k_1,s_2k_2}(2\pi)^4 \delta^{(4)}(q_1-q_2) \\\nonumber
&+\mathcal{D}^{-1}_{s_1k_1,s_2k_2}(q_1)(\mathbb{O}_{s_2k_2}(q_1-q_2)+ \tilde{\mathbb{O}}_{s_2k_2}(q_1-q_2)\Big)\\\nonumber
&- \frac{N^2-1}{2}\log \Det\Big(\delta_{s_1k_1,s_2k_2}(2\pi)^4 \delta^{(4)}(q_1-q_2) \\\nonumber
&+\mathcal{D}^{-1}_{s_1k_1,s_2k_2}(q_1)(\mathbb{O}_{s_2k_2}(q_1-q_2)- \tilde{\mathbb{O}}_{s_2k_2}(q_1-q_2)\Big)\\\nonumber
\Gamma_{conf}[\mathbb{S},\bar{\mathbb{S}}]=& -\frac{N^2-1}{2}\log \Det\Big(\delta_{s_1k_1,s_2k_2}(2\pi)^4\delta^{(4)}(q_1-q_2)\\
\qquad\qquad\qquad&-2\int \frac{d^4q}{(2\pi)^4}\sum_{sk}\mathcal{D}^{-1}_{s_1k_1,sk}(q_1)\bar{\mathbb{S}}_{sk}(q_1-q)\mathcal{D}^{-1}_{sk,s_2k_2}(q)\mathbb{S}_{s_2k_2}(q-q_2)\Big)\\\nonumber
\end{align}
with:
\begin{align}
\mathcal{D}^{-1}_{s_1k_1,s_2k_2}(p) &=\frac{i^{s_1}}{2}\frac{\Gamma(3)\Gamma(s_1+3)}{\Gamma(5)\Gamma(s_1+1)}{s_1\choose k_1}{s_2\choose k_2+2}(-ip_{+})^{s_1-k_1+k_2}\frac{-i}{\rvert p\rvert^2+i\epsilon} 
\end{align}
where the Fourier transform of the effective propagator in eq. \eqref{standardker} is computed by:
\begin{equation}
\int \frac{d^4p}{(2\pi)^4}\, e^{ip\cdot(x-y)}\, \frac{-i}{\rvert p\rvert^2+i\epsilon}= \frac{1}{4\pi^2}\frac{1}{\rvert x-y\rvert^2-i\epsilon}
\end{equation}
Expanding the logarithm of the determinant and defining $p_i = q_i-q_{i+1}$, we obtain:
\begin{align}
\nonumber
\Gamma^E_{conf}[\mathbb{O}]  =&-(N^2-1)\sum_{n=1}^{\infty}\frac{(-1)^{n+1}}{n}\int d^4p_1\ldots d^4p_n\,(2\pi)^4\delta^{(4)}(p_1+p_2+\ldots+p_n)\\\nonumber
&\int\frac{d^4q}{(2\pi)^4} \sum_{s_1k_1}\ldots\sum_{s_nk_n}\mathcal{D}^{-1}_{s_1k_1,s_2k_2}(q+p_n)\mathbb{O}_{s_2k_2}(p_1)\mathcal{D}^{-1}_{s_1k_1,s_2k_2}(q+p_1+p_n)\mathbb{O}_{s_2k_2}(p_2)\\
&\ldots \mathcal{D}^{-1}_{s_nk_n,s_1k_1}(q)\mathbb{O}_{s_1k_1}(p_n)
\end{align}
The functional derivatives yield:
\begin{align}
\nonumber
&\left(\frac{\Gamma(5)\Gamma(s_1+1)}{\Gamma(3)\Gamma(s_1+3)}\right)\ldots\left( \frac{\Gamma(5)\Gamma(s_n+1)}{\Gamma(3)\Gamma(s_n+3)}\right)\langle \mathbb{O}_{s_1}(p_1)\ldots \mathbb{O}_{s_n}(p_n)\rangle_{conn} \\\nonumber
&=\frac{N^2-1}{2^n}(2\pi)^4 i^n\delta^{(4)}(p_1+\ldots+p_n)\sum_{k_1=0}^{s_1-2}\ldots \sum_{k_n = 0}^{s_n-2}{s_1\choose k_1}{s_1\choose k_1+2}\ldots {s_n\choose k_n}{s_n\choose k_n+2}\\\nonumber
&\frac{1}{n}\sum_{\sigma\in P_n}\int \frac{d^4 q}{(2\pi)^4}\frac{(p_{\sigma(1)}+q)_{+}^{s_{\sigma(1)}-k_{\sigma(1)}+k_{\sigma(2)}}}{\rvert p_{\sigma(1)}+q\rvert^2}\frac{(p_{\sigma(1)}+p_{\sigma(2)}+q)_{+}^{s_{\sigma(2)}-k_{\sigma(2)}+k_{\sigma(3)}}}{\rvert p_{\sigma(1)}+p_{\sigma(2)}+q\rvert^2}\\
&\ldots\frac{(\sum_{l=1}^{n-1} p_{\sigma(l)}+q)_{+}^{s_{\sigma(n-1)}-k_{\sigma(n-1)}+k_{\sigma(n)}}}{\rvert\sum_{l=1}^{n-1} p_{\sigma(l)}+q\rvert^2}\frac{(q)_{+}^{s_{\sigma(n)}-k_{\sigma(n)}+k_{\sigma(1)}}}{\rvert q\rvert^2}
\end{align}
All the remaining correlators are obtained in a similar way.\par 
For the nonvanishing correlators in the balanced sector, we get:
 \begin{align}
\nonumber
&\left(\frac{\Gamma(5)\Gamma(s_1+1)}{\Gamma(3)\Gamma(s_1+3)}\right)\ldots\left( \frac{\Gamma(5)\Gamma(s_{n+2m}+1)}{\Gamma(3)\Gamma(s_{n+2m}+3)}\right)\\\nonumber
&\langle \mathbb{O}_{s_1}(p_1)\ldots \mathbb{O}_{s_n}(p_n)\tilde{\mathbb{O}}_{s_{n+1}}(p_{n+1})\ldots \tilde{\mathbb{O}}_{s_{n+2m}}(p_{n+2m})\rangle_{conn} \\\nonumber
&=\frac{N^2-1}{2^{n+2m}}(2\pi)^4 i^{n+2m}\delta^{(4)}(p_1+\ldots+p_{n+2m})\\\nonumber
&\sum_{k_1=0}^{s_1-2}\ldots \sum_{k_{n+2m} = 0}^{s_{n+2m}-2}{s_1\choose k_1}{s_1\choose k_1+2}\ldots {s_{n+2m}\choose k_{n+2m}}{s_{n+2m}\choose k_{n+2m}+2}\frac{1}{n+2m}\sum_{\sigma\in P_{n+2m}}\\\nonumber
&\int \frac{d^4 q}{(2\pi)^4}\frac{(p_{\sigma(1)}+q)_{+}^{s_{\sigma(1)}-k_{\sigma(1)}+k_{\sigma(2)}}}{\rvert p_{\sigma(1)}+q\rvert^2}\frac{(p_{\sigma(1)}+p_{\sigma(2)}+q)_{+}^{s_{\sigma(2)}-k_{\sigma(2)}+k_{\sigma(3)}}}{\rvert p_{\sigma(1)}+p_{\sigma(2)}+q\rvert^2}\\
&\ldots\frac{(\sum_{l=1}^{n+2m-1} p_{\sigma(l)}+q)_{+}^{s_{\sigma(n+2m-1)}-k_{\sigma(n+2m-1)}+k_{\sigma(n+2m)}}}{\rvert\sum_{l=1}^{n+2m-1} p_{\sigma(l)}+q\rvert^2}\frac{(q)_{+}^{s_{\sigma(n+2m)}-k_{\sigma(n+2m)}+k_{\sigma(1)}}}{\rvert q\rvert^2}
\end{align}
In the unbalanced sector, we obtain:
\begin{align}
\nonumber
&\left(\frac{\Gamma(5)\Gamma(s_1+1)}{\Gamma(3)\Gamma(s_1+3)}\right)\ldots\left(\frac{\Gamma(5)\Gamma(s_n+1)}{\Gamma(3)\Gamma(s_n+3)}\right)\left(\frac{\Gamma(5)\Gamma(s'_1+1)}{\Gamma(3)\Gamma(s'_1+3)}\right)\ldots\left( \frac{\Gamma(5)\Gamma(s'_{n}+1)}{\Gamma(3)\Gamma(s'_{n}+3)}\right)\\\nonumber
&\langle \mathbb{S}_{s_1}(p_1)\ldots \mathbb{S}_{s_n}(p_n)\bar{\mathbb{S}}_{s'_1}(p'_1)\ldots \bar{\mathbb{S}}_{s'_n}(p'_n)\rangle\\\nonumber
&=\frac{N^2-1}{2^{2n}}(2\pi)^4 i^{2n}\delta^{(4)}\left(\sum_{l=1}^n p_l+p'_l\right)\\\nonumber
&\sum_{k_1=0}^{s_1-2}\ldots \sum_{k_n = 0}^{s_n-2}\sum_{k'_1=0}^{s'_1-2}\ldots \sum_{k'_n = 0}^{s'_n-2}{s_1\choose k_1}{s_1\choose k_1+2}\ldots {s_n\choose k_n}{s_n\choose k_n+2}{s'_1\choose k'_1}{s'_1\choose k'_1+2}\ldots {s'_n\choose k'_n}{s'_n\choose k'_n+2}\\\nonumber
&\frac{2^{n-1}}{n}\sum_{\sigma\in P_n}\sum_{\rho\in P_n}
\int \frac{d^4 q}{(2\pi)^4}\frac{(p_{\sigma(1)}+q)_{+}^{s_{\sigma(1)}-k_{\sigma(1)}+k'_{\rho(1)}}}{\rvert p_{\sigma(1)}+q\rvert^2}\frac{(p_{\sigma(1)}+p'_{\rho(1)}+q)_{+}^{s'_{\rho(1)}-k'_{\rho(1)}+k_{\sigma(2)}}}{\rvert p_{\sigma(1)}+p'_{\rho(1)}+q\rvert^2}\\\nonumber
&\frac{(p_{\sigma(1)}+p_{\sigma(2)}+p'_{\rho(1)}+q)_{+}^{s_{\sigma(2)}-k_{\sigma(2)}+k'_{\rho(2)}}}{\rvert p_{\sigma(1)}+p_{\sigma(2)}+p'_{\rho(1)}+q\rvert^2}\frac{(p_{\sigma(1)}+p_{\sigma(2)}+p'_{\rho(1)}+p'_{\rho(2)}+q)_{+}^{s'_{\rho(2)}-k'_{\rho(2)}+k_{\sigma(3)}}}{\rvert p_{\sigma(1)}+p_{\sigma(2)}+p'_{\rho(1)}+p'_{\rho(2)}+q\rvert^2}\\
 &\ldots\frac{(\sum_{l=1}^{n-1} p_{\sigma(l)}+\sum_{l=1}^{n-2} p'_{\rho(l)} +q)_{+}^{s_{\sigma(n-1)}-k_{\sigma(n-1)}+k'_{\rho(n-1)}}}{\rvert\sum_{l=1}^{n-1} p_{\sigma(l)}+\sum_{l=1}^{n-2} p'_{\rho(l)}+q\rvert^2}\nonumber\\
 &\frac{(\sum_{l=1}^{n-1} p_{\sigma(l)}+\sum_{l=1}^{n-1} p'_{\rho(l)}+q)_{+}^{s'_{\rho(n-1)}-k'_{\rho(n-1)}+k_{\sigma(n)}}}{\rvert \sum_{l=1}^{n-1} p_{\sigma(l)}+\sum_{l=1}^{n-1} p'_{\rho(l)}+q\rvert^2}\nonumber\\
&\frac{(\sum_{l=1}^n p_{\sigma(l)}+\sum_{l=1}^{n-1} p'_{\rho(l)} +q)_{+}^{s_{\sigma(n)}-k_{\sigma(n)}+k'_{\rho(n)}}}{\rvert\sum_{l=1}^n p_{\sigma(l)}+\sum_{l=1}^{n-1} p'_{\rho(l)}+q\rvert^2}\frac{(q)_{+}^{s'_{\rho(n)}-k'_{\rho(n)}+k_{\sigma(1)}}}{\rvert q\rvert^2}
\end{align}

\subsection{Minkowskian extended basis}

Similarly:
\begin{align}
\nonumber
\Gamma_{conf}[\mathbb{A},\tilde{\mathbb{A}}]=& - \frac{N^2-1}{2}\log \Det\Big(\delta_{s_1k_1,s_2k_2} (2\pi)^4\delta^{(4)}(q_1-q_2) \\\nonumber
&+\mathcal{D}^{-1}_{s_1k_1,s_2k_2}(q_1)(\mathbb{A}_{s_2k_2}(q_1-q_2)+ \tilde{\mathbb{A}}_{s_2k_2}(q_1-q_2)\Big)\\\nonumber
& - \frac{N^2-1}{2}\log \Det\Big(\delta_{s_1k_1,s_2k_2}(2\pi)^4 \delta^{(4)}(q_1-q_2) \\\nonumber
&+\mathcal{D}^{-1}_{s_1k_1,s_2k_2}(q_1)(\mathbb{A}_{s_2k_2}(q_1-q_2)- \tilde{\mathbb{A}}_{s_2k_2}(q_1-q_2)\Big)\\\nonumber
\Gamma_{conf}[\mathbb{B},\bar{\mathbb{B}}]=& -\frac{N^2-1}{2}\log \det\Big(\delta_{s_1k_1,s_2k_2}(2\pi)^4\delta^{(4)}(q_1-q_2)\\
\qquad\qquad\qquad&-2\int \frac{d^4q}{(2\pi)^4}\sum_{sk}\mathcal{D}^{-1}_{s_1k_1,sk}(q_1)\bar{\mathbb{B}}_{sk}(q_1-q)\mathcal{D}^{-1}_{sk,s_2k_2}(q)\mathbb{B}_{s_2k_2}(q-q_2)\Big)\\\nonumber
\end{align}
with:
\begin{align}
\mathcal{D}^{-1}_{s_1k_1,s_2k_2}(p) =\frac{i^{s_1}}{2}{s_1\choose k_1}{s_2\choose k_2}(-ip_{+})^{s_1-k_1+k_2}\frac{-i}{\rvert p\rvert^2+i\epsilon} 
\end{align}
Explicitly, for the nonvanishing correlators in the balanced sector, we obtain in the momentum representation:
\begin{align}
\nonumber
&\langle \mathbb{A}_{s_1}(p_1)\ldots \mathbb{A}_{s_n}(p_n)\tilde{\mathbb{A}}_{s_{n+1}}(p_{n+1})\ldots \tilde{\mathbb{A}}_{s_{n+2m}}(p_{n+2m})\rangle_{conn} \\\nonumber
&=\frac{N^2-1}{2^{n+2m}}(2\pi)^4 i^{n+2m}\delta^{(4)}(p_1+\ldots+p_{n+2m})\\\nonumber
&\sum_{k_1=0}^{s_1}\ldots \sum_{k_{n+2m} = 0}^{s_{n+2m}}{s_1\choose k_1}{s_1\choose k_1}\ldots {s_{n+2m}\choose k_{n+2m}}{s_{n+2m}\choose k_{n+2m}}\frac{1}{n+2m}\sum_{\sigma\in P_{n+2m}}\\\nonumber
&\int \frac{d^4 q}{(2\pi)^4}\frac{(p_{\sigma(1)}+q)_{+}^{s_{\sigma(1)}-k_{\sigma(1)}+k_{\sigma(2)}}}{\rvert p_{\sigma(1)}+q\rvert^2}\frac{(p_{\sigma(1)}+p_{\sigma(2)}+q)_{+}^{s_{\sigma(2)}-k_{\sigma(2)}+k_{\sigma(3)}}}{\rvert p_{\sigma(1)}+p_{\sigma(2)}+q\rvert^2}\\
&\ldots\frac{(\sum_{l=1}^{n+2m-1} p_{\sigma(l)}+q)_{+}^{s_{\sigma(n+2m-1)}-k_{\sigma(n+2m-1)}+k_{\sigma(n+2m)}}}{\rvert\sum_{l=1}^{n+2m-1} p_{\sigma(l)}+q\rvert^2}\frac{(q)_{+}^{s_{\sigma(n+2m)}-k_{\sigma(n+2m)}+k_{\sigma(1)}}}{\rvert q\rvert^2}
\end{align}
In the unbalanced sector, we get:
\begin{align}
\nonumber
&\langle \mathbb{B}_{s_1}(p_1)\ldots \mathbb{B}_{s_n}(p_n)\bar{\mathbb{B}}_{s'_1}(p'_1)\ldots \bar{\mathbb{B}}_{s'_n}(p'_n)\rangle\\\nonumber
&=\frac{N^2-1}{2^{2n}}(2\pi)^4 i^{2n}\delta^{(4)}\left(\sum_{l=1}^n p_l+p'_l\right)\\\nonumber
&\sum_{k_1=0}^{s_1}\ldots \sum_{k_n = 0}^{s_n}\sum_{k'_1=0}^{s'_1}\ldots \sum_{k'_n = 0}^{s'_n}{s_1\choose k_1}{s_1\choose k_1}\ldots {s_n\choose k_n}{s_n\choose k_n}{s'_1\choose k'_1}{s'_1\choose k'_1}\ldots {s'_n\choose k'_n}{s'_n\choose k'_n}\\\nonumber
&\frac{2^{n-1}}{n}\sum_{\sigma\in P_n}\sum_{\rho\in P_n}
\int \frac{d^4 q}{(2\pi)^4}\frac{(p_{\sigma(1)}+q)_{+}^{s_{\sigma(1)}-k_{\sigma(1)}+k'_{\rho(1)}}}{\rvert p_{\sigma(1)}+q\rvert^2}\frac{(p_{\sigma(1)}+p'_{\rho(1)}+q)_{+}^{s'_{\rho(1)}-k'_{\rho(1)}+k_{\sigma(2)}}}{\rvert p_{\sigma(1)}+p'_{\rho(1)}+q\rvert^2}\\\nonumber
&\frac{(p_{\sigma(1)}+p_{\sigma(2)}+p'_{\rho(1)}+q)_{+}^{s_{\sigma(2)}-k_{\sigma(2)}+k'_{\rho(2)}}}{\rvert p_{\sigma(1)}+p_{\sigma(2)}+p'_{\rho(1)}+q\rvert^2}\frac{(p_{\sigma(1)}+p_{\sigma(2)}+p'_{\rho(1)}+p'_{\rho(2)}+q)_{+}^{s'_{\rho(2)}-k'_{\rho(2)}+k_{\sigma(3)}}}{\rvert p_{\sigma(1)}+p_{\sigma(2)}+p'_{\rho(1)}+p'_{\rho(2)}+q\rvert^2}\\
 &\ldots\frac{(\sum_{l=1}^{n-1} p_{\sigma(l)}+\sum_{l=1}^{n-2} p'_{\rho(l)} +q)_{+}^{s_{\sigma(n-1)}-k_{\sigma(n-1)}+k'_{\rho(n-1)}}}{\rvert\sum_{l=1}^{n-1} p_{\sigma(l)}+\sum_{l=1}^{n-2} p'_{\rho(l)}+q\rvert^2}\nonumber\\
 &\frac{(\sum_{l=1}^{n-1} p_{\sigma(l)}+\sum_{l=1}^{n-1} p'_{\rho(l)}+q)_{+}^{s'_{\rho(n-1)}-k'_{\rho(n-1)}+k_{\sigma(n)}}}{\rvert \sum_{l=1}^{n-1} p_{\sigma(l)}+\sum_{l=1}^{n-1} p'_{\rho(l)}+q\rvert^2}\nonumber\\
&\frac{(\sum_{l=1}^n p_{\sigma(l)}+\sum_{l=1}^{n-1} p'_{\rho(l)} +q)_{+}^{s_{\sigma(n)}-k_{\sigma(n)}+k'_{\rho(n)}}}{\rvert\sum_{l=1}^n p_{\sigma(l)}+\sum_{l=1}^{n-1} p'_{\rho(l)}+q\rvert^2}\frac{(q)_{+}^{s'_{\rho(n)}-k'_{\rho(n)}+k_{\sigma(1)}}}{\rvert q\rvert^2}
\end{align}

\subsection{Euclidean standard basis}

The Euclidean generating functionals read in the standard basis:
\begin{align}
\nonumber
\Gamma^E_{conf}[\mathbb{O}^E,\tilde{\mathbb{O}}^E]
=&- \frac{N^2-1}{2}\log \Det\Big(\delta_{s_1k_1,s_2k_2}(2\pi)^4 \delta^{(4)}(q_1-q_2) \\\nonumber
&+{\mathcal{D}_E^{-1}}_{s_1k_1,s_2k_2}(q_1)(\mathbb{O}^E_{s_2k_2}(q_1-q_2)+ \tilde{\mathbb{O}}^E_{s_2k_2}(q_1-q_2)\Big)\\\nonumber
& - \frac{N^2-1}{2}\log \Det\Big(\delta_{s_1k_1,s_2k_2} (2\pi)^4\delta^{(4)}(q_1-q_2)\\\nonumber
&+{\mathcal{D}_E^{-1}}_{s_1k_1,s_2k_2}(q_1)(\mathbb{O}^E_{s_2k_2}(q_1-q_2)- \tilde{\mathbb{O}}^E_{s_2k_2}(q_1-q_2)\Big)\\\nonumber
\Gamma^E_{conf}[\mathbb{S}^E,\bar{\mathbb{S}}^E]=&- \frac{N^2-1}{2}\log \Det\Big(\delta_{s_1k_1,s_2k_2} (2\pi)^4 \delta^{(4)}(q_1-q_2)\\
\qquad\qquad\qquad&-2\int \frac{d^4q}{(2\pi)^4}\sum_{sk}{\mathcal{D}_E^{-1}}_{s_1k_1,sk}(q_1)\bar{\mathbb{S}}^E_{sk}(q_1-q){\mathcal{D}_E^{-1}}_{sk,s_2k_2}(q)\mathbb{S}^E_{s_2k_2}(q-q_2)\Big)\\\nonumber
\end{align}
with:
\begin{align}
{\mathcal{D}^{-1}_E}_{s_1k_1,s_2k_2}(p) &=\frac{i^{s_1}}{2}\frac{\Gamma(3)\Gamma(s_1+3)}{\Gamma(5)\Gamma(s_1+1)}{s_1\choose k_1}{s_2\choose k_2+2}p_{z}^{s_1-k_1+k_2}\frac{1}{p^2} 
\end{align}
For the nonvanishing correlators on the balanced sector, we obtain:
\begin{align}
\nonumber
&\left(\frac{\Gamma(5)\Gamma(s_1+1)}{\Gamma(3)\Gamma(s_1+3)}\right)\ldots\left( \frac{\Gamma(5)\Gamma(s_{n+2m}+1)}{\Gamma(3)\Gamma(s_{n+2m}+3)}\right)\\\nonumber
&\langle \mathbb{O}^E_{s_1}(p_1)\ldots \mathbb{O}^E_{s_n}(p_n)\tilde{\mathbb{O}}^E_{s_{n+1}}(p_{n+1})\ldots \tilde{\mathbb{O}}^E_{s_{n+2m}}(p_{n+2m})\rangle_{conn} \\\nonumber
&=\frac{N^2-1}{2^{n+2m}}(2\pi)^4 i^{\sum_{l=1}^{n+2m} s_l} \delta^{(4)}(p_1+\ldots+p_{n+2m})\\\nonumber
&\sum_{k_1=0}^{s_1-2}\ldots \sum_{k_{n+2m} = 0}^{s_{n+2m}-2}{s_1\choose k_1}{s_1\choose k_1+2}\ldots {s_{n+2m}\choose k_{n+2m}}{s_{n+2m}\choose k_{n+2m}+2}\frac{1}{n+2m}\sum_{\sigma\in P_{n+2m}}\\\nonumber
&\int \frac{d^4 q}{(2\pi)^4}\frac{(p_{\sigma(1)}+q)_{z}^{s_{\sigma(1)}-k_{\sigma(1)}+k_{\sigma(2)}}}{( p_{\sigma(1)}+q)^2}\frac{(p_{\sigma(1)}+p_{\sigma(2)}+q)_{z}^{s_{\sigma(2)}-k_{\sigma(2)}+k_{\sigma(3)}}}{( p_{\sigma(1)}+p_{\sigma(2)}+q)^2}\\
&\ldots\frac{(\sum_{l=1}^{n+2m-1} p_{\sigma(l)}+q)_{z}^{s_{\sigma(n+2m-1)}-k_{\sigma(n+2m-1)}+k_{\sigma(n+2m)}}}{(\sum_{l=1}^{n+2m-1} p_{\sigma(l)}+q)^2}\frac{(q)_{z}^{s_{\sigma(n+2m)}-k_{\sigma(n+2m)}+k_{\sigma(1)}}}{(q)^2}
\end{align}
In the unbalanced sector, we get:
\begin{align}
\nonumber
&\left(\frac{\Gamma(5)\Gamma(s_1+1)}{\Gamma(3)\Gamma(s_1+3)}\right)\ldots\left(\frac{\Gamma(5)\Gamma(s_n+1)}{\Gamma(3)\Gamma(s_n+3)}\right)\left(\frac{8\pi}{N}\frac{\Gamma(5)\Gamma(s'_1+1)}{\Gamma(3)\Gamma(s'_1+3)}\right)\ldots\left( \frac{\Gamma(5)\Gamma(s'_{n}+1)}{\Gamma(3)\Gamma(s'_{n}+3)}\right)\\\nonumber
&\langle \mathbb{S}^E_{s_1}(p_1)\ldots \mathbb{S}^E_{s_n}(p_n)\bar{\mathbb{S}}^E_{s'_1}(p'_1)\ldots \bar{\mathbb{S}}^E_{s'_n}(p'_n)\rangle\\\nonumber
&=\frac{N^2-1}{2^{2n}}(2\pi)^4 i^{\sum_{l=1}^{n} s_l+s'_l}\delta^{(4)}\left(\sum_{l=1}^n p_l+p'_l\right)\\\nonumber
&\sum_{k_1=0}^{s_1-2}\ldots \sum_{k_n = 0}^{s_n-2}\sum_{k'_1=0}^{s'_1-2}\ldots \sum_{k'_n = 0}^{s'_n-2}{s_1\choose k_1}{s_1\choose k_1+2}\ldots {s_n\choose k_n}{s_n\choose k_n+2}
\\\nonumber
&{s'_1\choose k'_1}{s'_1\choose k'_1+2}\ldots {s'_n\choose k'_n}{s'_n\choose k'_n+2}\\\nonumber
&\frac{2^{n-1}}{n}\sum_{\sigma\in P_n}\sum_{\rho\in P_n}
\int \frac{d^4 q}{(2\pi)^4}\frac{(p_{\sigma(1)}+q)_{z}^{s_{\sigma(1)}-k_{\sigma(1)}+k'_{\rho(1)}}}{( p_{\sigma(1)}+q)^2}\frac{(p_{\sigma(1)}+p'_{\rho(1)}+q)_{z}^{s'_{\rho(1)}-k'_{\rho(1)}+k_{\sigma(2)}}}{( p_{\sigma(1)}+p'_{\rho(1)}+q)^2}\\\nonumber
&\frac{(p_{\sigma(1)}+p_{\sigma(2)}+p'_{\rho(1)}+q)_{z}^{s_{\sigma(2)}-k_{\sigma(2)}+k'_{\rho(2)}}}{( p_{\sigma(1)}+p_{\sigma(2)}+p'_{\rho(1)}+q)^2}\frac{(p_{\sigma(1)}+p_{\sigma(2)}+p'_{\rho(1)}+p'_{\rho(2)}+q)_{z}^{s'_{\rho(2)}-k'_{\rho(2)}+k_{\sigma(3)}}}{( p_{\sigma(1)}+p_{\sigma(2)}+p'_{\rho(1)}+p'_{\rho(2)}+q)^2}\\
 &\ldots\frac{(\sum_{l=1}^{n-1} p_{\sigma(l)}+\sum_{l=1}^{n-2} p'_{\rho(l)} +q)_{z}^{s_{\sigma(n-1)}-k_{\sigma(n-1)}+k'_{\rho(n-1)}}}{(\sum_{l=1}^{n-1} p_{\sigma(l)}+\sum_{l=1}^{n-2} p'_{\rho(l)}+q)^2}\nonumber\\
 &\frac{(\sum_{l=1}^{n-1} p_{\sigma(l)}+\sum_{l=1}^{n-1} p'_{\rho(l)}+q)_{z}^{s'_{\rho(n-1)}-k'_{\rho(n-1)}+k_{\sigma(n)}}}{( \sum_{l=1}^{n-1} p_{\sigma(l)}+\sum_{l=1}^{n-1} p'_{\rho(l)}+q)^2}\nonumber\\
&\frac{(\sum_{l=1}^n p_{\sigma(l)}+\sum_{l=1}^{n-1} p'_{\rho(l)} +q)_{z}^{s_{\sigma(n)}-k_{\sigma(n)}+k'_{\rho(n)}}}{(\sum_{l=1}^n p_{\sigma(l)}+\sum_{l=1}^{n-1} p'_{\rho(l)}+q)^2}\frac{(q)_{+}^{s'_{\rho(n)}-k'_{\rho(n)}+k_{\sigma(1)}}}{( q)^2}
\end{align}

\subsection{Euclidean extended basis}

Similarly, in the extended basis we get:
\begin{align}
\nonumber
\Gamma^E_{conf}[\mathbb{A}^E,\tilde{\mathbb{A}}^E]
=&- \frac{N^2-1}{2}\log \Det\Big(\delta_{s_1k_1,s_2k_2} (2\pi)^4\delta^{(4)}(q_1-q_2)\\\nonumber
&+{\mathcal{D}_E^{-1}}_{s_1k_1,s_2k_2}(q_1)(\mathbb{A}^E_{s_2k_2}(q_1-q_2)+ \tilde{\mathbb{A}}^E_{s_2k_2}(q_1-q_2)\Big)\\\nonumber
& - \frac{N^2-1}{2}\log \Det\Big(\delta_{s_1k_1,s_2k_2} (2\pi)^4\delta^{(4)}(q_1-q_2) \\\nonumber
&+{\mathcal{D}_E^{-1}}_{s_1k_1,s_2k_2}(q_1)(\mathbb{O}^E_{s_2k_2}(q_1-q_2)- \tilde{\mathbb{A}}^E_{s_2k_2}(q_1-q_2)\Big)\\\nonumber
\Gamma^E_{conf}[\mathbb{B}^E,\bar{\mathbb{B}}^E]=&- \frac{N^2-1}{2}\log \Det\Big(\delta_{s_1k_1,s_2k_2} (2\pi)^4 \delta^{(4)}(q_1-q_2)\\
\qquad\qquad\qquad&-2\int \frac{d^4q}{(2\pi)^4}\sum_{sk}{\mathcal{D}_E^{-1}}_{s_1k_1,sk}(q_1)\bar{\mathbb{B}}^E_{sk}(q_1-q){\mathcal{D}_E^{-1}}_{sk,s_2k_2}(q)\mathbb{B}^E_{s_2k_2}(q-q_2)\Big)\\\nonumber
\end{align}
with:
\begin{align}
{\mathcal{D}^{-1}_E}_{s_1k_1,s_2k_2}(p) &=\frac{i^{s_1}}{2}{s_1\choose k_1}{s_2\choose k_2}p_{z}^{s_1-k_1+k_2}\frac{1}{p^2} 
\end{align}
Explicitly, for the nonvanishing correlators in the balanced sector, we obtain in the momentum representation:
\begin{align}
\nonumber
&\langle \mathbb{A}^E_{s_1}(p_1)\ldots \mathbb{A}^E_{s_n}(p_n)\tilde{\mathbb{A}}^E_{s_{n+1}}(p_{n+1})\ldots \tilde{\mathbb{A}}^E_{s_{n+2m}}(p_{n+2m})\rangle_{conn} \\\nonumber
&=\frac{N^2-1}{2^{n+2m}}(2\pi)^4 i^{\sum_{l=1}^{n+2m} s_l} \delta^{(4)}(p_1+\ldots+p_{n+2m})\\\nonumber
&\sum_{k_1=0}^{s_1}\ldots \sum_{k_{n+2m} = 0}^{s_{n+2m}}{s_1\choose k_1}{s_1\choose k_1}\ldots {s_{n+2m}\choose k_{n+2m}}{s_{n+2m}\choose k_{n+2m}}\frac{1}{n+2m}\sum_{\sigma\in P_{n+2m}}\\
&\int \frac{d^4 q}{(2\pi)^4}\frac{(p_{\sigma(1)}+q)_{z}^{s_{\sigma(1)}-k_{\sigma(1)}+k_{\sigma(2)}}}{( p_{\sigma(1)}+q)^2}\frac{(p_{\sigma(1)}+p_{\sigma(2)}+q)_{z}^{s_{\sigma(2)}-k_{\sigma(2)}+k_{\sigma(3)}}}{( p_{\sigma(1)}+p_{\sigma(2)}+q)^2}\nonumber\\
&\ldots\frac{(\sum_{l=1}^{n+2m-1} p_{\sigma(l)}+q)_{z}^{s_{\sigma(n+2m-1)}-k_{\sigma(n+2m-1)}+k_{\sigma(n+2m)}}}{(\sum_{l=1}^{n+2m-1} p_{\sigma(l)}+q)^2}\frac{(q)_{z}^{s_{\sigma(n+2m)}-k_{\sigma(n+2m)}+k_{\sigma(1)}}}{(q)^2}
\end{align}
In the unbalanced sector, we get:
\begin{align}
\nonumber
&\langle \mathbb{B}^E_{s_1}(p_1)\ldots \mathbb{B}^E_{s_n}(p_n)\bar{\mathbb{B}}^E_{s'_1}(p'_1)\ldots \bar{\mathbb{B}}^E_{s'_n}(p'_n)\rangle\\\nonumber
&=\frac{N^2-1}{2^{2n}}(2\pi)^4 i^{\sum_{l=1}^{n} s_l+s'_l}\delta^{(4)}\left(\sum_{l=1}^n p_l+p'_l\right)\\\nonumber
&\sum_{k_1=0}^{s_1}\ldots \sum_{k_n = 0}^{s_n}\sum_{k'_1=0}^{s'_1}\ldots \sum_{k'_n = 0}^{s'_n}{s_1\choose k_1}{s_1\choose k_1}\ldots {s_n\choose k_n}{s_n\choose k_n}{s'_1\choose k'_1}{s'_1\choose k'_1}\ldots {s'_n\choose k'_n}{s'_n\choose k'_n}\\\nonumber
&\frac{2^{n-1}}{n}\sum_{\sigma\in P_n}\sum_{\rho\in P_n}
\int \frac{d^4 q}{(2\pi)^4}\frac{(p_{\sigma(1)}+q)_{z}^{s_{\sigma(1)}-k_{\sigma(1)}+k'_{\rho(1)}}}{( p_{\sigma(1)}+q)^2}\frac{(p_{\sigma(1)}+p'_{\rho(1)}+q)_{z}^{s'_{\rho(1)}-k'_{\rho(1)}+k_{\sigma(2)}}}{( p_{\sigma(1)}+p'_{\rho(1)}+q)^2}\\\nonumber
&\frac{(p_{\sigma(1)}+p_{\sigma(2)}+p'_{\rho(1)}+q)_{z}^{s_{\sigma(2)}-k_{\sigma(2)}+k'_{\rho(2)}}}{( p_{\sigma(1)}+p_{\sigma(2)}+p'_{\rho(1)}+q)^2}\frac{(p_{\sigma(1)}+p_{\sigma(2)}+p'_{\rho(1)}+p'_{\rho(2)}+q)_{z}^{s'_{\rho(2)}-k'_{\rho(2)}+k_{\sigma(3)}}}{( p_{\sigma(1)}+p_{\sigma(2)}+p'_{\rho(1)}+p'_{\rho(2)}+q)^2}\\
 &\ldots\frac{(\sum_{l=1}^{n-1} p_{\sigma(l)}+\sum_{l=1}^{n-2} p'_{\rho(l)} +q)_{z}^{s_{\sigma(n-1)}-k_{\sigma(n-1)}+k'_{\rho(n-1)}}}{(\sum_{l=1}^{n-1} p_{\sigma(l)}+\sum_{l=1}^{n-2} p'_{\rho(l)}+q)^2}\nonumber\\
 &\frac{(\sum_{l=1}^{n-1} p_{\sigma(l)}+\sum_{l=1}^{n-1} p'_{\rho(l)}+q)_{z}^{s'_{\rho(n-1)}-k'_{\rho(n-1)}+k_{\sigma(n)}}}{( \sum_{l=1}^{n-1} p_{\sigma(l)}+\sum_{l=1}^{n-1} p'_{\rho(l)}+q)^2}\nonumber\\
&\frac{(\sum_{l=1}^n p_{\sigma(l)}+\sum_{l=1}^{n-1} p'_{\rho(l)} +q)_{z}^{s_{\sigma(n)}-k_{\sigma(n)}+k'_{\rho(n)}}}{(\sum_{l=1}^n p_{\sigma(l)}+\sum_{l=1}^{n-1} p'_{\rho(l)}+q)^2}\frac{(q)_{+}^{s'_{\rho(n)}-k'_{\rho(n)}+k_{\sigma(1)}}}{( q)^2}
\end{align}

\appendix

\section{Notation and Wick rotation \label{appN}}

We mostly follow the notation in \cite{Braun:2003rp}. We define the Minkowskian metric as:
\begin{equation}
(g_{\mu\nu}) = \text{diag}(1,-1,-1,-1)
\end{equation}
The light-cone coordinates are:
\begin{equation}
x^{\pm} = \frac{x^0\pm x^3}{\sqrt{2}}=x_{\mp}
\end{equation}
The corresponding Minkowskian (squared) distance is:
\begin{equation}
\label{mod2}
\rvert x \rvert^2 = 2 x^+ x^- -x_{\perp}^2
\end{equation}
where:
\begin{equation}
x^2_\perp=(x^1)^2+(x^2)^2
\end{equation}
We denote the derivative with respect to $x^+$ by:
\begin{equation}
\partial_+ = \frac{\partial}{\partial x^+} = \partial_{x^+} =\frac{\partial}{\partial x_-} = \partial_{x_-}
\end{equation}
We define the light-like vectors $n^\mu$ and $\bar{n}^\mu$:
\begin{equation}
n_\mu n^\mu = \bar{n}_\mu \bar{n}^\mu = 0 \qquad n_\mu \bar{n}^\mu = 1
\end{equation}
that can be parametrized as $(n^{\mu}) = \frac{1}{\sqrt{2}}(1,0,0,1)$ and $(\bar{n}^{\mu}) = \frac{1}{\sqrt{2}}(1,0,0,-1)$.\par
The Minkowskian metric can be decomposed into orthogonal and longitudinal parts with respect to the light-like vectors:
\begin{equation}
g_{\mu\nu} =g^\perp_{\mu\nu}+n_\mu \bar{n}_\nu +n_\nu \bar{n}_\mu 
\end{equation} 
The Euclidean metric is:
\begin{equation}
(\delta_{\mu\nu}) =  \text{diag}(1,1,1,1)
\end{equation}
The corresponding Euclidean (squared) distance is:
\begin{equation}
x^2 = 2x^z x^{\bar{z}}+x_\perp^2
\end{equation}
with:
\begin{equation}
x^z= \frac{x^4+ix^3}{\sqrt{2}}=\frac{x_4+ix_3}{\sqrt{2}}=x_{\bar z}
\end{equation}
and:
\begin{equation}
x^{\bar z}=  \frac{x^4-ix^3}{\sqrt{2}}=\frac{x_4-ix_3}{\sqrt{2}}=x_z
\end{equation}
We define the Wick rotation by:
\begin{align}
\label{wick1}
&x^0=x_0 \rightarrow  - i x^4=-i x_4 
\end{align} 
and:
\begin{align}
\label{wick2}
&p_0=p^0 \rightarrow  i p_4= i p^4
\end{align} 
Eq. (\ref{wick1}) ensures that $\exp(iS_M)\rightarrow \exp(-S_E)$, where $S_M$ and $S_E$ are the Minkowskian and Euclidean actions respectively, with $S_E$ positive definite. \par
By defining $ p \cdot x = p_{\mu} x^{\mu}$ and $\langle p x \rangle= p_{\mu} x^{\mu}$ in Minkowskian and Euclidean space-time respectively, eq. (\ref{wick2}) ensures that, by the Wick rotation, $ p \cdot x  \rightarrow \langle p x \rangle$, in such a way that the pairings $p \cdot x$ and $\langle p x \rangle$ are actually independent of the Minkowskian and Euclidean metric respectively. \par
Therefore, by a slight abuse of notation, we also write $ p \cdot x$ in Euclidean space-time, instead of $\langle p x \rangle$.  
Besides, $|x|^2 \rightarrow - x^2$ and  $|p|^2 \rightarrow - p^2$.\par
As a consequence, the Wick rotation of the scalar propagator of mass $m$ in Minkowskian space-time:
\begin{equation}
\langle \phi(x) \phi(y)\rangle = \int \frac{d^4p}{(2\pi)^4}\, e^{i  p \cdot (x-y)}\, \frac{i}{|p|^2-m^2+i \epsilon}
\end{equation}
reads in Euclidean space-time:
\begin{equation}
\langle \phi^E(x) \phi^E(y)\rangle = \int \frac{d^4p}{(2\pi)^4}\, e^{i p \cdot(x-y)}\, \frac{i^2}{- p^2-m^2}= \int \frac{d^4p}{(2\pi)^4}\, e^{i p \cdot(x-y)}\, \frac{1}{p^2+m^2}
\end{equation}
as it should be.
Moreover, the Wick rotation of the light-cone coordinates is:
\begin{align}
&x^+=x_-\rightarrow -ix^z= -i x_{\bar z} \qquad 
\end{align}
and:
\begin{align}
&x^-=x_+\rightarrow -i x^{\bar{z}}= -i x_{z}\qquad
\end{align}
Correspondingly, the Wick rotation of the derivative with respect to $x^+$ is:
\begin{align}
\label{wickderivative}
\partial_+\rightarrow i \partial_{z} = i\frac{\partial}{\partial x^z}
\end{align}

\section{Minkowskian and Euclidean propagators\label{appA1}}

The gluon propagator in the light-cone gauge, $n\cdot A=A_+=0$, is:
\begin{equation}
\langle A^a_{\mu}(x)A^b_{\nu}(y)\rangle = \int \frac{d^4p}{(2\pi)^4}\, e^{ip\cdot(x-y)}\, \frac{-i\,\delta^{ab}}{\rvert p\rvert^2+i\epsilon}\left(g_{\mu\nu}-\frac{n_\mu p_\nu+n_\nu p_\mu}{p\cdot n}\right)
\end{equation}
and in the Feynman gauge:
\begin{equation}
\langle A^a_{\mu}(x)A^b_{\nu}(y)\rangle = \int \frac{d^4p}{(2\pi)^4}\, e^{ip\cdot(x-y)}\, \frac{-i\,\delta^{ab}}{\rvert p\rvert^2+i\epsilon}\,g_{\mu\nu}= \frac{\delta^{ab}}{4\pi^2}\frac{g_{\mu\nu}}{\rvert x-y\rvert^2-i\epsilon}
\end{equation}
Hence, in the light-cone gauge the transverse propagator is:
 \begin{align}
\label{axialprop}
&\langle A^a(x){A}^b(y)\rangle  = 0\\\nonumber
&\langle \bar{A}^a(x)\bar{A}^b(y)\rangle  = 0\\\nonumber
&\langle A^a(x)\bar{A}^b(y)\rangle  = -\frac{\delta^{ab}}{4\pi^2}\frac{1}{\rvert x-y\rvert^2-i\epsilon}
\end{align}
We employ eq. (\ref{fdef}) to work out the $2$-point correlators in the light-cone gauge:
 \begin{align}
\label{axialprop2}
&\langle f_{11}^a(x)f_{11}^b(y)\rangle  = 0\\\nonumber
&\langle f_{\dot{1}\dot{1}}^a(x) f_{\dot{1}\dot{1}}^b(y)\rangle  = 0\\\nonumber
&\langle f_{11}^a(x) f_{\dot{1}\dot{1}}^b(y)\rangle  = -\frac{\delta^{ab}}{4\pi^2}\partial_{x^+}\partial_{y^+}\frac{1}{\rvert x-y\rvert^2-i\epsilon}
\end{align}
and:
 \begin{align}
\label{axialpropeuc2}
&\langle\partial_{+}^{-1} f_{11}^{a}(x)\partial_{+}^{-1}f_{11}^{b}(y)\rangle  = 0\\\nonumber
&\langle \partial_{+}^{-1} f_{\dot{1}\dot{1}}^{a}(x)\partial_{+}^{-1} f_{\dot{1}\dot{1}}^{b}(y)\rangle  = 0\\\nonumber
&\langle \partial_{+}^{-1}f_{11}^{a}(x)\partial_{+}^{-1} f_{\dot{1}\dot{1}}^{b}(y)\rangle  = -\frac{\delta^{ab}}{4\pi^2}\frac{1}{\rvert x-y\rvert^2-i\epsilon}
\end{align}
The Euclidean propagator in the Feynman gauge follows from the Wick rotation (appendix \ref{appN}):
\begin{equation}
\langle A^{Ea}_{\mu}(x)A^{Eb}_{\nu}(y)\rangle = \int \frac{d^4p}{(2\pi)^4}\, e^{ip\cdot(x-y)}\, \frac{\delta^{ab}}{p^2}\,\delta_{\mu\nu}= \frac{\delta^{ab}}{4\pi^2}\frac{\delta_{\mu\nu}}{(x-y)^2}\,
\end{equation}
Moreover, by performing the Wick rotation of eq. (\ref{axialprop2}) and (\ref{axialpropeuc2}), we obtain in Euclidean space-time:
 \begin{align}
\label{axialpropeucl}
&\langle f_{11}^{Ea}(x)f_{11}^{Eb}(y)\rangle  = 0\\\nonumber
&\langle f_{\dot{1}\dot{1}}^{Ea}(x) f_{\dot{1}\dot{1}}^{Eb}(y)\rangle  = 0\\\nonumber
&\langle f_{11}^{Ea}(x) f_{\dot{1}\dot{1}}^{Eb}(y)\rangle  = -\frac{\delta^{ab}}{4\pi^2}\partial_{x^z}\partial_{y^z}\frac{1}{(x-y)^2}
\end{align}
and:
 \begin{align}
\label{axialpropeuc3}
&\langle\partial_{z}^{-1} f_{11}^{Ea}(x)\partial_{z}^{-1}f_{11}^{Eb}(y)\rangle  = 0\\\nonumber
&\langle \partial_{z}^{-1} f_{\dot{1}\dot{1}}^{Ea}(x)\partial_{z}^{-1} f_{\dot{1}\dot{1}}^{Eb}(y)\rangle  = 0\\\nonumber
&\langle \partial_{z}^{-1}f_{11}^{Ea}(x)\partial_{z}^{-1} f_{\dot{1}\dot{1}}^{Eb}(y)\rangle  = -\frac{\delta^{ab}}{4\pi^2}\frac{1}{(x-y)^2}
\end{align}

\section{Identities involving $\sigma^{\mu}$ and $\bar{\sigma}^{\mu}$} \label{appA200}

We define the matrix $(\sigma^{\mu}_{a\dot{a}})$:
\begin{equation}
(\sigma^{\mu}) = \left(1,\vec{\sigma}\right)
\end{equation} 
by means of the Pauli matrices that satisfy:
\begin{align} 
&\left[\sigma^i,\sigma^j\right] = 2i \epsilon^{ijk}\sigma^k \\\nonumber
&\{\sigma^i,\sigma^j\} = 2 \delta^{ij}\mathbb{I}
\end{align}
We also define:
\begin{equation}
(\bar{\sigma}^{\mu}) = (1,-\vec{\sigma})
\end{equation}
and:
\begin{equation}
\sigma^+ = \frac{1+\sigma_3}{\sqrt{2} } \quad{\sigma}^- = \frac{1-\sigma_3}{\sqrt{2}} \qquad\qquad(\sigma_\perp^{\mu}) = (\sigma^1,\sigma^2) \quad(\bar{\sigma}_\perp^{\mu}) = (-\sigma^1,-\sigma^2)
\end{equation}
By means of $(\sigma^{\mu}_{a\dot{a}})$ we may represent a vector, $V_{\mu}$, in matrix form:
\begin{equation}
\label{spinorvector}
\mathbb{V} = V_\mu \sigma^\mu = \sqrt{2}\begin{pmatrix}V_+ & \bar{V}\\ V & V_- \end{pmatrix}
\end{equation}
with:
\begin{align}
\nonumber
&V_+ = \frac{V_0+V_3}{\sqrt{2}} \qquad V_- = \frac{V_0-V_3}{\sqrt{2}}\\
&V = \frac{V_1+iV_2}{\sqrt{2}} \qquad\bar{V} = \frac{V_1-iV_2}{\sqrt{2}}
\end{align}
in such a way that:
\begin{equation}
\Det(\mathbb{V}_{a\dot{a}}) = 2(V_+V_--V\bar{V}) = V^\mu V_\mu 
\end{equation}
Hence, the Lorentz group is embedded into $SL(2,\mathbb{C})$, and a Lorentz transformation acts as:
\begin{equation}
\mathbb{V}' = L\mathbb{V}\bar{L}
\end{equation}
with $L\in SL(2,\mathbb{C})$, leaving the determinant invariant.\par
We introduce the antisymmetric symbols $\epsilon_{ab}, \epsilon^{ab}$ \cite{Dreiner:2008tw}:
\begin{equation}
	\epsilon^{12}=-\epsilon^{21}=\epsilon_{21}=-\epsilon_{12}=1
\end{equation}
with:
\begin{align}
&\epsilon_{ac}\epsilon^{cb} = \delta_a^{\,\,b}\nonumber\\
&\epsilon^{ab} = \epsilon^{ac}\epsilon_{cd}\epsilon^{db}
\end{align}
that are employed to lower and rise the spinor indices respectively. For example: 
\begin{equation}
\psi_a  =\epsilon_{ab}\psi^b\qquad \Phi^{ab} = \epsilon^{ac}\epsilon^{bd}\Phi_{cd}
\end{equation}
The following identities \cite{Dreiner:2008tw} hold:
\begin{align}
&\bar{\sigma}^{\mu\,\dot{a}a} = \epsilon^{ab}\epsilon^{\dot{a}\dot{b}}\sigma_{b\dot{b}}^\mu\\\nonumber
&\sigma^\mu_{a\dot{a}}\bar{\sigma}_\mu^{\dot{b}b} =2  \delta^{\,\,b}_a  \delta^{\dot{b}}_{\,\,\dot{a}}   \\\nonumber
&\sigma^\mu_{a\dot{a}}{\sigma}_{\mu\,b\dot{b}} = 2 \epsilon_{ab}\epsilon_{\dot{a}\dot{b}}\\\nonumber
&\bar{\sigma}^{\mu\,\dot{a}a}\bar{\sigma}_\mu^{\dot{b}b} = 2 \epsilon^{ab}\epsilon^{\dot{a}\dot{b}}
\end{align}
Besides, we define \cite{Dreiner:2008tw}:
\begin{align}
\label{sigmamunu}
&(\sigma^{\mu\nu})_{a}^{\,\, b} =\frac{i}{4}\left(\sigma^\mu_{a\dot{c}}\bar{\sigma}^{\nu\,\dot{c}b}-\sigma^\nu_{a\dot{c}}\bar{\sigma}^{\mu\,\dot{c}b}\right)\\\nonumber
&(\bar{\sigma}^{\mu\nu})^{\dot{a}}_{\,\,\dot{b}} =\frac{i}{4}\left(\bar{\sigma}^{\mu\,\dot{a}c}\sigma^\nu_{c\dot{b}}-\bar{\sigma}^{\nu\,\dot{a}c}\sigma^\mu_{c\dot{b}}\right)
\end{align}
with vanishing traces:
\begin{align}
&\Tr \sigma^{\mu\nu} = \epsilon_{ab}(\sigma^{\mu\nu})^{ab} = 0\\\nonumber
&\Tr \bar{\sigma}^{\mu\nu}=\epsilon_{\dot{a}\dot{b}}(\bar{\sigma}^{\mu\nu})^{\dot{a}\dot{b}} = 0
\end{align}
$\sigma_{\mu\nu}$ and $\bar{\sigma}_{\mu\nu}$ satisfy the duality relations:
\begin{align}
&\sigma^{\mu\nu} = -\frac{i}{2}\epsilon^{\mu\nu\rho\sigma}\sigma_{\rho\sigma}\\\nonumber
&\bar{\sigma}^{\mu\nu} = \frac{i}{2}\epsilon^{\mu\nu\rho\sigma}\bar{\sigma}_{\rho\sigma}
\end{align}
where $\epsilon^{0123}=1$. Moreover, the following identities \cite{Dreiner:2008tw} hold:
\begin{align}
&(\sigma^{\mu\nu})^{\,\,a}_b(\sigma_{\mu\nu})^{\,\,c}_d = 2\delta^{\,\,c}_b\delta^{\,\,a}_d-\delta^{\,\,a}_b\delta^{\,\,c}_d\\\nonumber
&(\bar{\sigma}^{\mu\nu})^{\dot{a}}_{\,\,\dot{b}}(\bar{\sigma}_{\mu\nu})^{\dot{c}}_{\,\,\dot{d}} =2\delta^{\dot{c}}_{\,\,\dot{b}} \delta^{\dot{a}}_{\,\,\dot{d}}-\delta^{\dot{a}}_{\,\,\dot{b}}\delta^{\dot{c}}_{\,\,\dot{d}}\\\nonumber
&(\sigma^{\mu\nu})^{\,\,a}_b(\bar{\sigma}_{\mu\nu})^{\dot{c}}_{\,\,\dot{d}} = 0
\end{align}

\section{Relation between the spinorial and vectorial bases in Minkowskian space-time} \label{appA2}

The components of $F_{\mu\nu}$, with their $s, j, \tau$ assignments, are:
\begin{align}
&F_{+\mu}  =  F^{\alpha\beta} {n}_{\alpha}g^{\perp}_{\beta\mu} \qquad \mu,\nu = 1,2 \qquad s = 1\,\, j= \frac{3}{2} \,\, \tau=1\\\nonumber
&F_{-\mu} =  F^{\alpha\beta} \bar{n}_{\alpha}g^{\perp}_{\beta\mu}\qquad \mu, \nu = 1,2\qquad  s = -1 \,\, j=\frac{1}{2} \,\,\tau=3 \\\nonumber
&F_{\mu\nu} = F^{\alpha\beta} g^{\perp}_{\alpha\mu}g^{\perp}_{\beta\nu} \qquad \mu,\nu = 1,2\qquad s = 0 \,\, j=1 \,\, \tau=2\\\nonumber
&F_{+-} = F^{\alpha\beta} n_\alpha\bar{n}_\beta \hspace{3.35cm} s = 0 \,\, j=1 \,\, \tau=2\\\nonumber
\end{align}
The component with maximal $s$, $F_{+\mu}$, is well suited (section \ref{00}) to build twist-$2$ operators that are primary \cite{conformalops,Ohrndorf:1981qv} for the collinear conformal subgroup. 
In the light-cone gauge:
\begin{equation}
F_{+ \mu} = \partial_+ A_\mu
\end{equation}
with $\mu=1,2$. Similarly, twist-$2$ primary conformal operators can also be built by means of $\tilde{F}_{+\mu}$:
\begin{align}
\tilde{F}_{+\mu} &= \tilde{F}^{\alpha\beta}{n}_\alpha g^{\perp}_{\beta\mu}=\frac{1}{2}\epsilon^{\alpha\beta\rho\sigma}n_{\alpha}g^{\perp}_{\beta\mu}F_{\rho\sigma} = \epsilon^{-\beta+\sigma}g^{\perp}_{\beta\mu}F_{+\sigma}\nonumber\\
& = \epsilon^{-\beta+\sigma}g^{\perp}_{\beta\mu}g^{\perp}_{\sigma\nu}F_{+}^{\,\nu} = - \epsilon^{\sigma\beta+-}g^{\perp}_{\beta\mu}g^{\perp}_{\sigma\nu}F_{+}^{\,\,\nu} = \epsilon^{\beta\sigma+-}g^{\perp}_{\beta\mu}g^{\perp}_{\sigma\nu}F_{+}^{\,\,\nu} 
\end{align}
with $s = 1$, $ j= \frac{3}{2}$ and $\tau=1$, where:
\begin{equation}
\tilde{F}^{\mu\nu} = \frac{1}{2}\epsilon^{\mu\nu\rho\sigma}F_{\rho\sigma}
\end{equation}
We define:
\begin{equation}
\epsilon_{\mu\nu} = \epsilon^{\alpha\beta\rho\sigma}g^\perp_{\alpha\mu}g^\perp_{\beta\nu}\bar{n}_{\rho}{n}_{\sigma} = \epsilon^{\alpha\beta+-}g^\perp_{\alpha\mu}g^\perp_{\beta\nu}
\end{equation}
with $\mu,\nu=1,2$ and $\epsilon_{12}=1$. Hence:
\begin{equation}
\tilde{F}_{+\mu} =\epsilon_{\mu\nu}F_{+}^{\,\,\nu}
\end{equation}
and:
\begin{equation}
\tilde{F}_{+1}= -F_{+2}\qquad \tilde{F}_{+2}= F_{+1}
\end{equation}
In the spinorial representation \cite{Braun:2008ia}:
\begin{equation}
\label{adota}
F_{a\dot{a}b\dot{b}} = \sigma_{a\dot{a}}^\mu\sigma_{b\dot{b}}^\nu F_{\mu\nu}
\end{equation}
It turns out that $F_{\mu\nu}$ decomposes \cite{Braun:2008ia} into the $(1,0)\oplus(0,1)$ representation of the Lorentz group:
\begin{equation}
\label{lorentza}
F_{a\dot{a}b\dot{b}} = 2(f_{ab}\epsilon_{\dot{a}\dot{b}} - \epsilon_{ab} f_{\dot{a}\dot{b}})
\end{equation}
where \footnote{We write eqs. \eqref{asd} and \eqref{sd} in the notation of appendix \ref{appA200}, as opposed to the one in \cite{Braun:2008ia}.}:
\begin{equation} \label{asd}
f_{ab} = \frac{i}{2}(\sigma^{\mu\nu})_{ab}F_{\mu\nu}
\end{equation}
and:
\begin{equation} \label{sd}
f_{\dot{a}\dot{b}} = -\frac{i}{2}(\bar{\sigma}^{\mu\nu})_{\dot{a}\dot{b}}F_{\mu\nu}
\end{equation}
with:
\begin{equation}
\label{dagger}
\bar{f}_{ab} = f_{\dot{a}\dot{b}}
\end{equation}
Indeed, since $\epsilon^{\dot{a}\dot{b}}f_{\dot{a}\dot{b}} = 0$ and $\epsilon_{\dot{a}\dot{b}} \epsilon_{\dot{a}\dot{b}} = -2$, we get from eqs. (\ref{adota}) and (\ref{lorentza}):
\begin{equation}
\epsilon^{\dot{a}\dot{b}}\sigma_{a\dot{a}}^\mu\sigma_{b\dot{b}}^\nu F_{\mu\nu} = 2 f_{ab}\epsilon^{\dot{a}\dot{b}}\epsilon_{\dot{a}\dot{b}}
\end{equation} 
that coincides with eq. \eqref{asd} by the antisymmetry of $F_{\mu\nu}$ and the definition of $\sigma^{\mu\nu}_{ab}$ in eq. (\ref{sigmamunu}). Similarly, we obtain eq. \eqref{sd}.
It follows that:
\begin{equation}
f_{11} =\frac{i}{2}(\sigma^{\mu\nu})_{11}F_{\mu\nu}
\end{equation}
where:
\begin{align}
(\sigma^{\mu\nu})_{11}&= \frac{i}{4}\epsilon^{\dot{c}\dot{d}}(\sigma^\mu_{1\dot{c}}\sigma^{\nu}_{1\dot{d}}-\sigma^\nu_{1\dot{c}}\sigma^{\nu}_{1\dot{d}})\nonumber\\
&=\frac{i}{4}(\sigma^\mu_{1\dot{1}}\sigma^{\nu}_{1\dot{2}}-\sigma^\mu_{1\dot{2}}\sigma^{\nu}_{1\dot{1}}-\sigma^\nu_{1\dot{1}}\sigma^{\mu}_{1\dot{2}}+\sigma^\nu_{1\dot{2}}\sigma^{\mu}_{1\dot{1}}) \nonumber \\
&=\frac{i}{2}(\sigma^\mu_{1\dot{1}}\sigma^{\nu}_{1\dot{2}}-\sigma^\mu_{1\dot{2}}\sigma^{\nu}_{1\dot{1}})
\end{align}
From the definition of the matrices $(\sigma^{\mu}_{a\dot{a}})$ (appendix \ref{appA200}):
\begin{align}
&(\sigma^{+}_{a\dot{a}})=\frac{1}{\sqrt{2}}\begin{pmatrix}2 & 0\\ 0 & 0 \end{pmatrix} = \frac{2}{\sqrt{2}}\left(\delta_{a1}\delta_{\dot{a}\dot{1}}\right)\nonumber\\
&(\sigma^{-}_{a\dot{a}})=\frac{1}{\sqrt{2}}\begin{pmatrix}0 & 0\\ 0 & 2 \end{pmatrix} = \frac{2}{\sqrt{2}}\left(\delta_{a2}\delta_{\dot{a}\dot{2}}\right)\nonumber\\
&(\sigma^{1}_{a\dot{a}})=\begin{pmatrix}0 & 1\\ 1 & 0 \end{pmatrix}= \left(\delta_{a1}\delta_{\dot{a}\dot{2}}+\delta_{a2}\delta_{\dot{a}\dot{1}}\right)\nonumber\\
&(\sigma^{2}_{a\dot{a}})=\begin{pmatrix}0 & -i\\ i & 0 \end{pmatrix}= i\left(-\delta_{a1}\delta_{\dot{a}\dot{2}}+\delta_{a2}\delta_{\dot{a}\dot{1}}\right)
\end{align}
it follows that $\sigma^{\mu}_{1\dot{1}}$ is nonvanishing only for $\mu=+$, and $\sigma^{\mu}_{1\dot{2}}$ is nonvanishing only for $\mu=1,2$.
Hence, by employing the antisymmetry of $F_{\mu\nu}$, we obtain:
\begin{equation}
f_{11} =\frac{i}{2}(2\frac{i}{2}\sigma^{+}_{1\dot{1}}\sigma^{\nu}_{1\dot{2}})F_{+\nu}=\frac{i}{2}2\frac{i}{2}\frac{2}{\sqrt{2}} (\sigma^{1}_{1\dot{2}}F_{+1}+\sigma^{2}_{1\dot{2}}F_{+2})  
\end{equation}
Therefore:
\begin{align} \label{ff}
\nonumber
&f_{11}= -\frac{1}{\sqrt{2}}\left(F_{+1}-i F_{+2}\right) \\
&f_{\dot{1}\dot{1}} =-\frac{1}{\sqrt{2}}\left(F_{+1}+i F_{+2}\right) 
\end{align}
We can now build the dictionary from the spinorial to the vectorial basis of the twist-$2$ operators:
\begin{align}
f_{11} f_{\dot{1}\dot{1}} &= \frac{1}{2}\left(F_{+1}-i F_{+2}\right)\left(F_{+1}+i F_{+2}\right)\\\nonumber
&=\frac{1}{2}\Big(F_{+1}F_{+1}+F_{+2}F_{+2}-i(F_{+1}F_{+2}-F_{+2}F_{+1})\Big)\\\nonumber
&=-\frac{1}{2}\Big(g_{\perp}^{\mu\nu}F_{+\mu}F_{+\nu}+i\epsilon^{\mu\nu}F_{+\mu}F_{+\nu}\Big)
\end{align}
and:
\begin{align}
\label{unba}
f_{11}f_{11} &=  \frac{1}{2}\left(F_{+1}-i F_{+2}\right)\left(F_{+1}-i F_{+2}\right)\\\nonumber
&=\frac{1}{2}\left(F_{+1}+i \tilde{F}_{+1}\right)\left(F_{+1}+i \tilde{F}_{+1}\right)
\end{align}
In principle, the unbalanced operators with $\tau=2$ in the vectorial basis should be constructed by means of the tensor:
\begin{equation}
\frac{1}{2} \left(F_{+\mu}+i \tilde{F}_{+\mu}\right)\left(F_{+\nu}+i \tilde{F}_{+\nu}\right)
\end{equation}
and its Hermitian conjugate, with $\mu,\nu=1,2$. However, a simple computation shows that all the components of the operators above are actually proportional to $f_{11}f_{11}$ and its Hermitian conjugate respectively. Indeed:
\begin{align}
&\frac{1}{2}\left(F_{+2}+i \tilde{F}_{+2}\right)\left(F_{+2}+i \tilde{F}_{+2}\right)=\frac{1}{2}\left(F_{+2}+i {F}_{+1}\right)\left(F_{+2}+i {F}_{+1}\right)\\\nonumber
&=-\frac{1}{2}\left(F_{+1}-i {F}_{+2}\right)\left(F_{+1}-i {F}_{+2}\right) =-f_{11}f_{11}
\end{align}
and:
\begin{align}
\frac{1}{2}\left(F_{+1}+i \tilde{F}_{+1}\right)\left(F_{+2}+i \tilde{F}_{+2}\right)=\frac{i}{2}\left(F_{+1}-i {F}_{+2}\right)\left(F_{+1}-i {F}_{+2}\right)=if_{11}f_{11}
\end{align}
It follows that in the standard basis:
\begin{align}
\nonumber
&\mathbb{O}_{s} = -\frac{1}{2}g^{\mu\nu}_{\perp}\Tr F_{+\mu}(x)(i\overrightarrow{D}_++i\overleftarrow{D}_+)^{s-2}C^{\frac{5}{2}}_{s-2}\left(\frac{\overrightarrow{D}_+-\overleftarrow{D}_+}{\overrightarrow{D}_++\overleftarrow{D}_+}\right)F_{+\nu}(x) \\\nonumber
&\tilde{\mathbb{O}}_{s} = -\frac{i}{2}\epsilon^{\mu\nu}\Tr F_{+\mu}(x)(i\overrightarrow{D}_++i\overleftarrow{D}_+)^{s-2}C^{\frac{5}{2}}_{s-2}\left(\frac{\overrightarrow{D}_+-\overleftarrow{D}_+}{\overrightarrow{D}_++\overleftarrow{D}_+}\right)F_{+\nu}(x)\\\nonumber
&\mathbb{S}_{s} =\frac{1}{2\sqrt{2}}\Tr\left(F_{+1}+i \tilde{F}_{+1}\right)(i\overrightarrow{D}_++i\overleftarrow{D}_+)^{s-2}C^{\frac{5}{2}}_{s-2}\left(\frac{\overrightarrow{D}_+-\overleftarrow{D}_+}{\overrightarrow{D}_++\overleftarrow{D}_+}\right)\left(F_{+1}+i \tilde{F}_{+1}\right)\\
&\bar{\mathbb{S}}_{s} =\frac{1}{2\sqrt{2}}\Tr\left(F_{+1}-i \tilde{F}_{+1}\right)(i\overrightarrow{D}_++i\overleftarrow{D}_+)^{s-2}C^{\frac{5}{2}}_{s-2}\left(\frac{\overrightarrow{D}_+-\overleftarrow{D}_+}{\overrightarrow{D}_++\overleftarrow{D}_+}\right)\left(F_{+1}-i \tilde{F}_{+1}\right) 
\end{align}
and in the extended basis:
\begin{align}
\nonumber
&\mathbb{A}_{s} = -\frac{1}{2}g^{\mu\nu}_{\perp}\Tr D_{+}^{-1}F_{+\mu}(x)(i\overrightarrow{D}_++i\overleftarrow{D}_+)^{s}C^{\frac{1}{2}}_{s}\left(\frac{\overrightarrow{D}_+-\overleftarrow{D}_+}{\overrightarrow{D}_++\overleftarrow{D}_+}\right)D_{+}^{-1}F_{+\nu}(x) \\\nonumber
&\tilde{\mathbb{A}}_{s} = -\frac{i}{2}\epsilon^{\mu\nu}\Tr D_{+}^{-1}F_{+\mu}(x)(i\overrightarrow{D}_++i\overleftarrow{D}_+)^{s}C^{\frac{1}{2}}_{s}\left(\frac{\overrightarrow{D}_+-\overleftarrow{D}_+}{\overrightarrow{D}_++\overleftarrow{D}_+}\right)D_{+}^{-1}F_{+\nu}(x) \\\nonumber
&\mathbb{B}_{s} =\frac{1}{2\sqrt{2}}\Tr D_{+}^{-1}\left(F_{+1}+i \tilde{F}_{+1}\right)(i\overrightarrow{D}_++i\overleftarrow{D}_+)^{s}C^{\frac{1}{2}}_{s}\left(\frac{\overrightarrow{D}_+-\overleftarrow{D}_+}{\overrightarrow{D}_++\overleftarrow{D}_+}\right)D_{+}^{-1}\left(F_{+1}+i \tilde{F}_{+1}\right)\\
&\bar{\mathbb{B}}_{s} =\frac{1}{2\sqrt{2}}\Tr D_{+}^{-1}\left(F_{+1}-i \tilde{F}_{+1}\right)(i\overrightarrow{D}_++i\overleftarrow{D}_+)^{s}C^{\frac{1}{2}}_{s}\left(\frac{\overrightarrow{D}_+-\overleftarrow{D}_+}{\overrightarrow{D}_++\overleftarrow{D}_+}\right)D_{+}^{-1}\left(F_{+1}-i \tilde{F}_{+1}\right)
\end{align}

\section{Complex basis} \label{appC}

The complex basis is defined by means of:
\begin{equation}
A = \frac{A_1+i A_2}{\sqrt{2}} \qquad\qquad \bar{A} = \frac{A_1-i A_2}{\sqrt{2}}
\end{equation}
In the light-cone gauge, it follows from eq. \eqref{ff} that:
\begin{align}
\label{fdef}
\nonumber
& f_{11}=-\partial_{+} \bar{A}\\
& f_{\dot{1}\dot{1}}=-\partial_{+} A
\end{align}
Hence, the operators in the standard basis are:
\begin{align}
\nonumber
&\mathbb{O}_{s} = \Tr \partial_{+} \bar{A}(x)(i\overrightarrow{\partial}_++i\overleftarrow{\partial}_+)^{s-2}C^{\frac{5}{2}}_{s-2}\left(\frac{\overrightarrow{\partial}_+-\overleftarrow{\partial}_+}{\overrightarrow{\partial}_++\overleftarrow{\partial}_+}\right)\partial_{+} {A}(x)\\\nonumber
&\tilde{\mathbb{O}}_{s} = \Tr \partial_{+} \bar{A}(x)(i\overrightarrow{\partial}_++i\overleftarrow{\partial}_+)^{s-2}C^{\frac{5}{2}}_{s-2}\left(\frac{\overrightarrow{\partial}_+-\overleftarrow{\partial}_+}{\overrightarrow{\partial}_++\overleftarrow{\partial}_+}\right)\partial_{+} {A}(x) \\\nonumber
&\mathbb{S}_{s} =\frac{1}{\sqrt{2}}\Tr \partial_{+} \bar{A}(x)(i\overrightarrow{\partial}_++i\overleftarrow{\partial}_+)^{s-2}C^{\frac{5}{2}}_{s-2}\left(\frac{\overrightarrow{\partial}_+-\overleftarrow{\partial}_+}{\overrightarrow{\partial}_++\overleftarrow{\partial}_+}\right)\partial_{+} \bar{A}(x)\\
&\bar{\mathbb{S}}_{s} =\frac{1}{\sqrt{2}}\Tr \partial_{+} A(x)(i\overrightarrow{\partial}_++i\overleftarrow{\partial}_+)^{s-2}C^{\frac{5}{2}}_{s-2}\left(\frac{\overrightarrow{\partial}_+-\overleftarrow{\partial}_+}{\overrightarrow{\partial}_++\overleftarrow{\partial}_+}\right)\partial_{+} A(x)
\end{align}
in the light-cone gauge, and analogously in the extended basis:
\begin{align}
\nonumber
&\mathbb{A}_{s} = \Tr  \bar{A}(x)(i\overrightarrow{\partial}_++i\overleftarrow{\partial}_+)^{s}C^{\frac{1}{2}}_{s}\left(\frac{\overrightarrow{\partial}_+-\overleftarrow{\partial}_+}{\overrightarrow{\partial}_++\overleftarrow{\partial}_+}\right) {A}(x)\\\nonumber
&\tilde{\mathbb{A}}_{s} = \Tr  \bar{A}(x)(i\overrightarrow{\partial}_++i\overleftarrow{\partial}_+)^{s}C^{\frac{1}{2}}_{s}\left(\frac{\overrightarrow{\partial}_+-\overleftarrow{\partial}_+}{\overrightarrow{\partial}_++\overleftarrow{\partial}_+}\right) {A}(x) \\\nonumber
&\mathbb{B}_{s} =\frac{1}{\sqrt{2}}\Tr  \bar{A}(x)(i\overrightarrow{\partial}_++i\overleftarrow{\partial}_+)^{s}C^{\frac{1}{2}}_{s}\left(\frac{\overrightarrow{\partial}_+-\overleftarrow{\partial}_+}{\overrightarrow{\partial}_++\overleftarrow{\partial}_+}\right)\bar{A}(x)\\
&\bar{\mathbb{B}}_{s} =\frac{1}{\sqrt{2}}\Tr  A(x)(i\overrightarrow{\partial}_++i\overleftarrow{\partial}_+)^{s}C^{\frac{1}{2}}_{s}\left(\frac{\overrightarrow{\partial}_+-\overleftarrow{\partial}_+}{\overrightarrow{\partial}_++\overleftarrow{\partial}_+}\right)A(x)
\end{align}

\section{Jacobi and Gegenbauer polynomials \label{appB}}

We work out the formulas for the Jacobi and Gegenbauer polynomials that are employed in the present paper. \par
For $x$ real, the Jacobi polynomials, $P^{(\alpha,\beta)}_l(x)$, admit the representation \cite{szego1959orthogonal}:
\begin{equation}
\label{jaco1}
P^{(\alpha,\beta)}_l(x) = \sum_{k = 0}^{l}{l+\alpha\choose k}{l+\beta\choose k+\beta}\left(\frac{x-1}{2}\right)^k\left(\frac{x+1}{2}\right)^{l-k}
\end{equation}
with $\alpha, \beta$ real and $l$ a natural number. Moreover, they satisfy the symmetry property:
\begin{align}
\label{jacox}
P^{(\alpha,\beta)}_l(-x) = (-1)^{l}P^{(\beta,\alpha)}_l(x)
\end{align} 
The Gegenbauer polynomials, $C^{\alpha'}_l(x)$, are a special case of the Jacobi polynomials:
\begin{equation} \label{GPol}
C^{\alpha'}_l(x) = \frac{\Gamma(l+2\alpha')\Gamma(\alpha'+\frac{1}{2})}{\Gamma(2\alpha')\Gamma(l+\alpha'+\frac{1}{2})}P_l^{(\alpha'-\frac{1}{2},\alpha'-\frac{1}{2})}(x)
\end{equation}
Therefore, they satisfy the symmetry property:
\begin{align}
\label{symmgegen}
C^{\alpha'}_l(-x) = (-1)^l C^{\alpha'}_l(x) 
\end{align}
From now on, we set:
\begin{equation}
x = \frac{b-a}{a+b}
\end{equation}
in such a way that:
\begin{equation}
\left(\frac{x-1}{2}\right)^k\left(\frac{x+1}{2}\right)^{l-k} = (-1)^{l-k} \frac{a^{l-k} b^k}{(a+b)^l}
\end{equation}
Hence, eq. (\ref{jaco1}) becomes:
\begin{equation}
\label{jaco2}
P^{(\alpha,\beta)}_l(x) = \sum_{k = 0}^{l}{l+\alpha\choose k}{l+\beta\choose k+\beta}(-1)^{l-k} \frac{a^{l-k} b^k}{(a+b)^l}
\end{equation}
Besides, setting $l = J-\alpha'+\frac{1}{2}$ in eq. \eqref{GPol} and $\alpha=\beta=\alpha'-\frac{1}{2}$ in eq. \eqref{jaco1}, we obtain: 
\begin{align}
C^{\alpha'}_{J-\alpha'+\frac{1}{2}}\left(x\right) =&\frac{\Gamma(J+\frac{1}{2}+\alpha')\Gamma(\alpha'+\frac{1}{2})}{\Gamma(2\alpha')\Gamma(J+1)}\nonumber\\
&\sum_{k=0}^{J-\alpha'+\frac{1}{2}} {J\choose k}{J\choose k+\alpha'-\frac{1}{2}}(-1)^{J-\alpha'+\frac{1}{2}-k} \frac{a^{J-\alpha'+\frac{1}{2}-k} b^k}{(a+b)^{J-\alpha'+\frac{1}{2}}}
\end{align}
Specializing the above equation to $J = s$ and $\alpha' = \frac{5}{2}$, we get:
\begin{equation}
\label{physicalgegen}
C^{\frac{5}{2}}_{s-2}\left(x\right) = \frac{\Gamma(s+3)\Gamma(3)}{\Gamma(5)\Gamma(s+1)}\sum_{k=0}^{s-2} {s\choose k}{s\choose k+2}(-1)^{s-k} \frac{a^{s-k-2} b^k}{(a+b)^{s-2}}
\end{equation}
Moreover, for $J = s$ and $\alpha' = \frac{1}{2}$, we obtain:
\begin{equation}
\label{physicalgegen2}
C^{\frac{1}{2}}_{s}\left(x\right) = \sum_{k=0}^{s} {s\choose k}{s\choose k}(-1)^{s-k} \frac{a^{s-k} b^k}{(a+b)^{s}}
\end{equation}
From now on, we restrict $\alpha, \beta$ to the natural numbers and, correspondingly, $\alpha'$ to the positive half-integers and $J$ to the natural numbers. \par
By employing the identity:
\begin{align}
{l+\alpha\choose k}{l+\beta\choose k+\beta}
=\frac{(l+\beta)!(l+\alpha)!}{l!(l+\alpha+\beta)!}{l\choose k}{l+\beta+\alpha\choose k+\beta}
\end{align}
it follows from eq. \eqref{jaco2} that:
\begin{equation}
\label{jaco3}
P^{(\alpha,\beta)}_l(x) = \frac{(l+\beta)!(l+\alpha)!}{l!(l+\alpha+\beta)!}\sum_{k = 0}^{l} {l\choose k}{l+\beta+\alpha\choose k+\beta}(-1)^{l-k} \frac{a^{l-k} b^k}{(a+b)^l}
\end{equation}
Corrispondingly, eq. \eqref{GPol} reads:
\begin{align}
C^{\alpha'}_l\left(x\right) = &\frac{\Gamma(l+2\alpha')\Gamma(\alpha'+\frac{1}{2})}{\Gamma(2\alpha')\Gamma(l+\alpha'+\frac{1}{2})}\frac{(l+\alpha'-\frac{1}{2})!(l+\alpha'-\frac{1}{2})!}{l!(l+2\alpha'-1)!}\\\nonumber
&\sum_{k = 0}^{l} {l\choose k}{l+2\alpha'-1\choose k+\alpha'-\frac{1}{2}}(-1)^{l-k} \frac{a^{l-k} b^k}{(a+b)^l}
\end{align}
which reduces to:
\begin{align} \label{bin}
C^{\alpha'}_l\left(x\right) = \frac{\Gamma(\alpha'+\frac{1}{2})\Gamma(l+\alpha'+\frac{1}{2})}{\Gamma(2\alpha')\Gamma(l+1)}
\sum_{k = 0}^{l} {l\choose k}{l+2\alpha'-1\choose k+\alpha'-\frac{1}{2}}(-1)^{l-k} \frac{a^{l-k} b^k}{(a+b)^l}
\end{align}
Specializing the above equation to $\alpha' = \frac{5}{2}$, we obtain:
\begin{align}
C^{\frac{5}{2}}_l\left(x\right) &= \frac{\Gamma(l+3)\Gamma(3)}{\Gamma(5)\Gamma(l+1)}\sum_{k=0}^{l} {l\choose k}{l+4\choose k+2}(-1)^{l-k} \frac{a^{l-k} b^k}{(a+b)^l} \nonumber \\
&= \frac{2(l+1)(l+2)}{4!}\sum_{k=0}^{l} {l\choose k}{l+4\choose k+2}(-1)^{l-k} \frac{a^{l-k} b^k}{(a+b)^l}
\end{align}
to be compared with eq. \eqref{physicalgegen}.

\section{Matching $2$- and $3$-point Minkowskian correlators with \cite{Kazakov:2012ar} \label{appcheck}}

We verify that our results for the $2$- and $3$-point Minkowskian correlators of the balanced operators with even collinear spin in the standard basis coincide with the ones in \cite{Kazakov:2012ar} up to the different normalization of the operators.\par
Starting from eq. (\ref{c2intro}):
\begin{align}
\mathcal{C}_{s}(x,y) = &\frac{1}{(4\pi^2)^2} \frac{N^2-1}{4} \frac{2^{2s+2}}{(4!)^2}(-1)^s(s-1)s(s+1)(s+2)(2s)!
\frac{(x-y)_+^{2s}}{(\rvert x-y\rvert^2)^{2s+2}}
\end{align}
we get:
\begin{align}
\langle \mathbb{O}_{s_1}(x) \mathbb{O}_{s_2}(y)\rangle=&\delta_{s_1 s_2}\frac{1}{(8\pi^2)^2} (N^2-1)2^{2s_1}(-1)^{s_1} \frac{1}{2^4}\frac{\Gamma(s_1+3)}{3^2\Gamma(s_1-1)}\frac{\Gamma(2s_1+2)}{2s_1+1}\frac{(x-y)_+^{2s_1}}{(\rvert x-y\rvert^2)^{2s_1+2}}
\end{align}
in terms of gamma functions.
Besides, we rewrite the above correlator in terms of $j=s-1$ \footnote{$j=s-1$ in this section should not be confused with the conformal spin in the rest of the present paper.} to match the notation of eq. (2.11) in \cite{Kazakov:2012ar}:
\begin{align}
\label{equiv}
\langle \mathbb{O}_{s_1}(x) \mathbb{O}_{s_2}(y)\rangle=&\delta_{j_1 j_2}\frac{1}{(8\pi^2)^2} (-1)^{j_1+1} (N^2-1)2^{2j_1-3} \frac{\Gamma(j_1+4)}{3^2\Gamma(j_1)}\frac{\Gamma(2j_1+4)}{j_1+\frac{3}{2}}\frac{(x-y)_+^{2j_1+2}}{(\rvert x-y\rvert^2)^{2j_1+4}}
\end{align}
This is the very same result in \cite{Kazakov:2012ar} up to the overall factor of $\sigma_{j_1}\sigma_{j_2}$, which is missing as -- contrary to eq. (2.2) in \cite{Kazakov:2012ar} -- we have defined the operators $ \mathbb{O}_{s}$ in eq. (\ref{basis}) without the factor of $\sigma_{j}$ in front.\par
Our $3$-point correlators of balanced operators with even collinear spin read in eq. \eqref{312}:
\begin{align}
\nonumber
&\langle {\mathbb{O}}_{s_1}(x){\mathbb{O}}_{s_2}(y){\mathbb{O}}_{s_3}(z)\rangle =-\frac{1}{(4\pi^2)^3}2\left(\frac{2}{4!}\right)^3\frac{N^2-1}{8}i^{s_1+s_2+s_3}2^{s_1+s_2+s_3}\\\nonumber
& (s_1+1)(s_1+2)(s_2+1)(s_2+2)(s_3+1)(s_3+2)\\\nonumber
&\sum_{k_1 = 0}^{s_1-2}\sum_{k_2 = 0}^{s_2-2}\sum_{k_3 = 0}^{s_3-2}{s_1\choose k_1}{s_1\choose k_1+2}{s_2\choose k_2}{s_2\choose k_2+2}{s_3\choose k_3}{s_3\choose k_3+2}\\\nonumber
&(s_1-k_1+k_2)!(s_2-k_2+k_3)!(s_3-k_3+k_1)!  \\
&\frac{(x-y)^{s_1-k_1+k_2}_+}{(\rvert x-y\rvert^2)^{s_1+1-k_1+k_2}}\frac{(y-z)^{s_2-k_2+k_3}_+}{(\rvert y-z\rvert^2)^{s_2+1-k_2+k_3}}\frac{(z-x)^{s_3-k_3+k_1}_+}{(\rvert z-x\rvert^2)^{s_3+1-k_3+k_1}}
\end{align}
Employing:
\begin{equation}
{s\choose k}{s\choose k+2}=\frac{s(s-1)}{(s+2)(s+1)}{s-2\choose k}{s+2\choose k+2}
\end{equation}
we obtain:
\begin{align}
\nonumber
&\langle {\mathbb{O}}_{s_1}(x){\mathbb{O}}_{s_2}(y){\mathbb{O}}_{s_3}(z)\rangle =-\frac{1}{(4\pi^2)^3}2\left(\frac{2}{4!}\right)^3\frac{N^2-1}{8}i^{s_1+s_2+s_3}2^{s_1+s_2+s_3}\\\nonumber
& s_1(s_1-1)s_2(s_2-1)s_3(s_3-1)\\\nonumber
&\sum_{k_1 = 0}^{s_1-2}\sum_{k_2 = 0}^{s_2-2}\sum_{k_3 = 0}^{s_3-2}{s_1-2\choose k_1}{s_1+2\choose k_1+2}{s_2-2\choose k_2}{s_2+2\choose k_2+2}{s_3-2\choose k_3}{s_3+2\choose k_3+2}\\\nonumber
&(s_1-k_1+k_2)!(s_2-k_2+k_3)!(s_3-k_3+k_1)!  \\
&\frac{(x-y)^{s_1-k_1+k_2}_+}{(\rvert x-y\rvert^2)^{s_1+1-k_1+k_2}}\frac{(y-z)^{s_2-k_2+k_3}_+}{(\rvert y-z\rvert^2)^{s_2+1-k_2+k_3}}\frac{(z-x)^{s_3-k_3+k_1}_+}{(\rvert z-x\rvert^2)^{s_3+1-k_3+k_1}}
\end{align}
that in terms of $j_1,j_2,j_3$ reads:
\begin{align}
\nonumber
&\langle {\mathbb{O}}_{s_1}(x){\mathbb{O}}_{s_2}(y){\mathbb{O}}_{s_3}(z)\rangle =-\frac{1}{(8\pi^2)^3}\frac{1}{2^2 3^3}(N^2-1)i^{j_1+j_2+j_3+3}2^{j_1+j_2+j_3}\\\nonumber
& j_1 (j_1+1)j_2(j_2+1)j_3(j_3+1)\\\nonumber
&\sum_{k_1 = 0}^{j_1-1}\sum_{k_2 = 0}^{j_2-1}\sum_{k_3 = 0}^{j_3-1}{j_1-1\choose k_1}{j_1+3\choose k_1+2}{j_2-1\choose k_2}{j_2+3\choose k_2+2}{j_3-1\choose k_3}{j_3+3\choose k_3+2}\\\nonumber
&(j_1+1-k_1+k_2)!(j_2+1-k_2+k_3)!(j_3+1-k_3+k_1)!  \\
&\frac{(x-y)^{j_1+1-k_1+k_2}_+}{(\rvert x-y\rvert^2)^{j_1+2-k_1+k_2}}\frac{(y-z)^{j_2+1-k_2+k_3}_+}{(\rvert y-z\rvert^2)^{j_2+2-k_2+k_3}}\frac{(z-x)^{j_3+1-k_3+k_1}_+}{(\rvert z-x\rvert^2)^{j_3+2-k_3+k_1}}
\end{align}
This is the very same result of eq. (2.22) in \cite{Kazakov:2012ar} up to the overall factor of $\sigma_{j_1} \sigma_{j_2} \sigma_{j_3}$, which is missing because of the aforementioned different normalization of the operators.

\section{Summation trick for $2$-point correlators \label{appA3}}

We compute the $2$-point correlators by means of the technique in \cite{Kazakov:2012ar}.

\subsection{Standard basis}

In the standard basis, we get:
\begin{align}
\nonumber
\langle \mathbb{O}_{s_1}(x) \mathbb{O}_{s_2}(y)\rangle &=  \mathcal{G}_{s_1-2}^{\frac{5}{2}}(\partial_{x_1^+},\partial_{x_2^+})\mathcal{G}_{s_2-2}^{\frac{5}{2}}(\partial_{y_1^+},\partial_{y_2^+})\\
&\langle\Tr f_{11}(x_1){f}_{\dot{1}\dot{1}}(y_2)\rangle\langle\Tr f_{11}(y_1){f}_{\dot{1}\dot{1}}(x_2)\rangle\Big\rvert_{x_1=x_2=x}^{y_1=y_2=y}
\end{align}
in the light-cone gauge. We restrict the correlators to $(x-y)_{\perp} = 0$, so that $\rvert x-y\rvert^2 = 2(x-y)_+(x-y)_-$. For $x_- > y_-$, we obtain:
\begin{equation}
\frac{\Gamma(k)}{(x-y)_-^k} = \int_{0}^{\infty}d\tau\, \tau^{k-1} e^{-\tau(x-y)_-}
\end{equation}
By the above formula, we convert derivatives into multiplications:
\begin{align}
\nonumber
&\partial_{x_1^+} \rightarrow -\tau_1\qquad \partial_{x_2^+}\rightarrow -\tau_2\\
&\partial_{y_1^+} \rightarrow \tau_2\qquad \partial_{y_2^+}\rightarrow \tau_1
\end{align}
By eq. (\ref{symmgegen}):
\begin{align}
C_{s_1-2}^{\frac{5}{2}}(-\frac{\tau_2-\tau_1}{\tau_1+\tau_2})=(-1)^{s_1}C_{s_1-2}^{\frac{5}{2}}(\frac{\tau_2-\tau_1}{\tau_1+\tau_2}) \nonumber \\
C_{s_2-2}^{\frac{5}{2}}(\frac{\tau_1-\tau_2}{\tau_2+\tau_1})=(-1)^{s_2}C_{s_2-2}^{\frac{5}{2}}(\frac{\tau_2-\tau_1}{\tau_1+\tau_2})
\end{align}
we obtain:
\begin{align}
&\langle \mathbb{O}_{s_1}(x) \mathbb{O}_{s_2}(y)\rangle= \frac{1}{(4\pi^2)^2} \frac{N^2-1}{4}i^{s_1+s_2-4}\frac{1}{4(x-y)^2_+} (-1)^{s_1+s_2}\\\nonumber
&\int_{0}^{\infty}d\tau_1d\tau_2\, (\tau_1+\tau_2)^{s_1+s_2-4}\tau_1^2 \tau_2^2 e^{-(\tau_1+\tau_2)(x-y)_-} C_{s_1-2}^{\frac{5}{2}}(\frac{\tau_2-\tau_1}{\tau_1+\tau_2})C_{s_2-2}^{\frac{5}{2}}(\frac{\tau_2-\tau_1}{\tau_1+\tau_2})
\end{align}
From the substitution:
\begin{equation}
\tau_1 = \tau \gamma \qquad \tau_2 = \tau (1-\gamma)
\end{equation}
it follows:
\begin{align}
&\langle \mathbb{O}_{s_1}(x) \mathbb{O}_{s_2}(y)\rangle= \frac{1}{(4\pi^2)^2} \frac{N^2-1}{4}i^{s_1+s_2-4}\frac{1}{4(x-y)^2_+} (-1)^{s_1+s_2}\\\nonumber
&\int_{0}^{\infty}d\tau \int_{0}^{1}d\gamma\, \tau^{s_1+s_2+1}\gamma^{2}(1-\gamma)^2 e^{-\tau(x-y)_-} C_{s_1-2}^{\frac{5}{2}}(1-2\gamma)C_{s_2-2}^{\frac{5}{2}}(1-2\gamma)
\end{align}
The $\tau$ integral is:
\begin{equation}
\int_{0}^{\infty}d\tau\, \tau^{s_1+s_2+1} e^{-\tau(x-y)_-} = \frac{\Gamma(s_1+s_2+2)}{(x-y)_-^{s_1+s_2+2}}
\end{equation}
while:
\begin{equation}
\int_{0}^{1}d\gamma\, \gamma^{2}(1-\gamma)^2 C_{s_1-2}^{\frac{5}{2}}(1-2\gamma)C_{s_2-2}^{\frac{5}{2}}(1-2\gamma)
\end{equation}
is rewritten as:
\begin{equation}
\int_{-1}^{1}\frac{du}{2}\, \left(\frac{1-u^2}{4}\right)^2 C_{s_1-2}^{\frac{5}{2}}(u)C_{s_2-2}^{\frac{5}{2}}(u)
\end{equation}
with $u = 1-2\gamma$. The orthogonality of Gegenbauer polynomials reads:
\begin{equation}
\int_{-1}^{1}dz\,(1-z^2)^{\alpha'-\frac{1}{2}} C_{n_1}^{\alpha'}(z)C_{n_2}^{\alpha'}(z) = \delta_{n_1 n_2}\frac{\pi 2^{1-2\alpha'} \Gamma(n_1+2\alpha')}{n_1!(n_1+\alpha')\Gamma(\alpha')^2}
\end{equation}
Hence, for $\alpha' = \frac{5}{2}$:
\begin{equation}
\int_{-1}^{1}du\, \left(1-u^2\right)^2 C_{s_1-2}^{\frac{5}{2}}(u)C_{s_2-2}^{\frac{5}{2}}(u)=\delta_{s_1 s_2} 2^{-4}\frac{16 (s_1+2)!}{9(s_1-2)!(2s_1+1)}
\end{equation}
Collecting all the factors, we get:
\begin{align}
\langle \mathbb{O}_{s_1}(x) \mathbb{O}_{s_2}(y)\rangle= &\delta_{s_1 s_2}\frac{1}{(4\pi^2)^2} \frac{N^2-1}{4}i^{s_1+s_2-4}(-1)^{s_1+s_2}\frac{1}{2^4}\frac{1 (s_1+2)!}{3^2(s_1-2)!}\frac{\Gamma(s_1+s_2+2)}{2s_1+1} \nonumber\\
& \frac{1}{4(x-y)^2_+(x-y)_-^{s_1+s_2+2}}
\end{align}
Going back outside the plane $(x-y)_{\perp} = 0$:
\begin{equation}
\frac{1}{4(x-y)^2_+(x-y)_-^{s_1+s_2+2}}  \rightarrow 2^{s_1+s_2}\frac{(x-y)_+^{s_1+s_2}}{(\rvert x-y\rvert^2)^{s_1+s_2+2}}
\end{equation}
we obtain:
\begin{align}
\label{resum}
\langle \mathbb{O}_{s_1}(x) \mathbb{O}_{s_2}(y)\rangle=&\delta_{s_1 s_2}\frac{1}{(4\pi^2)^2} \frac{N^2-1}{4}i^{s_1+s_2-4}2^{s_1+s_2} (-1)^{s_1+s_2}\frac{1}{2^43^2}\nonumber\\
&\frac{(s_1+2)!}{(s_1-2)!}\frac{\Gamma(s_1+s_2+2)}{2s_1+1}\frac{(x-y)_+^{s_1+s_2}}{(\rvert x-y\rvert^2)^{s_1+s_2+2}}\nonumber\\
=&\delta_{s_1 s_2}\frac{1}{(4\pi^2)^2} \frac{N^2-1}{4} \frac{2^{2s_1+2}}{(4!)^2}(-1)^{s_1}\nonumber\\
&(s_1-1)s_1(s_1+1)(s_1+2)(2s_1)!
\frac{(x-y)_+^{2s_1}}{(\rvert x-y\rvert^2)^{2s_1+2}}
\end{align}
Analogously, the above equation extends to the balanced operators $\tilde{\mathbb{O}}_{s}$ with odd $s$.

\subsection{Extended basis}

In the extended basis, we get:
\begin{align}
\nonumber
\langle \mathbb{A}_{s_1}(x) \mathbb{A}_{s_2}(y)\rangle &=  \mathcal{G}_{s_1}^{\frac{1}{2}}(\partial_{x_1^+},\partial_{x_2^+})\mathcal{G}_{s_2}^{\frac{1}{2}}(\partial_{y_1^+},\partial_{y_2^+})\\
&\langle\Tr \partial_{+}^{-1}f_{11}(x_1)\partial_{+}^{-1} f_{\dot{1}\dot{1}}(y_2)\rangle\langle\Tr \partial_{+}^{-1}f_{11}(y_1)\partial_{+}^{-1} f_{\dot{1}\dot{1}}(x_2)\rangle\Big\rvert_{x_1=x_2=x}^{y_1=y_2=y}
\end{align}
in the light-cone gauge. Performing the Wick contractions, we obtain:
\begin{align}
\langle \mathbb{A}_{s_1}(x) \mathbb{A}_{s_2}(y)\rangle &= \frac{1}{(4\pi^2)^2} \frac{N^2-1}{4}\mathcal{G}_{s_1}^{\frac{1}{2}}(\partial_{x_1^+},\partial_{x_2^+})\mathcal{G}_{s_2}^{\frac{1}{2}}(\partial_{y_1^+},\partial_{y_2^+})\frac{1}{\rvert x_1-y_2\rvert^2}\frac{1}{\rvert y_1-x_2\rvert^2}\Big\rvert_{x_1=x_2=x}^{y_1=y_2=y}
\end{align}
Similarly, we restrict the correlators to $(x-y)_{\perp} = 0$. 
By eq. (\ref{symmgegen}):
\begin{align}
C_{s_1}^{\frac{1}{2}}(-\frac{\tau_2-\tau_1}{\tau_1+\tau_2})=(-1)^{s_1}C_{s_1}^{\frac{1}{2}}(\frac{\tau_2-\tau_1}{\tau_1+\tau_2}) \nonumber \\
C_{s_2}^{\frac{1}{2}}(\frac{\tau_1-\tau_2}{\tau_2+\tau_1})=(-1)^{s_2}C_{s_2}^{\frac{1}{2}}(\frac{\tau_2-\tau_1}{\tau_1+\tau_2})
\end{align}
we get:
\begin{align}
&\langle \mathbb{A}_{s_1}(x) \mathbb{A}_{s_2}(y)\rangle = \frac{1}{(4\pi^2)^2} \frac{N^2-1}{4}i^{s_1+s_2}\frac{1}{4(x-y)^2_+} (-1)^{s_1+s_2}\\\nonumber
&\int_{0}^{\infty}d\tau_1d\tau_2\, (\tau_1+\tau_2)^{s_1+s_2} e^{-(\tau_1+\tau_2)(x-y)_-} C_{s_1}^{\frac{1}{2}}(\frac{\tau_2-\tau_1}{\tau_1+\tau_2})C_{s_2}^{\frac{1}{2}}(\frac{\tau_2-\tau_1}{\tau_1+\tau_2})
\end{align}
Changing variables:
\begin{equation}
	\tau_1 = \tau \gamma,\qquad \tau_2 = \tau (1-\gamma)
\end{equation}
we obtain:
\begin{align}
\langle \mathbb{A}_{s_1}(x) \mathbb{A}_{s_2}(y)\rangle = &\frac{1}{(4\pi^2)^2} \frac{N^2-1}{4}i^{s_1+s_2}\frac{1}{4(x-y)^2_+} (-1)^{s_1+s_2}\\\nonumber
&\int_{0}^{\infty}d\tau \int_{0}^{1}d\gamma\, \tau^{s_1+s_2+1} e^{-\tau(x-y)_-} C_{s_1}^{\frac{1}{2}}(1-2\gamma)C_{s_2}^{\frac{1}{2}}(1-2\gamma)
\end{align}
The $\tau$ integral is:
\begin{equation}
\int_{0}^{\infty}d\tau\, \tau^{s_1+s_2+1} e^{-\tau(x-y)_-} = \frac{\Gamma(s_1+s_2+2)}{(x-y)_-^{s_1+s_2+2}}
\end{equation}
while:
\begin{equation}
\int_{0}^{1}d\gamma\,  C_{s_1}^{\frac{1}{2}}(1-2\gamma)C_{s_2}^{\frac{1}{2}}(1-2\gamma)
\end{equation}
is rewritten as:
\begin{equation}
\int_{-1}^{1}\frac{du}{2}\, C_{s_1}^{\frac{1}{2}}(u)C_{s_2}^{\frac{1}{2}}(u)
\end{equation}
with $u = 1-2\gamma$. The orthogonality of the Gegenbauer polynomials reads:
\begin{equation}
\int_{-1}^{1}dz\,(1-z^2)^{\alpha'-\frac{1}{2}} C_{s_1}^{\alpha'}(z)C_{s_2}^{\alpha'}(z) = \delta_{s_1s_2}\frac{\pi 2^{1-2\alpha'} \Gamma(s_1+2\alpha')}{s_1!(s_1+\alpha')\Gamma(\alpha')^2}
\end{equation}
Hence, for $\alpha' = \frac{1}{2}$:
\begin{equation}
\int_{-1}^{1}\frac{du}{2}\, C_{s_1}^{\frac{1}{2}}(u)C_{s_2}^{\frac{1}{2}}(u) = \delta_{s_1 s_2}\frac{1}{2s_1+1}\,
\end{equation}
Collecting all the factors, we get:
\begin{align}
\langle \mathbb{A}_{s_1}(x) \mathbb{A}_{s_2}(y)\rangle = &\delta_{s_1 s_2}\frac{1}{(4\pi^2)^2} \frac{N^2-1}{4}i^{s_1+s_2} (-1)^{s_1+s_2}\frac{\Gamma(s_1+s_2+2)}{2s_1+1} \nonumber \\
 &\frac{1}{4(x-y)^2_+(x-y)_-^{s_1+s_2+2}}
\end{align}
Going back outside the plane $(x-y)_{\perp} = 0$:
\begin{equation}
\frac{1}{4(x-y)^2_+(x-y)_-^{s_1+s_2+2}}  \rightarrow 2^{s_1+s_2}\frac{(x-y)_+^{s_1+s_2}}{(\rvert x-y\rvert^2)^{s_1+s_2+2}}
\end{equation}
we obtain:
\begin{align}
\langle \mathbb{A}_{s_1}(x) \mathbb{A}_{s_2}(y)\rangle = &\delta_{s_1 s_2}\frac{1}{(4\pi^2)^2} \frac{N^2-1}{4}i^{s_1+s_2}(-1)^{s_1+s_2}2^{s_1+s_2} \frac{\Gamma(s_1+s_2+2)}{2s_1+1}\frac{(x-y)_+^{s_1+s_2}}{(\rvert x-y\rvert^2)^{s_1+s_2+2}}\nonumber\\
= &\delta_{s_1 s_2}\frac{1}{(4\pi^2)^2}\frac{N^2-1}{4}(-1)^{s_1}  2^{2s_1}(2s_1)!
\frac{(x-y)_+^{2s_1}}{(\rvert x-y\rvert^2)^{2s_1+2}}
\end{align}
Analogously, the above equation extends to the balanced operators $\tilde{\mathbb{A}}_{s}$ with odd $s$.
\bibliographystyle{JHEP}
\bibliography{mybib} 

\providecommand{\href}[2]{#2}\begingroup\raggedright\begin{thebibliography}{10}

\bibitem{Kazakov:2012ar}
V.~Kazakov and E.~Sobko, \emph{{Three-point correlators of twist-2 operators in
  N=4 SYM at Born approximation}},
  \href{http://dx.doi.org/10.1007/JHEP06(2013)061}{\emph{JHEP} {\bf 06} (2013)
  061}, [\href{https://arxiv.org/abs/1212.6563}{{\tt 1212.6563}}].

\bibitem{Beisert:2004fv}
N.~Beisert, G.~Ferretti, R.~Heise and K.~Zarembo, \emph{{One-loop QCD spin
  chain and its spectrum}},
  \href{http://dx.doi.org/10.1016/j.nuclphysb.2005.04.004}{\emph{Nucl. Phys. B}
  {\bf 717} (2005) 137--189}, [\href{https://arxiv.org/abs/hep-th/0412029}{{\tt
  hep-th/0412029}}].

\bibitem{Robertson:1990bf}
D.~G. Robertson and F.~Wilczek, \emph{{Anomalous dimensions of anisotropic
  gauge theory operators}},
  \href{http://dx.doi.org/10.1016/0370-2693(90)90731-K}{\emph{Phys. Lett. B}
  {\bf 251} (1990) 434--438}.

\bibitem{Bochicchio:2013tfa}
M.~Bochicchio and S.~P. Muscinelli, \emph{{Ultraviolet asymptotics of glueball
  propagators}}, \href{http://dx.doi.org/10.1007/JHEP08(2013)064}{\emph{JHEP}
  {\bf 08} (2013) 064}, [\href{https://arxiv.org/abs/1304.6409}{{\tt
  1304.6409}}].

\bibitem{Bochicchio:2013eda}
M.~Bochicchio, \emph{{Glueball and meson propagators of any spin in large-N
  QCD}}, \href{http://dx.doi.org/10.1016/j.nuclphysb.2013.07.023}{\emph{Nucl.
  Phys. B} {\bf 875} (2013) 621--649},
  [\href{https://arxiv.org/abs/1305.0273}{{\tt 1305.0273}}].

\bibitem{Bochicchio:2016toi}
M.~Bochicchio, \emph{{An asymptotic solution of Large-N QCD, for the glueball
  and meson spectrum and the collinear S-matrix}},
  \href{http://dx.doi.org/10.1063/1.4949387}{\emph{AIP Conf. Proc.} {\bf 1735}
  (2016) 030004}.

\bibitem{Braun:2003rp}
V.~M. Braun, G.~P. Korchemsky and D.~Mueller, \emph{{The Uses of conformal
  symmetry in QCD}},
  \href{http://dx.doi.org/10.1016/S0146-6410(03)90004-4}{\emph{Prog. Part.
  Nucl. Phys.} {\bf 51} (2003) 311--398},
  [\href{https://arxiv.org/abs/hep-ph/0306057}{{\tt hep-ph/0306057}}].

\bibitem{Braun2}
V.~M. Braun, A.~N. Manashov, S.~Moch and M.~Strohmaier, \emph{{Three-loop
  evolution equation for flavor-nonsinglet operators in off-forward
  kinematics}}, \href{http://dx.doi.org/10.1007/JHEP06(2017)037}{\emph{JHEP}
  {\bf 06} (2017) 037}, [\href{https://arxiv.org/abs/1703.09532}{{\tt
  1703.09532}}].

\bibitem{Braun3}
V.~M. Braun, Y.~Ji and A.~N. Manashov, \emph{{Two-loop evolution equation for
  the B-meson distribution amplitude}},
  \href{http://dx.doi.org/10.1103/PhysRevD.100.014023}{\emph{Phys. Rev. D} {\bf
  100} (2019) 014023}, [\href{https://arxiv.org/abs/1905.04498}{{\tt
  1905.04498}}].

\bibitem{Ohrndorf:1981qv}
T.~Ohrndorf, \emph{{Constraints From Conformal Covariance on the Mixing of
  Operators of Lowest Twist}},
  \href{http://dx.doi.org/10.1016/0550-3213(82)90542-9}{\emph{Nucl. Phys. B}
  {\bf 198} (1982) 26--44}.

\bibitem{conformalops}
N.~Craigie, V.~Dobrev and I.~Todorov, \emph{Conformally covariant composite
  operators in quantum chromodynamics},
  \href{http://dx.doi.org/https://doi.org/10.1016/0003-4916(85)90118-6}{\emph{Annals
  of Physics} {\bf 159} (1985) 411--444}.

\bibitem{BUKHVOSTOV1985601}
A.~Bukhvostov, G.~Frolov, L.~Lipatov and E.~Kuraev, \emph{Evolution equations
  for quasi-partonic operators},
  \href{http://dx.doi.org/https://doi.org/10.1016/0550-3213(85)90628-5}{\emph{Nucl.
  Phys. B} {\bf 258} (1985) 601 -- 646}.

\bibitem{Belitsky:2003sh}
A.~V. Belitsky, S.~E. Derkachov, G.~P. Korchemsky and A.~N. Manashov,
  \emph{{Superconformal operators in N=4 superYang-Mills theory}},
  \href{http://dx.doi.org/10.1103/PhysRevD.70.045021}{\emph{Phys. Rev. D} {\bf
  70} (2004) 045021}, [\href{https://arxiv.org/abs/hep-th/0311104}{{\tt
  hep-th/0311104}}].

\bibitem{Belitsky:2004sc}
A.~V. Belitsky, S.~E. Derkachov, G.~P. Korchemsky and A.~N. Manashov,
  \emph{{Dilatation operator in (super-)Yang-Mills theories on the
  light-cone}},
  \href{http://dx.doi.org/10.1016/j.nuclphysb.2004.11.034}{\emph{Nucl. Phys. B}
  {\bf 708} (2005) 115--193}, [\href{https://arxiv.org/abs/hep-th/0409120}{{\tt
  hep-th/0409120}}].

\bibitem{Braun:2008ia}
V.~M. Braun, A.~N. Manashov and J.~Rohrwild, \emph{{Baryon Operators of Higher
  Twist in QCD and Nucleon Distribution Amplitudes}},
  \href{http://dx.doi.org/10.1016/j.nuclphysb.2008.08.012}{\emph{Nucl. Phys. B}
  {\bf 807} (2009) 89--137}, [\href{https://arxiv.org/abs/0806.2531}{{\tt
  0806.2531}}].

\bibitem{Braun1}
V.~M. Braun, Y.~Ji and A.~N. Manashov, \emph{{Two-photon processes in conformal
  QCD: Resummation of the descendants of leading-twist operators}},
  \href{http://dx.doi.org/10.1007/JHEP03(2021)051}{\emph{JHEP} {\bf 51} (2021)
  51}, [\href{https://arxiv.org/abs/2011.04533}{{\tt 2011.04533}}].

\bibitem{Dreiner:2008tw}
H.~K. Dreiner, H.~E. Haber and S.~P. Martin, \emph{{Two-component spinor
  techniques and Feynman rules for quantum field theory and supersymmetry}},
  \href{http://dx.doi.org/10.1016/j.physrep.2010.05.002}{\emph{Phys. Rept.}
  {\bf 494} (2010) 1--196}, [\href{https://arxiv.org/abs/0812.1594}{{\tt
  0812.1594}}].

\bibitem{szego1959orthogonal}
G.~Szeg{\"o}, \emph{Orthogonal Polynomials}, vol.~23 of \emph{American
  Mathematical Society colloquium publications}.
\newblock American Mathematical Society, 1959.

\end{thebibliography}\endgroup
\end{document}